\documentclass[longauth]{aa} 
\pdfoutput=1
\usepackage{graphicx}
\usepackage{longtable,rotating}
\usepackage{booktabs}
\usepackage{amsmath}
\usepackage{natbib}
\usepackage{placeins}
\usepackage{color}

\usepackage{txfonts}
\def\ltsima{$\; \buildrel < \over \sim \;$}
\def\simlt{\lower.5ex\hbox{\ltsima}}
\def\gtsima{$\; \buildrel > \over \sim \;$}
\def\simgt{\lower.5ex\hbox{\gtsima}}

\begin{document}
    \title{A census of dense cores in the Aquila cloud complex: SPIRE/PACS observations from the Herschel\thanks{Herschel is an ESA space observatory
           with science instruments provided by European-led Principal Investigator consortia and with important participation from NASA.} 
           Gould Belt survey}

   \subtitle{}

  \author{
           V. K\"onyves\inst{1,2}
          \and
           Ph. Andr\'e\inst{1}
	  \and 
	   A. Men'shchikov\inst{1}
          \and
           P. Palmeirim\inst{1}	  	  
	  \and
	   D. Arzoumanian\inst{2,1}
	  \and 
	   N. Schneider\inst{3,4,1}
	  \and
	   A. Roy\inst{1}
	  \and
	   P. Didelon\inst{1}
	  \and
	   A. Maury\inst{1}
	  \and
	   Y. Shimajiri\inst{1}
	  \and
	   J. Di Francesco\inst{5,6}
          \and
	   S. Bontemps\inst{3,4}
	  \and
	   N. Peretto\inst{7,1}
	  \and 
	   M. Benedettini\inst{8}
	  \and
	  J.-Ph. Bernard\inst{9,10}
          \and
	   D. Elia\inst{8}
	  \and
	   M. J. Griffin\inst{7}
	  \and
	   T. Hill\inst{11,1}
	  \and
	   J. Kirk\inst{12}
	  \and
	   B. Ladjelate\inst{1}
	  \and
	   K. Marsh\inst{7}
	  \and
	   P. G. Martin\inst{13}
	  \and
	   F. Motte\inst{1}
	  \and
	   Q. Nguyen Luong\inst{13,14,1}
	  \and	  
	   S. Pezzuto\inst{8}
	  \and
	   H. Roussel\inst{15}
	  \and
	   K. L. J. Rygl\inst{16,17}
	  \and
	   S. I. Sadavoy\inst{18,5,6}
	  \and
	   E. Schisano\inst{8}
	  \and
	   L. Spinoglio\inst{8}
	  \and
	   D. Ward-Thompson\inst{12}	 
	  \and
	   G. J. White\inst{19,20}
          }

   \institute{Laboratoire AIM, CEA/DSM-CNRS-Universit\'e Paris Diderot, IRFU/Service d'Astrophysique, CEA Saclay,
             91191 Gif-sur-Yvette, France
             \email{vera.konyves@cea.fr, pandre@cea.fr}
         \and
	     Institut d'Astrophysique Spatiale, UMR8617, CNRS/Universit\'e Paris-Sud 11, 91405 Orsay, France
         \and
             Univ. Bordeaux, LAB, UMR5804, F-33270 Floirac, France 
         \and
             CNRS, LAB, UMR5804, F-33270 Floirac, France 
         \and 
             Department of Physics and Astronomy, University of Victoria, P.O. Box 355, STN CSC, Victoria, BC, V8W 3P6, Canada
         \and
             National Research Council Canada, 5071 West Saanich Road, Victoria, BC, V9E 2E7, Canada
         \and
             School of Physics \& Astronomy, Cardiff University, The Parade, Cardiff CF24 3AA, UK
         \and
             Istituto di Astrofisica e Planetologia Spaziali-INAF, Via Fosso del Cavaliere 100, I-00133 Roma, Italy
         \and
	     CNRS, IRAP, 9 Av.~colonel Roche, BP 44346, F-31028 Toulouse cedex 4, France
         \and    
	     Universit\'e de Toulouse, UPS-OMP, IRAP, F-31028 Toulouse cedex 4, France
	 \and
	     Joint ALMA Observatory, Alonso de Cordova 3107, Vitacura, Santiago, Chile
         \and
	     Jeremiah Horrocks Institute, University of Central Lancashire, Preston, Lancashire, PR1 2HE, UK
         \and
	     Canadian Institute for Theoretical Astrophysics, University of Toronto, 60 St. George Street, Toronto, ON M5S 3H8, Canada
	 \and
             National Astronomical Observatory of Japan, 2-21-1 Osawa, Mitaka, Tokyo 181-8588, Japan
         \and
	     Institut d'Astrophysique de Paris, Sorbonne Universit\'es, UPMC Univ. Paris 06, CNRS UMR 7095, 75014 Paris, France
         \and
	     Scientific Support Office, Directorate of Science and Robotic Exploration, European Space Research and Technology Centre (ESA/ESTEC), 
	     Keplerlaan 1, 2201 AZ Noordwijk, The Netherlands
         \and
	     INAF-IRA, Via P. Gobetti 101, I-40129 Bologna, Italy
	 \and	 
             Max-Planck-Institut f\"ur Astronomie (MPIA), K\"onigstuhl 17, D-69117, Heidelberg, Germany	     
         \and 
             RALSpace, The Rutherford Appleton Laboratory, Chilton, Didcot, Oxfordshire, OX11 0QX, England
         \and
             Department of Physics and Astronomy, The Open University, Walton Hall, Milton Keynes, MK7 6AA, England
	     }

   \date{Received 11 February 2015; Accepted 13 July 2015}

   \abstract{
    We present and discuss the results of the {\it Herschel} Gould Belt survey (HGBS) observations in a $\sim$11~deg$^2$ area of the Aquila molecular cloud 
    complex at $d \sim260$~pc, imaged with the SPIRE and PACS photometric cameras in parallel mode from 70~$\mu$m to 500~$\mu$m. 
    Using the multi-scale, multi-wavelength source extraction method \textsl{getsources}, we identify a complete sample of starless dense cores and embedded 
    (Class 0-I) protostars in this region, and analyze their global  properties and spatial distributions. We find a total of  651 starless cores, 
    $\sim\, 60\% \pm 10\% $ of which are gravitationally bound prestellar cores, and they will likely form stars in the future. We also detect 58 protostellar cores. 
    The core mass function (CMF) derived for the large population of prestellar cores is very similar in shape to the stellar initial mass function (IMF), 
    confirming earlier findings on a much stronger statistical basis and supporting the view that there is a 
    close physical link between the stellar IMF and the prestellar CMF. 
    The global shift in mass scale observed between the CMF and the IMF is consistent with a typical star formation efficiency of $\sim$40\% 
    at the level of an individual core. 
    By comparing the numbers of starless cores in various density bins to the number of young stellar objects (YSOs), we estimate that the lifetime of prestellar 
    cores is $\sim 1$~Myr, which is typically $\sim 4$ times longer than the core free-fall time, and that it decreases with average core density.  
    We find a strong correlation between the spatial distribution of prestellar cores and the densest filaments observed in the Aquila complex. 
    About 90\% of the {\it Herschel}-identified prestellar cores are located above a background column density corresponding to $A_{\rm V} \sim 7$, 
    and $\sim 75\%$ of them lie within filamentary structures with supercritical masses per unit length $\simgt$16~$M_\odot$/pc.
    These findings support a picture wherein the cores making up the peak of the CMF (and probably responsible for the base of the IMF) 
    result primarily from the gravitational fragmentation of marginally supercritical filaments. 
    Given that filaments appear to dominate the mass budget of dense gas at $A_{\rm V} > 7$, our findings also suggest 
    that the physics of prestellar core formation within filaments is responsible for a characteristic ``efficiency'' 
    ${\rm SFR}/M_{\rm dense} \sim 5^{+2}_{-2} \times 10^{-8}\, {\rm yr}^{-1}$ for the star formation process in dense gas.
    }


   \titlerunning{\emph{Herschel} Gould Belt survey for prestellar cores in Aquila}

   \maketitle
%

\section{Introduction: The {\it Herschel} Gould Belt survey}\label{sec:intro}

Understanding how dense cloud cores and protostars form out of the diffuse interstellar medium (ISM) 
is a fundamental question in contemporary astrophysics 
(e.g., \citealp{McKeeOstriker2007} and other recent reviews in \citealp{Beuther+2014}).
Much progress is being made on this front thanks to imaging surveys with the {\it Herschel} Space Observatory 
\citep{Pilbratt2010}. Its far-infrared and submillimeter cameras PACS \citep{Poglitsch2010}
and SPIRE \citep{Griffin2010} provide unprecedented sensitivity and dynamic range at 
wavelengths around the peak of the spectral energy distributions (SEDs) of
starless cores and protostars.

In particular, the bulk of nearby ($d \simlt 500$~pc) molecular clouds, 
mostly located in the Gould Belt \citep[e.g.,][]{Guillot2001, PerrotGrenier2003}, have been imaged 
at five wavelengths between 70~$\mu$m and 500~$\mu$m as part of  
the {\it Herschel} Gould Belt survey \citep[HGBS --][]{Andre+2010}.  
Observationally, the molecular clouds of the Gould Belt are the best laboratories at our disposal for investigating the star 
formation process in detail, at least as far as low-mass stars are concerned. 
They are the only clouds for which the $\sim 15\arcsec $ angular resolution of {\it Herschel} around $\lambda \sim 200\, \mu$m 
is sufficient to resolve the typical Jeans length $\sim 0.03$~pc in cluster-forming clumps \citep[e.g.,][]{Larson1985, Myers1998}. 

The $15$ or so nearby clouds covered by the HGBS  span a wide range of physical and environmental conditions, 
from very active, cluster-forming complexes 
such as the Orion A \& B giant molecular clouds (GMCs) or the Aquila Rift cloud complex \citep[e.g.,][]{Dame+2001, Gutermuth+2008} 
to quiescent regions 
with no star formation activity whatsoever, such as the Polaris flare translucent cloud \citep[e.g.,][]{Heithausen+2002,Ward+2010}.
The total surface area covered by the survey exceeds 160~deg$^2$ (cf. http://gouldbelt-herschel.cea.fr/ for the list of 
all target regions). The HGBS will eventually provide an essentially complete census of (solar-type) prestellar cores and 
Class~0 protostars with well-characterized luminosity and mass functions in most nearby star-forming regions. 

The main scientific goals of the HGBS are to clarify the nature of the relationship between the prestellar core mass 
function (CMF) and the stellar initial mass function (IMF) and to elucidate the physical mechanisms responsible for 
the growth of structure in the cold interstellar medium (ISM), leading to the formation of prestellar cores and protostars 
in molecular clouds. 

Initial results from the HGBS have already been presented in several ``first-look'' papers   
and may be summarized as follows. The HGBS observations  
confirm the omnipresence of  filaments in nearby molecular clouds 
and suggest an intimate connection between the filamentary structure 
of the cold ISM and the formation process of prestellar cores \citep{Andre+2010, Menshchikov+2010}.
While molecular clouds were already known to exhibit large-scale filamentary 
structures long before {\it Herschel} \citep[e.g.,][]{SchneiderElmegreen1979, Hartmann2002, 
Myers2009}, the {\it Herschel} observations from 
the HGBS \citep[e.g.,][]{Menshchikov+2010, MAMD+2010} 
and other imaging surveys such as HiGAL \citep{Molinari+2010,Schisano+2014}, HOBYS  \citep{Motte+2010, Hill+2011}, 
and EPoS \citep{Henning+2010} now demonstrate that these filaments are truly ubiquitous in the cold ISM, 
present a high degree of universality \citep[e.g.,][]{Arzoumanian+2011}, and likely play a central role in the star formation 
process \citep[see][for a recent review]{Andre+2014}. 
In any given cloud, {\it Herschel} imaging reveals a whole network of filaments, 
and a detailed analysis of the radial column density profiles 
of the nearby, resolved filaments observed in the HGBS
shows that they are characterized by a very narrow distribution of central widths with a typical full width at half maximum (FWHM) 
value $\sim 0.1$~pc and a dispersion of less than a factor of 2 
\citep{Arzoumanian+2011, Palmeirim+2013, AdOliveira+2014}. 
Other groups have reported results in broad agreement with our HGBS finding of a common filament width. 
\citet{Juvela+2012} found a typical FWHM width of $\sim\, $0.2--0.3~pc for a number of filaments mapped 
as part of the {\it Herschel} ``Galactic Cold Cores'' project in clouds with (rather uncertain) distances ranging from $\sim 100\, $~pc to a few kpc.
\citet{Ysard+2013} reported a mean FWHM width $\sim\, $0.1~pc for the L1506 filament in Taurus but found significant variations -- by up to a factor of $\sim 2$ 
on either side of the mean width -- along the length of the filament. 
\citet{Smith+2014} explored filament properties in a set of numerical hydrodynamic simulations and 
found a range of filament widths rather than a constant value. Recent magneto-hydrodynamic (MHD) simulations by \citet{Ntormousi+2015}, however,  
suggest that non-ideal MHD turbulence can account for the properties of observed filaments much better than hydrodynamic turbulence does 
\citep[see also][]{Hennebelle2013}.

The origin of the common inner width of interstellar filaments in nearby clouds 
is not yet well understood.
A possible interpretation is that filaments result 
from planar intersecting shock waves due to supersonic interstellar turbulence \citep[e.g.][]{Pudritz+2013}.
and that the filament width corresponds to the sonic scale below which the turbulence becomes subsonic in diffuse, non-star-forming molecular gas 
\citep[cf.][]{Padoan+2001}. 
Alternatively, a characteristic width may arise if interstellar filaments are formed as quasi-equilibrium structures 
in pressure balance with a typical ambient ISM pressure $P_{\rm ext} {\sim} 2$$-$5$\times$$10^4 \, \rm{K\, cm}^{-3} $ 
\citep[][S. Inutsuka, private communication]{FischeraMartin2012}. 
Yet another possibility is that the filament inner width may be set by the dissipation mechanism of MHD waves 
due to ion-neutral friction \citep{Hennebelle2013}.

The early results from the HGBS further suggest that prestellar cores and protostars form primarily in the densest filaments 
\citep[e.g.,][]{Andre+2010, Polychroni+2013}, 
for which the mass per unit length exceeds the critical line mass of nearly isothermal, long cylinders  
\citep[e.g.,][]{InutsukaMiyama1997}, $M_{\rm line, crit} = 2\, c_s^2/G \sim 16\, M_\odot$/pc,  
where $c_{\rm s} \sim 0.2$~km/s is the isothermal sound speed for molecular gas at $T \sim 10$~K. 
They also confirm the existence of a close relationship between the prestellar CMF  
and the stellar IMF in the regime of low to intermediate stellar masses \citep[$\sim \, $0.1--5$\, M_\odot $ --][]{Konyves+2010}.
These {\it Herschel} findings support a scenario according to which  
the formation of solar-type stars occurs in two main steps \citep{Andre+2014}: 
first, the dissipation of kinetic energy in 
large-scale magneto-hydrodynamic (MHD) flows (turbulent or not) generates a quasi-universal web-like 
filamentary structure in the ISM; second, the densest filaments fragment into 
prestellar cores (and ultimately protostars) by gravitational instability. 

In this paper, we present the ``first-generation'' catalog of dense cores obtained from HGBS data in the Aquila Rift cloud complex
and discuss the global properties of these dense cores in relation to the filamentary structure of the complex. 
In particular, we use these results to quantify the role of filaments in the star formation process.  
The present study extends and reinforces our early {\it Herschel} findings in Aquila \citep[][]{Konyves+2010,Andre+2010} on the basis 
of a more advanced examination of the data with improved data reduction, source extraction, and source characterization. 
The paper is organized as follows. 
Section~\ref{sec:AqlRift} introduces the Aquila Rift region.
Section~\ref{sec:obs_data} provides details about the {\it Herschel} imaging observations and the data reduction.
Section~\ref{sec:res_analys} presents the dust temperature and column density maps derived from {\it Herschel} data, describes  
the filamentary structure seen in these maps, and explains how dense cores were extracted, prestellar cores selected, 
and their properties measured from the maps. 
In Sect.~\ref{sec:discuss}, we discuss estimates of prestellar core lifetimes,  
the observational evidence of a column density threshold for prestellar core formation, 
the spatial distribution of extracted dense cores, and the strong connection with the filamentary 
structure of the Aquila cloud. 
We also compare the CMF of the Aquila sample of prestellar cores with the IMF, and link the global star formation rate 
of the complex with the total mass of dense gas above the column density threshold for star formation.
Finally, Sect.~\ref{sec:conclusions} concludes the paper by summarizing the HGBS results in the Aquila region and 
discussing possible implications for our understanding of star formation on GMC scales.

\section{The Aquila Rift region}\label{sec:AqlRift}

The Aquila Rift molecular cloud complex corresponds to a large extinction feature (see Prato et al. 2008), located 
above the Galactic plane ($b \simlt 4^\circ$) at galactic longitudes between $l=30^\circ$ and $l=50^\circ$. 
The portion of the cloud complex mapped with {\it Herschel} as part of the HGBS 
corresponds to the western high-extinction area of the Aquila Rift at $l<35^\circ$  \citep[see][]{Bontemps+2010}.

While the northern part of the Aquila high-extinction area harbors the well-documented Serpens Main star-forming region, 
the properties of the southern part  (the focus of the present paper) remained largely unexplored until 
{\it Spitzer} infrared observations \citep[e.g.,][]{Gutermuth+2008}. This extinction-defined area \citep[see][]{Bontemps+2010}, rich in 
gas but initially thought to be almost devoid of star formation \citep{Prato+2008}, is now known to harbor two cluster-forming clumps \citep{Maury+2011}:  
Serpens South, a young protostellar cluster showing very active recent star formation and   
embedded in a dense filamentary cloud 
\citep{Gutermuth+2008, Bontemps+2010, Nakamura+2011, Teixeira+2012, Friesen+2013, KirkH+2013, Tanaka+2013}, and 
W40 a young star cluster associated with the eponymous H{\small II} region, also known as Sharpless~2-64 
\citep{Smith+1985, Vallee1987, Kuhn+2010, Pirogov+2013}.

Whether or not the southern part of the Aquila high-extinction region and the Serpens Main cloud are at the same distance is still a 
matter of debate \citep{Bontemps+2010, Maury+2011, Loinard2013}. 
Based on stellar photometry, \citet{Straizys+2003} concluded that the front edge of the Aquila molecular cloud was at 255$\pm$55~pc. 
Using VLBI observations, however, \citet{Dzib+2010} measured the trigonometric parallax of the binary system EC95 in the Serpens 
Main region and obtained a distance of 415$\pm$15~pc.
From the extinction maps obtained by \citet{Bontemps+2010}, the respective extinction features toward the eastern Aquila 
Rift region (containing Serpens South, W40, and Sh2-62) and the Serpens Main cloud are seen as clearly distinct  
regions. It is therefore possible that the two clouds are not physically associated, but located along neighboring lines of sight.
While the method used by \citet{Straizys+2003, Straizys+1996} would naturally be sensitive to the first dust extinction screen along the 
line of sight, the larger VLBI-based distance of the Serpens Main core by \citet{Dzib+2010} suggests that Serpens Main 
is located behind the extinction wall associated with the Aquila clouds (Serpens South being the highest extinction region found 
inside the Aquila Rift complex).
A distance of 260~pc for the Aquila Rift complex also suits the MWC297/Sh2-62 region since the young star MWC297 itself has 
an accepted distance of 250 pc \citep{Drew+1997}. It is finally worth noting that the visual extinction map 
derived by \citet{Cambresy1999} from optical star counts and only tracing the first layer of the extinction wall has exactly the same global 
aspect as the 2MASS extinction map of \citet{Bontemps+2010}, suggesting that both Serpens South and the W40/Aquila rift/MWC297 
region are associated with this extinction wall at 260~pc. We will thus adopt a distance $d = 260$~pc for the entire Aquila complex, 
throughout this paper (see Appendix~\ref{sec:appendix_distance}, however, for a brief discussion of how our results would change
had we adopted a distance of 415~pc instead).

Rescaled to our adopted distance, the molecular mass of the entire Aquila Rift has been estimated from CO and extinction data  to be $2-5 \times 10^{5}$ $M_{\odot}$ within 
a 25~pc-radius region \citep{Dame+1987, Straizys+2003}. 
Rescaled to the same distance, the virial mass for the entire Aquila Rift estimated by \citet{DameThaddeus1985} is $\sim 3.3 \times 10^{5}$ $M_{\odot}$, 
suggesting that the whole complex is close to virial balance on large scales. 
More recently, \citet{Tanaka+2013} obtained a virial parameter $\sim $0.08--0.24 for the Serpens South filament (again rescaled to a distance of 260~pc) 
on $\sim 0.5$~pc scales \citep[see also][]{KirkH+2013}, and
\citet{Maury+2011} derived a high star formation rate of $\sim$23 $M_{\odot}$ Myr$^{-1}$pc$^{-2}$ for the protocluster 
associated with the filament (of total mass $\sim$610 $M_{\odot}$, also using $d = 260$~pc). 
Altogether, these results suggest that the Aquila Rift complex is globally gravitationally bound on scales of  $\sim$25~pc 
and includes a few highly unstable (sub-virial) clumps on the verge of forming rich star clusters on sub-parsec scales.

\section{Observations and data reduction}\label{sec:obs_data}

The {\it Herschel} Gould Belt survey observations of the Aquila Rift complex were taken on 24 October 2009 during the Science 
Demonstration Phase of {\it Herschel} \citep{Pilbratt2010}. 
The SPIRE/PACS parallel-mode scan maps covered a common $\sim$11~deg$^2$ area both by SPIRE \citep{Griffin2010} and PACS \citep{Poglitsch2010}. 
With one repetition in two orthogonal observing directions (OBSIDs: 1342186277, 1342186278), the scanning speed was 60$\arcsec$s$^{-1}$, and 
the total duration of the mapping took $\sim$12 hours. The above strategy is similar for all the parallel-mode SPIRE/PACS observations of the HGBS.

\vspace{4mm}
\noindent
\textit{PACS data reduction}
\vspace{2mm}
\newline \noindent
The individual scan directions of the parallel-mode PACS data at 70~$\mu$m and 160~$\mu$m were reduced with HIPE \citep{Ott2011} version 9.0.3063, 
provided by the {\it Herschel} Science Center. 

Starting from the raw data (level-0) and up to the level-1 stage, standard steps 
of the default pipeline were applied. 
The PACS photometer 
flux calibration scheme was applied using the up-to-date responsivity and 
correction factors (PACS ICC report, Balog et al.)\footnote{http://herschel.esac.esa.int/twiki/bin/view/Public/PacsCalibrationWeb} of the 
executed HIPE version with the calibration file set PACS\_CAL\_45\_0. 
During the actual processing of the data, we created masks to avoid bad and saturated pixels, calibration blocks and their unexpected
transient effect on the subsequent frames. Besides the flat-field correction, we applied a non-linearity correction to the data (PACS ICC 
report, Billot et al.)$^1$.
The PACS bolometers enter a non-linear regime for point sources above $\sim$100 Jy/beam in all bands (70/100/160~$\mu$m), and the 
flux densities of brighter targets are underestimated by typically a few percent. 
The applied non-linearity correction of the PACS bolometer signal had a very minor effect on the Aquila data.
Cosmic ray hits on the detectors were removed with the "second-level deglitching" method of HIPE. 
To make best use of the deglitcher, we took special care to prepare its input data. First, a high-pass filtering with a scan-leg length outside of a 
protective object mask was performed. Next, the second-level deglitching was then applied on these temporary data. Baseline subtraction was only used for 
deglitching purposes, but not on the resulting level-1 frames. The slew/turn-around data at the end of the scan legs were also preserved in the processing.

Further treatment of the flux- and pointing-calibrated level-1 time series and the projection of the combined scans were performed with 
an IDL-based map-maker, Scanamorphos, version 20 \citep{Roussel2013}\footnote{The documentation and repository of the software can be found at: 
http://www2.iap.fr/users/roussel/herschel}. 
The processing is fully automated with some user-defined options. 
It consists of the main functionalities of subtracting both the thermal and non-thermal 
components of the brightness drifts,
as well as detecting and masking remaining glitches and brightness discontinuities in the PACS data. 
In the final map projection, we adopted a spatial grid of 3$\arcsec$/pixel. 
Scanamorphos also provides associated maps of error, total drift, and weight
(see Sect.~3.7 of  \citealp{Roussel2013} for details). 
The error map provides the error on the mean brightness in 
each pixel. In the case of two scan directions, an additional ``clean'' map is produced, which is a signal map weighted 
so that noisy scans are excluded for each pixel. 
The clean map is only used for diagnostic purposes. 
In the PACS map processing, a final step 
was performed to remove long artifact glitches, which remained mainly in 
the 70~$\mu$m map, due to a jump in the brightness of the PACS data that could affect whole array rows. 
Thanks to the various planes of the output map, 
we could replace only affected pixels by ``clean map'' pixels. 
Our PACS output (level-2) fits files were produced in Jy/3\arcsec-pixel units.

For PACS data, the absolute flux accuracy of point sources is 3\% in the blue band  (70~$\mu$m)  
and better than 5\% in the red band  (160~$\mu$m)
(cf. PACS ICC report by M\"uller et al.$^1$). 
The extended source calibration is more uncertain. 
In this paper, we conservatively adopted 10\% and 20\% absolute calibration uncertainties for 
the integrated source flux densities measured in the 70~$\mu$m and 160~$\mu$m bands, respectively (see 
also Sect.~\ref{sec:deriv_core_prop} below).

\vspace{4mm}
\noindent
\textit{SPIRE data reduction}
\vspace{2mm}
\newline \noindent
The SPIRE 250~$\mu$m, 350~$\mu$m, and 500~$\mu$m data were reduced with HIPE version 10.0.2751 using modified pipeline scripts.   
The nominal and orthogonal scan directions were processed individually, and combined in a second step. 
Data taken during the turnarounds of the satellite were not included in the final maps. 

The raw level-0 data (in engineering units) were processed to level-0.5 (in physical units) using the relevant 
calibration trees (SPIRE\_CAL\_10\_1) built in HIPE. The following pipeline steps to level-1 \citep[cf.][]{Dowell+2010} consist of: 
1) converting detector timelines to angles on the sky, 2) creating the pointing product for the observation, 3) correcting for thermistor-bolometer 
electrical crosstalk, 4) correcting temperature drifts and detecting temperature jumps, 5) identifying glitches caused by cosmic 
rays, for which the assumption was that all glitches affect all bolometers of SPIRE simulataneously, 
6) applying the low-pass filter response correction, 7) applying the flux conversion, and 8) searching and correcting  
for cooler burps by recalculating the temperature drift calibration table. (A cooler burp is a steep temperature rise 
which reaches a stable plateau $\sim$6--7 hours after the cooler recycle ends.)

As the Aquila region is dominated by extended emission from the interstellar medium, 
relative gain factors appropriate to extended sources were applied to the bolometer timelines. 
These gains, determined by the SPIRE ICC, 
represent the ratio between the response of each bolometer to the extended emission and the average response. 
Variations in the specific response of each bolometer arise due to variations in the beam area among bolometers.  

The destriper module of the pipeline was used in an iterative manner.
The iterative process starts with level-1 timelines for both scan directions, and 
reconstructs an initial na\"ive map which is only corrected for a median offset. 
The destriper then fits a constant level to the difference between each input timeline 
and the corresponding map timeline, subtracts the fit from the original timeline, and reconstructs another map.
By default, bright sources are excluded during baseline fitting. These steps are iterated until convergence.
We adopted default grid pixel sizes of 6\arcsec, 10\arcsec, 14\arcsec ~for the SPIRE 250~$\mu$m, 350~$\mu$m, 500~$\mu$m wavelengths,
respectively. The output (level-2) fits files for each SPIRE wavelength were in Jy/beam units. 
For SPIRE data, the absolute flux accuracy is better than $\sim 5 \% $ for point sources  \citep{Bendo+2013} 
and  better than $\sim 10 \% $ for extended sources \citep[cf.][]{Griffin+2013} in the three bands.

\vspace{4mm}
\noindent
\textit{Map-making tests and consistency of the SPIRE vs. PACS maps}
\vspace{2mm}
\newline \noindent
SPIRE and PACS map-making tests and benchmarks were carried out in early 2012 by SPIRE/PACS ICC members, map-maker developers, 
and {\it Herschel} Key Program representatives. 
The public SPIRE\footnote{https://nhscsci.ipac.caltech.edu/sc/index.php/Spire/SPIREMap-MakingTest2013} and
PACS\footnote{http://herschel.esac.esa.int/twiki/bin/view/Public/PacsCalibrationWeb} results of this test campaign, 
which compared the performance of several publicly available map-making methods, justify our choice of the destriper pipeline 
with 0$^{th}$-order baseline removal (P0) for SPIRE data reduction 
and the choice of Scanamorphos for PACS map-making.  
In particular, the destriper P0, the default map-maker in the SPIRE scan-map pipeline since HIPE v9, performed remarkably well and compared favorably among 
all map makers in all test cases except for those suffering from the ``cooler burp'' effect. 
Furthermore, the destriper can handle observations with complex 
extended emission structures and with large-scale background gradients very well.
Power-spectrum tests carried out on SPIRE scan maps 
by the SPIRE ICC \citep[see also][for the case of the HGBS images of the Polaris flare cirrus cloud]{MAMD+2010} 
demonstrate that large SPIRE maps such as the HGBS maps trace a wide range of angular scales reliably, 
from $\simgt 30\arcmin $ or more 
down to the SPIRE angular resolution (e.g. $\sim 18\arcsec $ at 250~$\mu$m).
This high spatial dynamic range is a key advantage of the {\it Herschel}/SPIRE images (compared to, e.g., ground-based 
submillimeter continuum data), which makes our {\it Herschel} survey  simultaneously sensitive to both large-scale structures 
in molecular clouds (e.g. filaments) and small-scale structures such as individual prestellar and protostellar cores. 

As for PACS maps, comparison metrics showed that the photometry of both point-like and extended sources carried out on Scanamorphos maps 
is highly consistent with the results obtained on maps produced with other map-makers. 

The relative astrometry between the SPIRE and PACS images was tested by cross-correlating the various maps after reprojecting them on the same grid. 
Using the evolved pointing products in the {\it Herschel} system, 
a good match was found between the intensity peaks seen in the SPIRE and PACS maps on a resampled 2\arcsec/pixel scale grid.  

The 3\arcsec/pixel maps used in the present analysis are thus well registered and have a relative astrometric consistency better than 2\arcsec.
The absolute astrometry of the {\it Herschel} images was also compared with publicly-available {\it Spitzer} data, as well as high-positional accuracy 
($<$1$\arcsec$) 3~mm IRAM Plateau de Bure observations of a small field at the center of the Serpens South filament \citep{Maury+2011}. 
The final absolute astrometric accuracy of the {\it Herschel} maps is estimated to be better than 3\arcsec.

The parallel-mode PACS and SPIRE maps used in this paper were all converted to MJy/sr units and 
reprojected to a common 3$\arcsec$ pixel grid.
The conversion of the PACS maps from Jy/3\arcsec-pixel units to MJy/sr units was obtained using a square pixel area of 9~arcsec$^2$. 
For the SPIRE unit conversion from Jy/beam to MJy/sr units, we assumed the beam areas measured 
in 1\arcsec-pixel beam maps by the SPIRE ICC, as given in Table~5.2 of the SPIRE Observer's Manual v.2.2 (29 Nov. 2010), 
namely 426~arcsec$^2$, 771~arcsec$^2$, 1626~arcsec$^2$ at 250~$\mu$m, 350~$\mu$m, 500~$\mu$m, respectively.
The half-power beam width (HPBW) resolutions of the maps are 8.4$\arcsec$, 13.5$\arcsec$, 18.2$\arcsec$, 24.9$\arcsec$, 
and 36.3$\arcsec$ at 70\,$\mu$m, 160\,$\mu$m, 250\,$\mu$m, 350\,$\mu$m, and 500\,$\mu$m, respectively.
These high-quality maps are publicly available from the {\it Herschel} Gould Belt Survey 
Archive\footnote{http://gouldbelt-herschel.cea.fr/archives}.

\section{Results and analysis}\label{sec:res_analys}

\subsection{Dust temperature and column density maps}\label{sec:cd_t_maps}

\begin{figure*}[]
   \centering
 \begin{minipage}{1.0\linewidth}
     \resizebox{0.495\hsize}{!}{\includegraphics[angle=0]{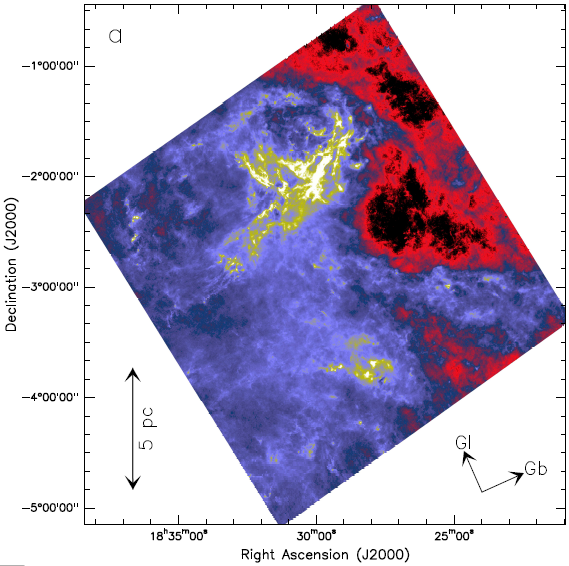}}
     \resizebox{0.495\hsize}{!}{\includegraphics[angle=0]{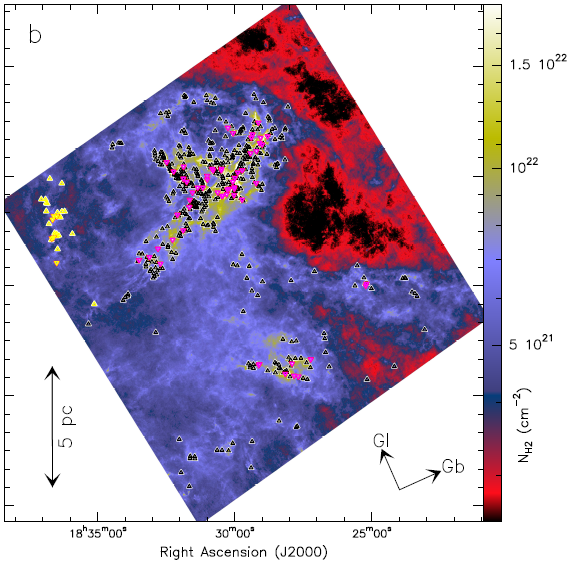}}
 \end{minipage}
   \caption{{\bf a)} H$_2$ column density map of the Aquila region at $18.2\arcsec $ angular resolution, as derived from HGBS data 
   using the method described in Sect.~\ref{sec:cd_t_maps}.
   {\bf b)} Same map as in the left panel with the positions of the 446 candidate prestellar cores and 58 protostellar 
   cores identified in the {\it Herschel} images with \textsl{getsources} (see Sect.~\ref{sec:getsources} and Sect.~\ref{sec:core_selection})
   shown as black and magenta triangles, respectively. 
   Yellow triangles locate additional prestellar and protostellar cores which    
   were excluded from the analysis and discussion of this paper, due to likely contamination by more distant objects belonging to 
   background CO clouds at significantly higher LSR velocities than the bulk of the Aquila complex (cf. Sect.~\ref{sec:mass_distrib}). 
   The orientation of the galactic coordinate axes is indicated at the lower right of each panel. 
   The lower left edge of the map is oriented almost parallel to the galactic longitude axis at $Gb\sim 2\degr $ above the Galactic plane.
   }
   \label{fig_cd}%
   \end{figure*}

\begin{figure*}[]
   \centering
 \begin{minipage}{0.49\linewidth}
    \resizebox{1.0\hsize}{!}{\includegraphics[angle=0]{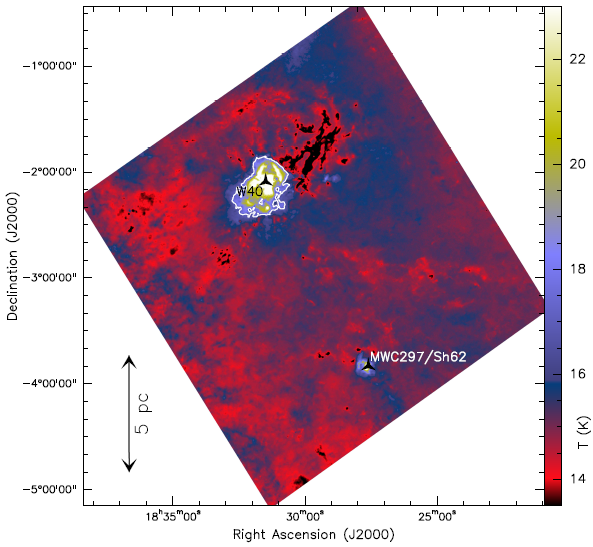}}
 \end{minipage}
   \caption{Dust temperature map of the Aquila region at 36.3\arcsec resolution, as derived from HGBS data (see Sect.~\ref{sec:cd_t_maps}).
   The white contour outlines the W40 HII region which has $T_{\rm d} > 17$~K.}
              \label{fig_temp}%
    \end{figure*}

We used the {\it Herschel} images to construct an H$_2$ column density map ($N_{\rm H_2}$, Fig.~\ref{fig_cd}) 
and a dust temperature map ($T_{\rm d}$, Fig.~\ref{fig_temp}) of the Aquila field.
We first smoothed all {\it Herschel} images (reprojected to the same 3$\arcsec$ pixel grid -- see above)
to the 36.3\arcsec~HPBW resolution of the SPIRE  500~$\mu$m data. 

A zero-level offset, obtained by correlating the {\it Herschel} data with $Planck$ and $IRAS$ data \citep[cf.][]{Bernard2010}, 
was also added at this stage to each {\it Herschel} map. 
The added offset values were 27.7, 159.8, 169.6, 94.4, and 41.4~MJy/sr at
70, 160, 250, 350, and 500~$\mu$m, respectively. 
Assuming optically thin dust emission at a single temperature $T_{\rm d}$
for each map pixel, we then fitted a modified blackbody function of the form  $I_{\nu} = B_{\nu}(T_{\rm d}) \kappa_{\nu} \Sigma$ 
to the four observed data points from 160~$\mu$m to 500~$\mu$m on a pixel-by-pixel basis, 
where $I_{\nu}$ is the surface brightness at frequency $\nu$ and $B_{\nu}(T_{\rm d})$ is the Planck blackbody function.
Each SED data point was weighted by 1/$\sigma_{\rm cal}^2$, where $\sigma_{\rm cal}$ corresponds to the 
absolute calibration error (20\% of the intensity at 160~$\mu$m and 10\% for the SPIRE bands). 
%
%
We adopted a power law approximation to the dust opacity law per unit mass (of dust$+$gas) 
at submillimeter wavelengths, namely   
$\kappa_{\lambda} = 0.1 \times (\lambda/\rm 300\mu m)^{-\beta}$ cm$^2$/g, 
and fixed the dust emissivity index $\beta$ to 2 \citep[cf.][]{Hildebrand1983}. 
Based on a detailed comparison of the {\it Herschel} results  
with the near-infrared extinction study of  \citet{Alves+2001} for the starless core B68, \citet{Roy+2014} concluded that 
these dust opacity assumptions are likely appropriate to better than 50\% accuracy over the whole 
range of H$_2$ column densities between $\sim 3 \times 10^{21}$ cm$^{\rm -2}$ and $\sim 10^{23}$ cm$^{\rm -2}$.

In the SED fitting procedure, the gas surface density distribution ($\Sigma$) and the dust temperature were left as two free parameters. 
The H$_2$ column density ($N_{\rm H_2}$) was then calculated 
from $\Sigma = \mu_{\rm H_2} m_{\rm H} N_{\rm H_2}$, adopting a mean molecular weight per hydrogen molecule 
$\mu_{\rm H_2} = 2.8$ \citep[e.g.,][]{Kauffmann+2008}.
Based on this SED-fitting method, we derived both a standard column density map at the $\sim 36.3\arcsec $ resolution of the 
SPIRE 500~$\mu$m data and a `high-resolution'  column density map at the $\sim 18.2\arcsec $ resolution of the SPIRE 250~$\mu$m data. 
The procedure used to construct the `high-resolution'  column density map is based on a multi-scale decomposition of the 
imaging data and described in detail in Appendix~A of \citet{Palmeirim+2013}.

Both the standard and the high-resolution 
column density maps were tested against a near-infrared extinction map of the Aquila/Serpens region derived from 
2MASS data \citep[see][]{Bontemps+2010,Schneider+2011}, the latter with a FWHM spatial resolution of $\sim $120$\arcsec$. 
To do this, the {\it Herschel} column density maps were smoothed to 120$\arcsec$ and converted to visual extinction units assuming 
$N_{\rm H_2}\, ({\rm cm}^{-2}) = 0.94 \times 10^{21}\, A_{\rm V} \,({\rm mag }) $ 
\citep{Bohlin+1978}. We then derived ratio maps of the converted {\it Herschel} maps to the A$_V$ map from 2MASS on the same grid. 
In most of the field covered by Fig.~\ref{fig_cd}, the ratio maps are within $\sim$10\%  of unity, indicating excellent agreement 
(see also Appendix of K\"onyves et al. 2010).

\subsection{Filamentary structure of the Aquila cloud complex}\label{sec:filam}

As emphasized by  \citet{Menshchikov+2010} and \citet{Andre+2010} and 
mentioned in Sect.~\ref{sec:intro}, filaments are widespread in the {\it Herschel} images of the Aquila region. 
Conceptually, an interstellar {\it filament} may be defined as any elongated structure in the ISM 
which is significantly denser than its surroundings. 
For the purposes of this paper, we adopt a minimum aspect ratio of $\sim 3$ and a minimum 
column density excess 
of $\sim 10\% $ with respect to the local background, i.e., $\Delta N_{\rm H_2}^{\rm fil}/N_{\rm H_2}^{\rm back} > 0.1 $, 
when averaged along the length of the structure.
For more mathematical and algorithmic definitions of a filament, the reader is referred to \citet{Sousbie2011} and \citet{Menshchikov2013}, 
respectively. 

\begin{figure*}[!ht]
   \centering
 \begin{minipage}{0.52\linewidth}
    \resizebox{1.0\hsize}{!}{\includegraphics[angle=0]{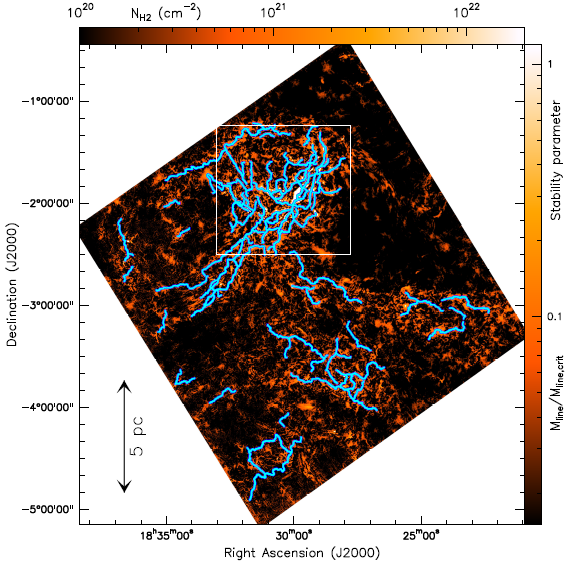}}
 \end{minipage}
 \begin{minipage}{0.43\linewidth}
    \hspace{2mm}
    \vspace{6mm}
    \resizebox{1.0\hsize}{!}{\includegraphics[angle=0]{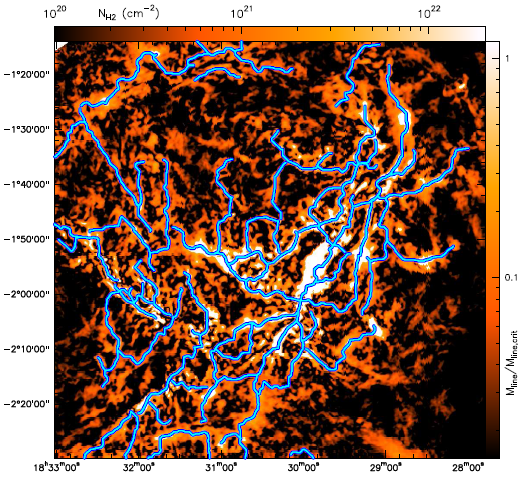}}
 \end{minipage}
   \caption{{\it Left:} Network of filaments in the Aquila cloud complex as traced by the curvelet transform component \citep[cf.][]{Starck+2003} 
   of the {\it Herschel} high-resolution column density map shown in Fig.~\ref{fig_cd}.
   Given the typical filament width $W_{\rm fil} \sim 0.1$~pc \citep{Arzoumanian+2011} and the relation $M_{\rm line} \approx \Sigma_0 \times W_{\rm fil}$ 
   between the central gas surface density $\Sigma_0$ of a filament, this curvelet column density map provides 
   information on the mass per unit length along the filaments \citep[cf.][]{Andre+2010}, as indicated by the color bar on the right. 
   The white areas highlight regions of the map where 
   $\Sigma \times W_{\rm fil}$ 
   exceeds half the critical mass per unit length $M_{\rm line, crit} = 2 c^2_{\rm s} /G$ \citep[cf.][]{InutsukaMiyama1997} 
   and the filaments are likely supercritical ($\Sigma_0 \times W_{\rm fil} > M_{\rm line, crit} $ on the filament crest).
   The overplotted blue skeleton marks the crests of the 
   filaments selected with the DisPerSE algorithm of \citet{Sousbie2011} 
   (see Sect.~\ref{sec:filam} for details). {\it Right:} Blow-up of the Aquila main subfield marked by the white square in the left panel, using the same color scale. 
   }
              \label{fig_fil-cur}%
   \end{figure*}

In order to identify filaments in the high-resolution column density map of the Aquila field, several
methods were employed and compared. 
First, the contrast of elongated features was enhanced using a ``morphological component analysis''  (MCA) decomposition of the map 
on a basis of curvelets and wavelets \citep[e.g.,][]{Starck+2003}. In such a decomposition, filamentary features are contained in the curvelet 
components, while roundish structures (e.g. dense cores) are contained in the wavelet components. 
Summing up all curvelet components led to the image shown in Fig.~\ref{fig_fil-cur}, 
which provides a high-contrast view of the filaments after subtraction of core-like and other non-elongated structures 
(e.g. non-filamentary background). 
Given the typical filament width $W_{\rm fil} \sim 0.1$~pc \citep{Arzoumanian+2011} and the relation $M_{\rm line} \approx \Sigma_0 \times W_{\rm fil}$ 
between the central gas surface density $\Sigma_0$ of a filament 
and its mass per unit length $M_{\rm line}$ 
\citep[cf. Appendix A of][]{Andre+2010},  
this curvelet component of the column density map is equivalent to a map of mass per unit length along the filaments. 
The white areas trace regions of the map where $\Sigma \times W_{\rm fil}$ is larger than half the critical value $M_{\rm line, crit} = 2 c^2_{\rm s} /G$ 
\citep[cf.][]{InutsukaMiyama1997} and the filaments are likely to be gravitationally unstable, i.e., 
supercritical\footnote{Throughout this paper, 
by supercritical or subcritical filament, we mean a filament with a supercritical or subcritical mass per unit length 
($M_{\rm line} > M_{\rm line, crit}$ or $M_{\rm line} < M_{\rm line, crit}$ -- see Sect.~\ref{sec:intro}), respectively. 
This notion should not be confused with the concept of a magnetically supercritical or subcritical cloud/core 
\citep[e.g.,][]{Mouschovias1991}.}  
with $M_{\rm line} \sim \Sigma_0 \times W_{\rm fil} > M_{\rm line, crit} $ on the filament crest. 

\begin{figure*}[!ht]
   \centering
 \begin{minipage}{0.52\linewidth}
    \resizebox{1.0\hsize}{!}{\includegraphics[angle=0]{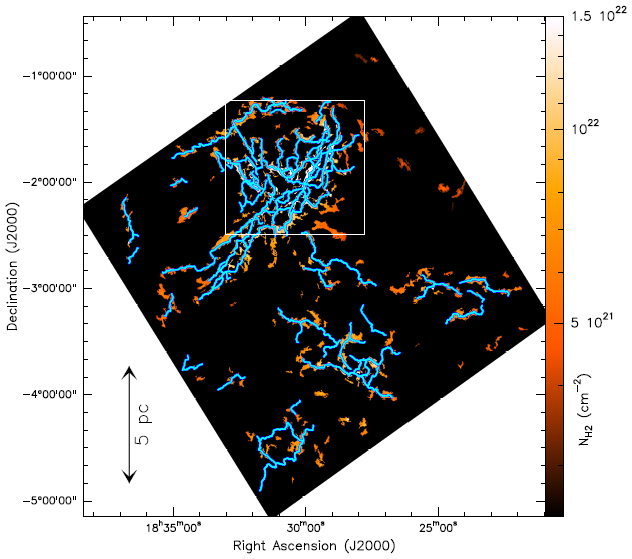}}
 \end{minipage}
 \begin{minipage}{0.43\linewidth}
    \hspace{2mm}
    \vspace{3mm}
    \resizebox{1.0\hsize}{!}{\includegraphics[angle=0]{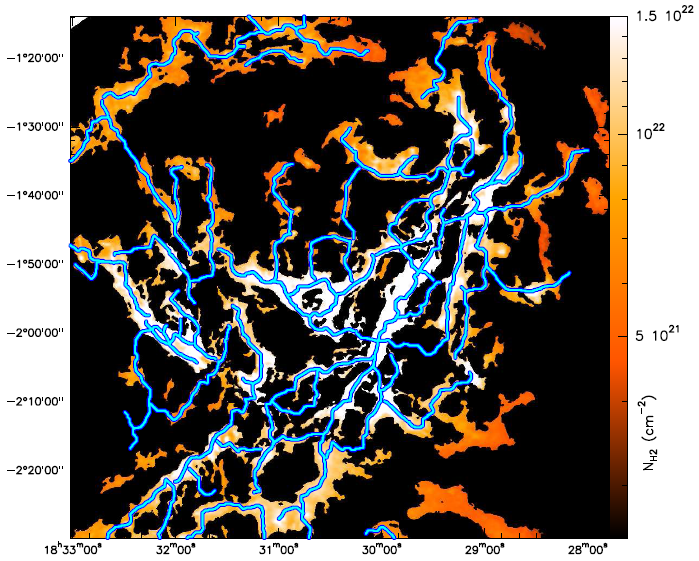}}
 \end{minipage} 
   \caption{{\it Left:} Mask of the filamentary network traced by \textsl{getfilaments} \citep{Menshchikov2013} in 
   the {\it Herschel} high-resolution column density map of the Aquila cloud complex. 
   For better visualization, only angular scales up to $320\arcsec $ (i.e., $\sim$0.4~pc at $d = 260$~pc) are shown. 
   The color scale displayed within the filamentary mask corresponds to column density values in the column density map (Fig.~\ref{fig_cd}).
   The crests of the 
   filaments traced by DisPerSE \citep{Sousbie2011} are overlaid in blue 
   (see Sect.~\ref{sec:filam} for details). {\it Right:} Blow-up of the subfield marked by the white square in the left panel, using the same color scale. 
   }
              \label{fig_fil-gf}%
    \end{figure*}

A second, independent method 
used to trace filamentary structures in the mapped region 
was the multi-scale algorithm \textsl{getfilaments} \citep{Menshchikov2013}. 
Instead of tracing filaments directly in the observed images, \textsl{getfilaments} analyzes highly-filtered spatial decompositions of 
them (called ``single-scale'' images) across a wide range of scales
\citep[Sect.~2.3 of][]{Menshchikov2013}.
Using an automated iterative thresholding algorithm \citep[Sect.~2.4.1 of][]{Menshchikov2013},
\textsl{getfilaments} analyzes single-scale images and finds $1\sigma$ intensity levels (on each 
spatial scale) that separate significant elongated structures from noise and background fluctuations.
Setting to zero those pixels whose intensities are below the thresholds, the algorithm effectively
``cleans'' the single-scale images from noise and background. 
Fine spatial decomposition allows the algorithm to identify filaments as significantly
elongated clusters of connected pixels \emph{on each spatial
scale} \citep[Sect.~2.4.2 of][]{Menshchikov2013}, separating them from other
(roundish) clusters of non-filamentary nature (e.g. sources or cores, noise peaks, isotropic
backgrounds).
Having produced the clean single-scale images of filamentary
structures on each spatial scale, \textsl{getfilaments} reconstructs the
intrinsic intensity distribution of the filamentary component of the images
(largely free of sources, noise, and background) by accumulating the clean decomposed
images over all (or a range of) spatial scales \citep[Sect.~2.4.3
of][]{Menshchikov2013}. 
Finally, the algorithm generates mask images of filaments up to various transverse 
angular scales, as well as skeletons of the filament spines in the format of fits images 
\citep[see Sect.~2.4.4 of][]{Menshchikov2013}. 
Filament extraction with \textsl{getfilaments} is fully automated 
and there are no free parameters involved. 
Figure~\ref{fig_fil-gf} displays the filamentary network obtained by applying  \textsl{getfilaments} 
to the high-resolution column density map shown in Fig.~\ref{fig_cd}. 
For better visualization, Fig.~\ref{fig_fil-gf} shows a mask image corresponding to 
elongated structures with transverse angular scales up to 320\arcsec, equivalent to $\sim$0.4~pc at $d = 260$~pc. 
The color scale displayed within the filamentary mask corresponds to the column density values in 
the input column density map (i.e., Fig.~\ref{fig_cd}).
The network of filaments outlined in this way (Fig.~\ref{fig_fil-gf}) 
is very similar to that traced by the curvelet transform (Fig.~\ref{fig_fil-cur}).

      
As a third, independent method to trace filaments, we also applied the DisPerSE algorithm\footnote{see http://www2.iap.fr/users/sousbie/web/html/indexd41d.html} 
\citep{Sousbie2011}.
DisPerSE is a general tool to identify persistent topological 
features such as peaks, voids, and filamentary structures in astrophysical data sets. 
It traces filaments by connecting saddle points to maxima with integral lines, following the gradient in a map. 
This method has already been used successfully to trace filamentary networks in {\it Herschel} images of 
nearby star-forming clouds \citep[e.g.,][]{Arzoumanian+2011,Hill+2011,Peretto+2012,Schneider+2012,Palmeirim+2013}.
To trace filaments in the Aquila field, DisPerSE was run on the standard column density map (at 36.3\arcsec resolution) 
on a 6\arcsec/pixel scale where this pixel scale sets the resolution of the filament skeleton sampling.
We used DisPerSE with a relative ``persistence'' threshold of 4.8$\times$10$^{20}$~cm$^{-2}$, which corresponds to $\sim 3$ times the 
rms level of background column density fluctuations in the low density portion ($A_{\rm V} \sim 2$) of the column density image. 
``Persistence'' is a measure of the robustness of topological features in the map \citep[see][for details]{Sousbie2011}.
Segments of filaments found by DisPerSE
were assembled into longer filaments, with the constraint that assembled segments did not form an angle larger than 65$\degr$. 
The DisPerSE filaments were also trimmed to ensure that
the minimum column density along the resulting skeleton was $5 \times10^{21}$~cm$^{-2}$ everywhere. 
This choice of DisPerSE parameters was adopted to facilitate the clean identification of dense, 
supercritical filaments, which are most relevant to the problem of core formation and the present paper 
\citep[see][and Sect.~\ref{sec:spatial_distrib} below]{Andre+2010}.
From the output of DisPerSE, we then built a 1-pixel-wide mask or skeleton 
image of the filament crests in the same way as \citet{Arzoumanian+2011} did, 
after removing filamentary features shorter than $3 \times 0.1$~pc long (or $\sim 80$ pixels of 3$\arcsec$). 
The resulting DisPerSE skeleton, which comprises a total of 90 filaments, 
is overlaid in blue in both Fig.~\ref{fig_fil-cur} and Fig.~\ref{fig_fil-gf}. 
Owing to the adopted minimum column density, 
this DisPerSE skeleton is biased toward filaments which are either entirely or at least partly supercritical along their length.
It nevertheless contains a dozen subcritical filaments.
As can be seen by comparing Fig.~\ref{fig_fil-cur} and Fig.~\ref{fig_fil-gf}, the three above-mentioned methods 
trace very similar sets of filamentary structures. The agreement is particularly good in the case of 
supercritical filaments going over white areas in the image panels.

The same filament profile analysis as described in \citet{Arzoumanian+2011} was performed on the sample of 
filaments identified here with DisPerSE, resulting in the distribution of filament inner widths shown in online Fig.~\ref{fig_fil_width}. 
A median FWHM width of $0.12 \pm 0.04$~pc was found, which is very similar 
to the median width reported by \citet{Arzoumanian+2011} for a smaller sample of 32 filaments in Aquila. 
%
   \begin{figure*}[!ht]
   \begin{center}
 \begin{minipage}{1.0\linewidth}
    \resizebox{0.49\hsize}{!}{\includegraphics[angle=0]{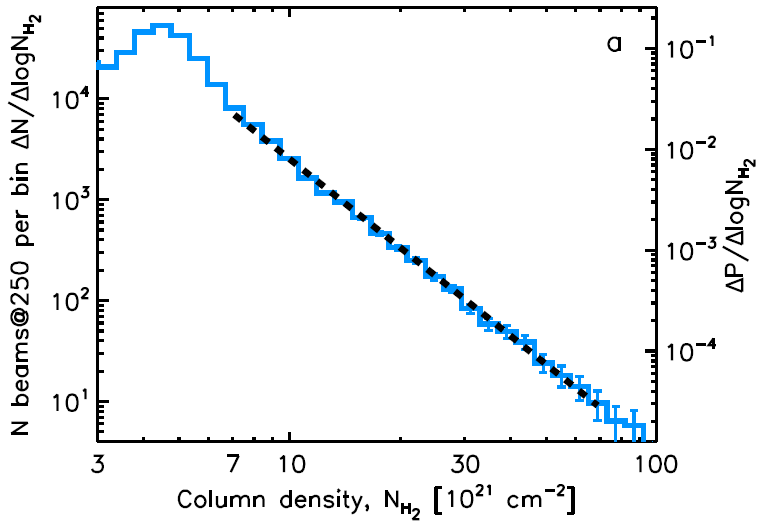}}
    \hspace{8mm}
    \resizebox{0.45\hsize}{!}{\includegraphics[angle=0]{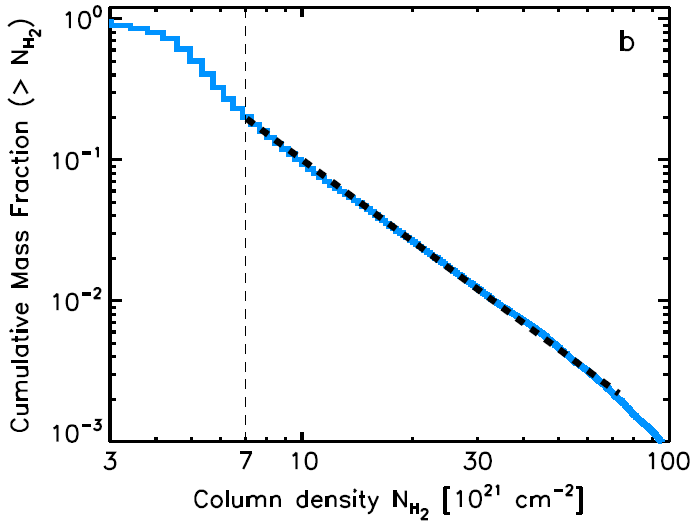}} 
 \end{minipage}
   \end{center}
   \caption{{\bf a)} Probability density function of column density ($N$-PDF) in the Aquila cloud, derived from the 18.2\arcsec -resolution column density 
            image shown in Fig.~\ref{fig_cd}. 
            The left and right axes give the actual and normalized numbers of independent beams per logarithmic bin in the column density map, respectively.
            (The right axis thus provides an estimate of a proper $N$-PDF whose integral is 1.)
	    A power-law fit to the high column density tail at $A_{\rm V} > 7$ gives $dN$/$d$log$N_{\rm H_2} \propto N_{\rm H_2}^{-2.9 \pm 0.1}$. 
	    The CO high-$V_{\rm LSR}$ area in the eastern corner of the field (see Sect.~\ref{sec:mass_distrib}) has been excluded from this PDF. 
            {\bf b)} Normalized cumulative mass fraction as a function of column density for the Aquila cloud (excluding the CO high-$V_{\rm LSR}$ area 
	    -- see Sect.~\ref{sec:mass_distrib}), based on the $Herschel$ column density map shown in Fig.~\ref{fig_cd}. 
            The dense material to the right of the dashed vertical line (equivalent to $A_{\rm V} \sim $~7--8) 
	    represents only $\sim$17\% 
	    of the total cloud mass,  while the majority of the mass ($\sim$83\%)
	    corresponds to
	    lower-density gas.  A power-law fit to the cumulative mass fraction for $A_{\rm V} > 7$ gives
	    $M(> N_{\rm H_2}) \propto N_{\rm H_2}^{-1.9 \pm 0.1}$. 
	   }
	      \label{fig_cdPDF}%
    \end{figure*}
\subsection{Distribution of mass in the Aquila cloud}\label{sec:mass_distrib}

Figure~\ref{fig_cdPDF}a shows the probability density function (PDF) of column density in the Aquila cloud complex as 
derived from the high-resolution column density map displayed in Fig.~\ref{fig_cd}. 
(Likewise, the distribution of dust temperatures corresponding to the dust temperature map in Fig.~\ref{fig_temp}
is shown in online Fig.~\ref{fig_map-core_temp}a.) 
The column density PDF is well fit by a log-normal distribution at low column densities (i.e., $3 \simlt A_{\rm V} \simlt 7$) 
and by a power-law distribution at high column densities (i.e., $A_{\rm V} \simgt 7$). 
Similar column density PDFs have already been reported in the literature for other star-forming complexes 
based on near-infrared extinction data \citep[e.g.,][]{Kainulainen+2009} and {\it Herschel} observations \citep[e.g.,][]{Schneider+2013}. 
As discussed by, e.g., \citet{Kainulainen+2011}, column density PDFs are a powerful tool to characterize molecular 
cloud structure and the transition from turbulence-dominated to collapsing, star-forming gas. The slope of the power-law distribution 
at high column densities can be readily related to the logarithmic slope of the equivalent radial density profile 
expected in cloud collapse models \citep[see, e.g.,][]{FederrathKlessen2013}.  
The power-law tail of the PDF is particularly well developed in the case of the Aquila complex (see Fig.~\ref{fig_cdPDF}a). 
The best power-law fit for $A_{\rm V} > 7$ corresponds to $dN$/$d$log$N_{\rm H_2} \propto N_{\rm H_2}^{-2.9\pm 0.1}$. 

The total mass of the Aquila cloud imaged with {\it Herschel} was derived  
from the column density map (Fig.~\ref{fig_cd}) as:
$$ M_{\rm cl} = \delta A_{\rm pixel}\, \mu_{\rm H_2}\, m_{\rm H} \sum_{i} N_{\rm H_2}, $$
where $ \delta A_{\rm pixel}$ is the surface area subtended by one pixel at the adopted distance $d = 260$~pc 
of the cloud, 
$\mu_{\rm H_2} = 2.8$ is the mean molecular weight,
$m_{\rm H}$ is the hydrogen atom mass, and the column density values in the map 
are summed up over all pixels. 
This procedure gave a total cloud mass of $\sim$2.4$\times 10^4$~$M_\odot$. 	
This estimate of the total cloud mass from {\textit Herschel} data is in very good agreement 
with the total mass of $\sim$2.0$\times 10^4$~$M_\odot$ derived from the extinction map of \citet{Bontemps+2010} 
and with the total gas mass of $\sim$2.5$\times 10^4$~$M_\odot$ 
derived from the CO(1--0) observations of \citet{Dame+2001} over the same area.  

The same mass calculation was repeated for the pixels above a given column density, which led to the 
cumulative mass fraction of gas mass in the cloud as a function of column density shown in Fig.~\ref{fig_cdPDF}b. 
For future reference, the fraction of dense gas mass above $A_{\rm V}$=7--8~mag in visual extinction represents 
only $\sim$24--17\% (5800--4200~$M_\odot$) of the total cloud mass, respectively. 			
A similar fraction of cloud mass at $A_{\rm V} > 7$~mag was reported by \citet{Johnstone+2004} in the case 
of the Ophiuchus main cloud.
Clearly, the low (column) density regions in the map shown in Fig.~\ref{fig_cd} account for most of the cloud mass.

As the Aquila Rift lies quite close to the Galactic Plane, we have to consider the potential contamination of the column density 
map (Fig.~\ref{fig_cd}) by background clouds along some lines of sight.
To assess the importance of this potential contamination, we used the CO database of \citet{Dame+2001}. 
Throughout the whole field shown in Fig.~\ref{fig_cd}, the most significant CO(1--0) emission was found 
in the same 5--7~kms$^{-1}$ LSR velocity range.
This correspondence suggests that the bulk of the CO emission comes from the same cloud complex at $d \sim 260$~pc.
Two isolated patches with significantly higher LSR velocities  (30--40~km$\,$s$^{-1}$), however, are present in the CO data of \citet{Dame+2001}, 
in the eastern corner and the relatively empty central part of the {\it Herschel} field of Fig.~\ref{fig_cd}, respectively. Given that these patches are very 
local and contribute only modest CO emission, their influence on the above column density and mass estimates is minor. 
Based on the fraction of CO emission observed at high LSR velocities and the column densities derived from {\it Herschel} data, 
we estimate that the background patches cannot change the value of the total cloud mass given above -- which excludes these patches -- by more than 4\%.
We also stress that the exclusion of the CO high-$v_{\rm LSR}$ areas from the distributions shown in Fig.~\ref{fig_cdPDF} 
has very little impact on the power-law slopes of the $N$-PDF and cumulative mass fraction plots since the background patches occupy only a small 
surface area. (Including the two patches in the $N$-PDF would change the power-law slope of $-2.9\pm 0.1$ by much less than the quoted error bar.)
The presence of these background clouds will nevertheless be taken into account when selecting dense cores 
belonging to the Aquila complex in Sect.~\ref{sec:core_selection}.

\subsection{Multiwavelength core extraction with \textsl{getsources}}\label{sec:getsources}

Conceptually, a {\it dense core}  is a single star-forming entity which may potentially form a star or a multiple system by gravitational
collapse \citep[e.g.,][]{Myers1983,Ward+1994,Andre+2000,DiFrancesco+2007}. 
In practice, a core can be defined as the immediate vicinity of a
local peak in the {\it Herschel} column density maps. In more mathematical terms, a dense core corresponds to a \emph{descending
2-manifold} \citep[cf.][]{Sousbie2011} associated with a local peak in column density. This manifold defines a region in projection
to the plane of sky whose map pixels are connected to the peak by lines following the gradient of the column density distribution.

To generate an extensive catalog of dense cores 
from HGBS data in the Aquila region, 
the parallel-mode SPIRE/PACS images were processed with \textsl{getsources}, a multi-scale, multi-wavelength source extraction
algorithm \citep{Menshchikov2012}\footnote{The HGBS first-generation catalog of cores presented in this paper (see Appendix A) 
was produced with the ``November 2013'' major release of \textsl{getsources} (v1.140127), 
which is publicly available from http://gouldbelt-herschel.cea.fr/getsources.}. 
This algorithm was designed primarily for extracting dense cores and young stellar objects (YSOs) 
in far-infrared/submillimeter surveys of
Galactic molecular clouds with {\it Herschel}. 
The main features of the source extraction method, which may be conveniently 
divided into a detection and a measurement stage, can be summarized as follows 
\citep[see][for full details]{Menshchikov2012}. 

At the detection stage, in contrast to the usual approach of detecting sources directly in the observed images,
\textsl{getsources} analyzes `single-scale' images (i.e., fine spatial decompositions of the original images -- cf. Sect.~\ref{sec:filam}) 
across a wide range of scales and across all observed wavebands. 
This decomposition filters out irrelevant spatial
scales and improves source detectability, especially in crowded regions and for extended sources. 
Using an automated
iterative thresholding method \citep[see Sect.~2.3 of][]{Menshchikov2012}, \textsl{getsources} analyzes single-scale
images and finds $3\sigma$ to $6\sigma$ intensity levels (on each spatial scale) that separate
signals of significant sources from noise and background fluctuations. Setting to zero those pixels whose intensities
are below the thresholds, the algorithm effectively 'cleans' the single-scale images from noise and background
(including the filamentary component of the images).
For detecting sources, \textsl{getsources} constructs a set of
wavelength-independent single-scale detection images that preserve information
in both spatial and wavelength dimensions \citep[Sect.~2.4 of][]{Menshchikov2012}.
This multi-wavelength design combines data over all
wavebands and thus naturally produces a wavelength-independent detection
catalog with invariant source positions for all wavebands. Besides eliminating
the need and problems of matching independent monochromatic extraction
catalogs, the method also improves the detectability of weak sources and
enables substantial super-resolution at wavelengths with lower spatial
resolution.
Sources are detected by \textsl{getsources} in the combined
single-scale detection images by analyzing the evolution of their peak
intensities and segmentation masks across all spatial scales \citep[Sect.~2.5
of][]{Menshchikov2012}. 
The spatial scale on which a source is brightest
determines its characteristic size and corresponding footprint size. 
The latter is defined as the entire area that would give a non-negligible
contribution to the integrated flux. 
The peak position of each source is determined from the wavelength-combined single-scale
detection images using the first moments of intensity  
\citep[Appendix F of][]{Menshchikov2012} measured  
over a range of spatial scales between the smallest scale on which the source appears 
and the characteristic scale on which the source is brightest. 
In effect,  source coordinates are largely determined by the wavebands with
higher angular resolution and unaffected by large-scale emission.

At the measurement stage, properties of detected
sources are measured in the original observed images at each wavelength. 
These measurements go beyond simple aperture photometry since they are done together with 
background subtraction and deblending of overlapping sources \citep[Sect.~2.6
of][]{Menshchikov2012}. 
Background is subtracted by linear interpolation under
the source footprints found at the detection stage, constrained by different
angular resolutions in each waveband. The footprints must be at least as large 
as the beam size and their elongation must correspond to that of the source 
intensity distribution at that wavelength \citep[Eq.~20 of][]{Menshchikov2012}. 
Overlapping sources are deblended in an iterative procedure that splits pixel
intensity between blended sources, assuming a simple shape for their intensity
distributions. The deblending shape has a Gaussian-like circular profile with
somewhat stronger power-law wings \citep[see Eq.~14 of][]{Menshchikov2012} that
should approximate the intensity profiles of observed sources.
Local uncertainties of the peak intensities and integrated
fluxes are given by the standard deviations estimated in elliptical annuli
(covering areas of 20 observational beams) just outside the footprints. 
In crowded areas, the standard deviations are estimated in expanded annuli outside 
of any of the overlapping sources \citep[see Sect.~2.6 of][]{Menshchikov2012}.
Aperture corrections are applied by \textsl{getsources}
using tables of the encircled energy fraction values for the actual PSFs
provided by the PACS and SPIRE ICCs \citep{Balog+2014,Bendo+2013}. 

Source extraction with \textsl{getsources} is fully automated and there are no
free parameters involved: default configuration parameters have been
extensively tested and fine-tuned to work in most practical cases. 
For the production of the `first-generation' catalogs of starless and protostellar cores 
from the HGBS, the following two-pronged extraction strategy has been adopted. Two sets
of dedicated \textsl{getsources} extractions are performed, optimized for the
detection of dense cores and YSOs/protostars, respectively.  

In the first set, all of the {\it Herschel} data tracing column density are combined at the detection stage,  
to improve the detectability of dense cores.
The detection image is thus combined from the clean 160\,$\mu$m, 250\,$\mu$m, 350\,$\mu$m, and 500\,$\mu$m maps, 
together with the high-resolution column density image (see Sect.~\ref{sec:cd_t_maps}) used as an additional ``wavelength''. 
The latter is added to the combined detection image to ensure that detected sources 
correspond to genuine column density peaks. 
Furthermore, the 160~$\mu$m component to the detection image is ``temperature-corrected'' 
to reduce the effects of strong, anisotropic temperature gradients present in parts of the observed fields, 
such as in the vicinity of the W40 HII region in Aquila
\footnote{In the presence of an anisotropic radiation field, due to a closeby HII region for instance,  
radiative transfer calculations show that the far-infrared emission expected from a starless core at, e.g., 160~$\mu$m 
is not centered on the column density peak but is shifted toward the source of illumination. 
Using a ``temperature-corrected'' 160\,$\mu$m map instead of the original 160\,$\mu$m map at the detection stage 
in \textsl{getsources} alleviates this problem and helps to better trace the intrinsic position 
of the underlying column density core.}. 
The temperature-corrected 160\,$\mu$m map is obtained by converting the original observed 160\,$\mu$m map 
($13.5\arcsec $ resolution) to an approximate column density image, using the color-temperature map derived from the 
intensity ratio between 160\,$\mu$m and 250\,$\mu$m (at the $18.2\arcsec $ resolution of the 250\,$\mu$m map). 
Simulations on synthetic emission maps including model cores (see, e.g., Sect.~\ref{sec:completeness} below and 
Appendix~\ref{sec:appendix_simulations}) confirm the validity of this approach to detecting dense cores.

A second set of \textsl{getsources} extractions is performed to trace the presence of self-luminous YSOs/protostars 
and discriminate between protostellar and starless cores. Here, the only {\it Herschel} data used at the detection 
stage come from the 70\,$\mu$m image. Indeed, the presence of point-like 70~$\mu$m emission traces 
the internal luminosity of a protostar very well  \citep[e.g.,][]{Dunham+2008}, and {\it Herschel}/PACS observations 
of nearby ($d < 500$~pc) clouds even have the sensitivity to detect candidate ``first hydrostatic cores'', the very first 
and lowest-luminosity stage of protostars \citep[cf.][]{Pezzuto+2012}.

At the measurement stage of both sets of extractions, source properties are measured at the detected positions 
of either cores or YSOs/protostars,  using the observed, background-subtracted, and deblended images 
at all five {\it Herschel} wavelengths, plus the high-resolution column density map. 
The advantage of this two-pronged extraction strategy is that it provides 
more reliable detections and measurements of 
column-density cores and 70\,$\mu$m luminous YSOs/protostars, respectively.

\subsection{Selection and classification of reliable core detections}\label{sec:core_selection}

Here, we summarize the criteria adopted to select various types of dense cores 
from the raw source lists produced by the two sets of multi-wavelength \textsl{getsources} extractions 
described at the end of Sect.~\ref{sec:getsources}.   
For each source type, the following prescribed criteria should be met at the same time. 

\vspace{4mm}
\noindent
\textit{Selection of candidate dense cores (either starless or protostellar) from the ``core'' set of extractions}
\vspace{-3.5mm}
\newline \noindent
\begin{itemize}
\item Column density {\it detection} significance 
greater than 5, where detection significance here refers 
to a single-scale analog to a classical signal-to-noise ratio (S/N) [see Eq.~(17) of Men'shchikov et al. 2012] 
in the high-resolution column density map; 
\item Global detection significance over all wavelengths [see Eq.~(18) of Men'shchikov et al. 2012] greater than 10; 
\item Global ``goodness'' 
$\ge 1$, where goodness is an output quality parameter of  \textsl{getsources}, 
combining global signal-to-noise ratio and source reliability, and defined in Eq.~(19) of Men'shchikov et al. (2012);
\item Column density {\it measurement} signal-to-noise ratio\footnote{The measurement S/N is estimated 
at the measurement step of the \textsl{getsources} extractions (see Sect.~\ref{sec:getsources}) 
and characterizes the flux measurement uncertainties. In crowded situations, the measurement S/N 
of a source with a high ``detection significance'' at the detection step can be low 
because of large deblending and background-subtraction uncertainties.} 
(S/N) greater than 1 in the high-resolution column density map; 
\item Monochromatic detection significance greater than 5 in at least two bands between 160~$\mu$m and 500~$\mu$m; and 
\item Flux measurement with S/N $>$ 1 in at least one band between 160~$\mu$m and 500~$\mu$m for which the monochromatic 
detection significance is simultaneously greater than 5.
\end{itemize}

\vspace{4mm}
\noindent
\textit{Selection of candidate YSOs from the ``protostellar'' set of extractions}
\vspace{-3.5mm}
\newline \noindent
\begin{itemize}
\item Monochromatic detection significance greater than 5 in the 70~$\mu$m band; 
\item Positive peak and integrated flux density at 70~$\mu$m;  
\item Global ``goodness'' greater than or equal to 1; 
\item Flux measurement with S/N $>$ 1.5 in the 70~$\mu$m band; 
\item FWHM source size at 70~$\mu$m smaller than 1.5 times the 70~$\mu$m beam size (i.e., $< 1.5 \times 8.4\arcsec$ or $< 12.6\arcsec$); and 
\item Estimated source elongation $<$ 1.30 at 70~$\mu$m, where source elongation is defined as the ratio of the major and minor FWHM sizes.
\end{itemize}
The discussion of the {\it Herschel}-identified sample of protostars and YSOs in Aquila will be presented 
in a complementary paper (K\"onyves et al., in prep.; see Maury et al. 2011 for a preliminary subsample around W40 and Serpens-South).

\vspace{4mm}
\noindent
\textit{Selection of candidate starless cores and protostellar cores}
\vspace{-3.5mm}
\newline \noindent
\begin{itemize}

\item After cross-matching the selected dense cores with the candidate YSOs/protostars, a selected dense core is classified as 'starless' if there 
is no candidate 70~$\mu$m YSO within its half-power (high-resolution) column density contour.

\item 
Conversely, a selected dense core is classified as 'protostellar' if there 
is a candidate 70~$\mu$m YSO within its half-power column density contour. 

\item The most reliable SED of a selected protostellar core 
is obtained by combining the 70~$\mu$m flux density from the ``protostellar''
extractions with the 160~$\mu$m, 250~$\mu$m, 350~$\mu$m, and 500~$\mu$m flux densities from the ``core'' extractions.

\end{itemize}

\vspace{4mm}
\noindent
\textit{Post-selection checks}
\vspace{3.5mm}
\newline \noindent
All of the cores automatically selected according to the above criteria were visually inspected
in the SPIRE/PACS and column density images (see blow-up maps in Figs.~\ref{fig_zooms1} \& \ref{fig_zooms2}).
Any dubious source was removed from the final catalog of cores presented online Table~A.1 (see below). 

To eliminate from our discussion of Aquila cores extragalactic contaminants that may be misidentified as cores or YSOs, 
we also cross-matched all selected sources 
with the NASA Extragalactic Database\footnote{https://ned.ipac.caltech.edu/forms/nearposn.html} (NED), 
but no close match (within 6\arcsec) was found.

Likewise, we checked likely associations between the selected {\it Herschel} cores and objects in the SIMBAD database 
or the combined c2d and Gould Belt {\it Spitzer} database (Dunham et al. 2013; Allen et al., in prep.). 
Any matches are reported in the online catalog (Table~A.1).
In particular, 27 associations with a {\it Spitzer} source were found using a 6$\arcsec$ matching radius.

In the eastern corner of the field shown in Fig.~\ref{fig_cd}, there are two known dense clumps 
(ISOSS J18364-0221 SMM1/SMM2) with $> 30$~km/s LSR velocities from molecular line measurements \citep{Birkmann+2006}. 
Their LSR velocities correspond to a kinematical distance of $\sim$2.2~kpc. 
We therefore excluded from our Aquila discussion 23 candidate prestellar cores and 6 protostellar cores 
(shown as yellow triangles in Fig.~\ref{fig_cd}) 
lying in the high-$V_{\rm LSR}$ CO area of the {\it Herschel} field 
mentioned at the end of Sect.~\ref{sec:mass_distrib}. 
These cores are nevertheless listed (with appropriate comments) in the online catalogs.

   \begin{figure*}[!ht]
   \begin{center}
 \begin{minipage}{0.47\linewidth}
    \resizebox{1.0\hsize}{!}{\includegraphics[angle=0]{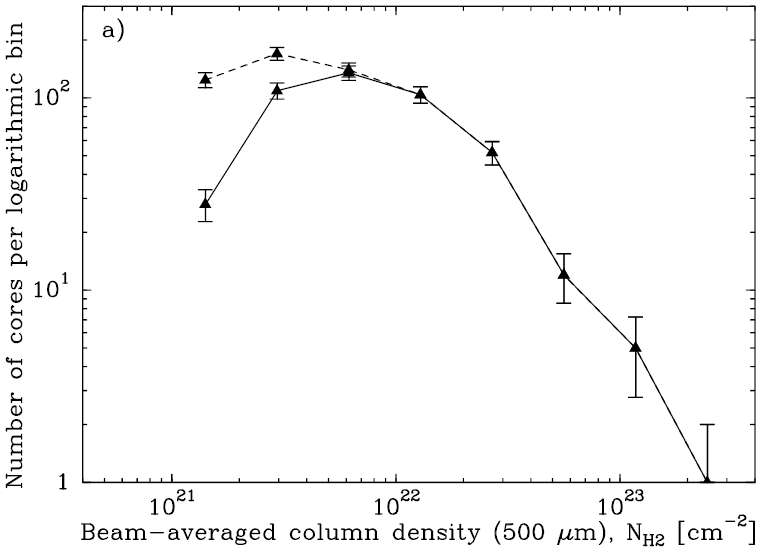}}
 \end{minipage}
 \hspace{0.4cm}
 \begin{minipage}{0.47\linewidth}
    \resizebox{1.0\hsize}{!}{\includegraphics[angle=0]{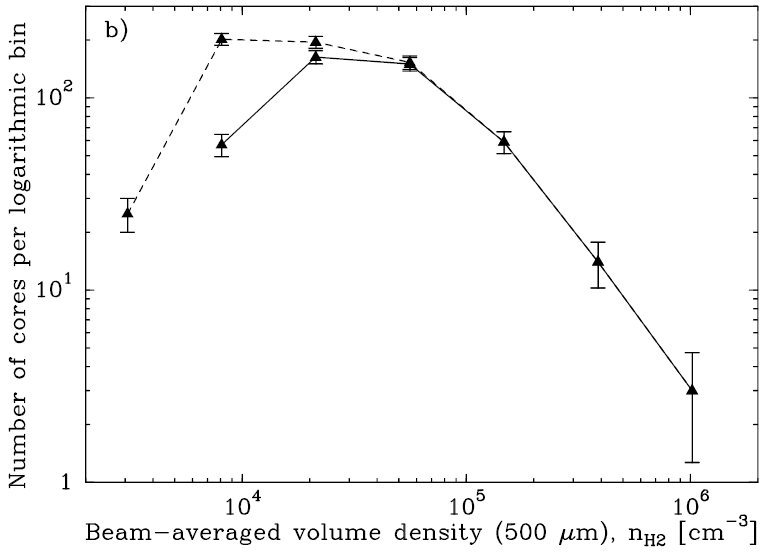}}
 \end{minipage} 
   \end{center}
   \caption{Distributions of beam-averaged column densities ({\it left}) and beam-averaged volume densities ({\it right}) 
   at the resolution of the SPIRE 500~$\mu$m observations 
   for the population of 446 candidate prestellar cores in Aquila (solid curves). In both panels, the dashed curves show the corresponding distributions 
   for all 651 selected starless cores.}
              \label{fig_dens}%
    \end{figure*}

In the post-selection phase, we also used 
another source extraction method to generate an ``alternative-algorithm''
flag for our \textsl{getsources} 
master source catalog entries. 
For this purpose, we used CSAR \citep[Cardiff Sourcefinding AlgoRithm -- ][]{KirkJ+2013}, a 
hierarchical source-finding algorithm, which we applied to the high-resolution column density map.
In the Aquila entire field, the fraction of matches between selected starless cores and CSAR-detected sources 
is $\sim$45\%, adopting a matching separation of less than 6$\arcsec$ between peak positions. 
The resulting 'CSAR'-flag appears in the online catalogs to indicate if a given \textsl{getsources} core was independently detected 
by CSAR. Several reasons explain the relatively low fraction of matches with CSAR-detected sources based on peak positions. 
First, CSAR is a very conservative source-finding algorithm, especially in crowded areas where 
the multi-scale nature of \textsl{getsources} makes it easier to detect blended objects. 
Second, CSAR is a `monochromatic' algorithm which detects sources in a single observed image (here the column density map, 
intrinsically noisier than the {\it Herschel} images)  
and does not benefit from the multi-wavelength design of \textsl{getsources} (significantly improving the detectability of weak sources). 
Third, the present core sample is dominated by starless cores which tend to have flat-topped density profiles \citep[][]{Ward+1994}
and whose peak positions are not as well defined as the peak positions of YSOs. 
Relaxing our matching condition somewhat, we note that $\sim 65\% $ of all selected cores 
include the peak position of a CSAR source within their FWHM contours. 
Moreover, as much as $\sim 85\% $ of the surface area occupied by the FWHM ellipses of our selected cores 
in the high-resolution column density map is also covered by the outer footprints of CSAR sources. 
Based on our visual inspection checks, we are confident that all \textsl{getsources} detections reported in online Table~A.1 are robust. 
(The reader can judge by looking at the blow-up maps provided online for each source -- see examples in Figs.~\ref{fig_zooms1} \& \ref{fig_zooms2}). 
For the sake of robustness at this `first-generation' stage, $\sim 20\%$ of the automatically-selected \textsl{getsources} 
cores were visually rejected for reasons such as map artifacts, 
sources seen only at some {\it Herschel} bands but not in the column density image, 
or sources seen only in the column density map but not at any of the {\it Herschel} bands.

\vspace{4mm}
Our \textsl{getsources} selection and classification procedure resulted in a final sample of 709 Aquila cores 
(not counting 40 objects -- 34 starless cores, including 23 candidate prestellar cores, 
and 6 protostellar cores -- in the high-$V_{\rm LSR}$ CO area), comprising 651 starless cores and 
58 protostellar cores. 
A total of 446 starless cores were classified as candidate prestellar cores on the basis of their locations 
in a mass versus size diagram (see Sect.~\ref{sec:bound_selection} and Fig.~\ref{fig_massSize} below).
The spatial distribution of the prestellar and protostellar cores is shown in Fig.~\ref{fig_cd}, overplotted on the high-resolution column 
density map of the cloud.

The observed properties of all selected cores are given in the accompanying online catalog (cf. Table~A.1 in Appendix~A).
The contents of Table~A.1 are as follows: 
core running number (Col.~{\bf 1}), HGBS source name (Col.~{\bf 2}), J2000 equatorial coordinates (Cols. {\bf 3} \& {\bf 4}), 
then,  for each {\it Herschel} wavelength, detection significance {\bf(5)}, peak flux density and error [{\bf(6)} \& {\bf(7)}], contrast over local background {\bf(8)}, 
peak flux density in a $36.3\arcsec $ beam {\bf(9)}, total integrated flux density and error [{\bf(10)} and {\bf(11)}], major \& minor FWHM diameters [{\bf(12)} \& {\bf(13)}], 
and position angle of the major axis {\bf(14)}, 
followed by detection significance in the high-resolution column density map {\bf(15)},  peak H$_2$ column density at $18.2\arcsec $ resolution {\bf(16)}, 
column density contrast over local background {\bf(17)}, peak column density in a $36.3\arcsec $ beam {\bf(18)}, 
column density of local background {\bf(19)}, major \& minor FWHM diameters and position angle of the major axis in the high-resolution column density map [{\bf(20)}, {\bf(21)}, \& {\bf(22)}], 
number of {\it Herschel} bands at which the core is significant {\bf(23)}, 'CSAR'-flag {\bf(24)}, core type {\bf(25)}, SIMBAD counterpart(s) {\bf(26)}, \textit{Spitzer}-c2d counterpart if any {\bf(27)}, 
and comments {\bf(28)}.

\subsection{Derived core properties}\label{sec:deriv_core_prop}

The SED fitting procedure used to derive core properties was similar to the procedure described in Sect.~\ref{sec:cd_t_maps}  
for the production of the column density map. Here, the SEDs were constructed
from the integrated flux densities measured by \textsl{getsources} for each extracted core  (see Fig.~\ref{fig_SED}) 
and the SED data points were weighted by 1/$\sigma_{\rm err}^2$, where $\sigma_{\rm err}$ corresponds 
to the flux {\it measurement} error estimated by \textsl{getsources} for each point. 
(In contrast to Sect.~\ref{sec:cd_t_maps} where the dominant source of error was the calibration uncertainty, 
the errors on source flux estimates are primarily driven by uncertain background subtraction.)
The modified blackbody fits to the observed SEDs were performed with the MPCURVEFIT routine \citep{Markwardt2009} in IDL. 
These SED fits provided direct estimates of the mass and line-of-sight-averaged (SED) dust temperature for most of the selected cores. 
The core masses were derived assuming the same dust opacity law as in Sect.~\ref{sec:cd_t_maps} and a distance $d = 260$~pc for the Aquila complex. 
The angular FWHM size estimate returned by \textsl{getsources} for each core 
(as measured at $18.2\arcsec $ resolution 
in the high-resolution column density map) 
was converted to a physical core radius assuming the same distance. 
Two estimates of the core radius are provided (see online Table~A.2). 
The first estimate is a deconvolved radius, calculated as 
$R_{\rm deconv} = (\overline{FWHM}^2_{\rm N_{H_2}} - \overline{HPBW}^2)^{1/2}$, where $\overline{FWHM}$  
and $\overline{HPBW}$ denote the physical sizes corresponding to the FWHM angular size of the core and 
the HPBW resolution of  the high-resolution column density map, respectively. 
The second estimate simply corresponds to the observed average FWHM size of the core 
(geometrical average between the major and minor FWHM sizes). 
In principle the first value provides a more accurate estimate of the intrinsic core radius, but it is affected 
by significantly larger uncertainties than the second value in the case of marginally resolved cores. 
In the case of a self-gravitating {\it prestellar} core, 
both values provide estimates of the core outer {\it radius} under the assumption 
that such a core can be approximately described 
as a critical Bonnor-Ebert (BE) sphere \citep[e.g.,][]{Bonnor1956}. 
(Indeed, a critical BE sphere of outer radius $R_{{\rm BE}}$
has a column density profile approaching that of a Gaussian distribution of FWHM diameter $\sim R_{{\rm BE}}$.)
A peak (or central beam) column density, an average column density, 
a  central-beam volume density, and an average volume density were also derived for each core based on its estimated mass and radius. 
The central-beam column density was estimated from the peak flux densities of the core 
at the resolution of the SPIRE 500~$\mu$m observations 
($HPBW = 36.3\arcsec $ or $\sim 0.046$~pc at $d=260$~pc) using an SED fitting procedure similar to that 
described in Sect.~\ref{sec:cd_t_maps}. 
The central-beam volume density $n_0$ (at the same resolution) was derived from the central-beam column density $N_0$
assuming a Gaussian spherical distribution, for which $n_0  = N_0/(\sqrt{2\pi}\, \sigma )$,  where $\sigma $ is the 
standard  deviation of the Gaussian distribution.
The distributions of column densities and volume densities for the population of starless cores  
are shown in Fig.~\ref{fig_dens}.

All of the derived properties are provided in online Table~A.2 for the whole sample of selected {\it Herschel} cores.
The contents of Table~A.2 are as follows: 
core running number (Col.~{\bf 1}), HGBS core name (Col.~{\bf 2}), J2000 equatorial coordinates (Cols. {\bf 3 \& 4}), 
deconvolved and observed core radius (Cols. {\bf 5} \& {\bf 6}), estimated core mass and corresponding error (Cols.~{\bf 7} and {\bf 8}), 
SED dust temperature and corresponding error  (Cols.~{\bf 9} and {\bf 10}), peak column density at $36.3\arcsec$ resolution (Col.~{\bf 11}), 
average column density measured  before and after deconvolution (Cols.~{\bf 12} and {\bf 13}), 
beam-averaged peak volume density at $36.3\arcsec$ resolution (Col.~{\bf 14}), 
average volume density derived before and after deconvolution (Cols.~{\bf 15} and {\bf 16}), 
Bonnor-Ebert mass ratio (Col.~{\bf 17} -- see Sect.~\ref{sec:bound_selection}), 
core type (Col.~{\bf 18}), and comments (Col.~{\bf 19}).

Since color correction factors are usually small, we did not apply any color corrections to the measured flux densities. 
Instead, like \citet{Kelly+2012}, we adopted an elevated calibration uncertainty representing multiple sources of uncertainties.
Our adopted calibration uncertainties for the SED data points were 10--20\% for the PACS 70--160~$\mu$m bands  and 10\% for the 
SPIRE 250/350/500~$\mu$m bands, respectively, which are conservative values compared to the HSC-recommended point source calibration 
uncertainties\footnote{The photometric point-source calibration uncertainty is less than 7\% for the PACS bands (Balog et al. 2014) 
and $\sim 5\% $ for the SPIRE bands (Bendo et al. 2013).}.

The robustness of the SED fits was assessed by using and comparing two successive runs of the fitting routine with slightly different weighting schemes 
for each source.
In the first run the 70~$\mu$m data point was included in the fit  and the \textsl{getsources} {\it detection} errors 
were used to weigh the SED data points, while in the second run the 70~$\mu$m point was not fitted and 
the (more conservative) {\it measurement} errors were used to weigh the SED data points.
The {\it detection} errors on significant data points were typically $\sim 15\% $ (comparable to the absolute 
calibration uncertainty), while the median {\it measurement} errors ranged from $\sim 30\%$ to $\sim 70\%$
depending on wavelength (being typically higher at 160$\mu$m).
The results of the SED fits were accepted for a given source if 1) significant flux measurements exist for this source in at least three {\it Herschel} bands,  
2) the source has a larger integrated flux density at 350~$\mu$m than at 500~$\mu$m,  
and 3) there was less than a factor of 2 difference between the core mass estimates derived from the two fit runs. 
About 68\% of the starless cores had reliable SED fits.
The corresponding distribution of SED dust temperatures is shown in online Fig.~\ref{fig_map-core_temp}b. 
Comparison with the distribution of dust temperatures in the background cloud (online Fig.~\ref{fig_map-core_temp}a) 
indicates that the Aquila starless cores are somewhat colder than the parent cloud, as expected \citep[cf.][]{Roy+2014}.

The masses of the starless cores for which the SED fit results were rejected were directly estimated 
from the measured integrated flux density at the longest significant wavelength in each case, 
assuming optically thin dust emission at the median dust temperature found 
for starless cores with reliable SED fits (i.e., $11.5 \pm 2$~K outside the W40/Sh62 areas 
and $14.5 \pm 3$~K within the higher radiation field areas W40 and Sh62). 
The corresponding cores have more uncertain properties and are marked as 
having ``unreliable SED fits'' in the last column of online Table~A.2.
 
\vspace{2mm}
\noindent
\textit{Accuracy of the core mass estimates}
\vspace{2mm}
\newline \noindent
Uncertainties in the dust opacity law alone 
induce uncertainties of up to a factor $\sim \,$1.5--2 
in the core mass estimates.
As mentioned in Sect.~\ref{sec:cd_t_maps},  the dust opacity law adopted here and in 
other HGBS papers, namely $\kappa_{\lambda} = 0.1 \times (\lambda/\rm 300\mu m)^{-\beta}$ cm$^2$/g, 
is likely appropriate to better than 50\% in the 160--500~$\mu$m range for column densities between 
$\sim 3 \times 10^{21}$ cm$^{\rm -2}$ and $\sim 10^{23}$ cm$^{\rm -2}$ \citep[cf.][]{Roy+2014}.

In addition to the dust opacity, another systematic effect affects the accuracy of our simple SED mass estimates.  
A single-temperature graybody fit to the integrated flux densities can only provide an average value of the 
dust temperature for each source and neglects any variation in the dust temperature within the source. 
In reality, starless dense cores, which are externally heated objects, are known from both 
radiative transfer calculations \citep[e.g.,][]{Evans+2001, Stamatellos+2007} 
and, e.g., {\it Herschel} observations \citep[e.g.,][]{Nielbock+2012, Roy+2014} to have a stratified temperature 
structure with a significant drop in dust temperature toward core center. 
In such a situation, the average dust temperature derived from a global SED fit can sometimes significantly 
overestimate the mass-averaged dust temperature within a starless core, leading to an underestimate of the core mass.
The magnitude of this effect is very modest ($< 20\% $) for low column density cores such as B68 \citep{Roy+2014}
but increases to up to a factor $\sim 2$ for high-density cores with average column densities $\simgt 10^{23}$ cm$^{\rm -2}$. 
In the case of spatially-resolved cores with good signal-to-noise data, techniques such as the 
Abel-inversion method \citep{Roy+2014} or the COREFIT method \citep{Marsh+2014} can help to retrieve the 
intrinsic temperature structure and derive more accurate mass estimates. 
We did not attempt to use such techniques here. 
Based on the results of the simulations performed to estimate the completeness of the survey (see Sect.~\ref{sec:completeness} below 
and Appendix~\ref{sec:appendix_simulations}), 
however, we estimate that the SED masses listed in Table~\ref{tab_der_cat_cores} 
for starless cores are likely underestimated by $\sim \, $20--30\% 
on average compared to the intrinsic core masses, mainly due to the fact that the SED dust temperatures tend to slightly 
overestimate the intrinsic mass-averaged temperatures of starless cores. 
The column densities and volume densities listed in Table~\ref{tab_der_cat_cores} and used in Fig.~\ref{fig_dens} 
(see also Fig.~\ref{fig_lifetime} below) have {\it not} been corrected for this small effect.

\subsection{Selecting self-gravitating prestellar cores}\label{sec:bound_selection}

Conceptually, a dense core is deemed to be {\it prestellar} if it is both starless 
{\it and} self-gravitating 
\citep[cf.][]{Andre+2000,DiFrancesco+2007,Ward+2007}. 
Such starless cores will 
most likely form (proto)stars in the future.
Lacking spectroscopic observations for most of the $Herschel$ cores, we used the 
thermal value of the 
critical Bonnor-Ebert (BE) mass ($M_{\rm BE,crit}$ -- Bonnor 1956) 
to assess whether a starless core was self-gravitating 
or not 
based on the value 
of its BE mass ratio $\alpha_{\rm BE} = M_{\rm BE,crit} / M_{\rm obs} $. 
The critical BE mass can be expressed as
$$M_{{\rm BE,crit}}~\approx~2.4~R_{{\rm BE}}~c_s^2 / G, $$
where $R_{{\rm BE}}$ is the BE radius, $c_s$ the isothermal sound speed, and $G$ the gravitational constant. 
In the presence of significant nonthermal motions, the thermal BE mass should be replaced 
by a modified BE mass obtained by substituting the total (thermal $+$ non thermal) one-dimensional velocity dispersion 
for the isothermal sound speed in the above formula. 
The simplified approach adopted here, where the nonthermal component of the velocity dispersion is neglected, 
is justified by observations of nearby cores in 
dense gas molecular tracers such as NH$_3$ and N$_2$H$^+$ lines, which show that nonthermal motions 
are negligible in low-mass (and intermediate-mass) starless cores \citep[e.g.,][]{Myers1983,Andre+2007}.
For each object, we estimated 
the thermal BE mass,  $M_{\rm BE}$, from the  
deconvolved core radius $R_{\rm deconv}$ measured in the high-resolution column density map
(see Sect.~\ref{sec:deriv_core_prop}) 
assuming a typical gas temperature of 10~K.

   \begin{figure}[!h]
   \begin{center}
 \begin{minipage}{1.0\linewidth}
    \resizebox{1.0\hsize}{!}{\includegraphics[angle=0]{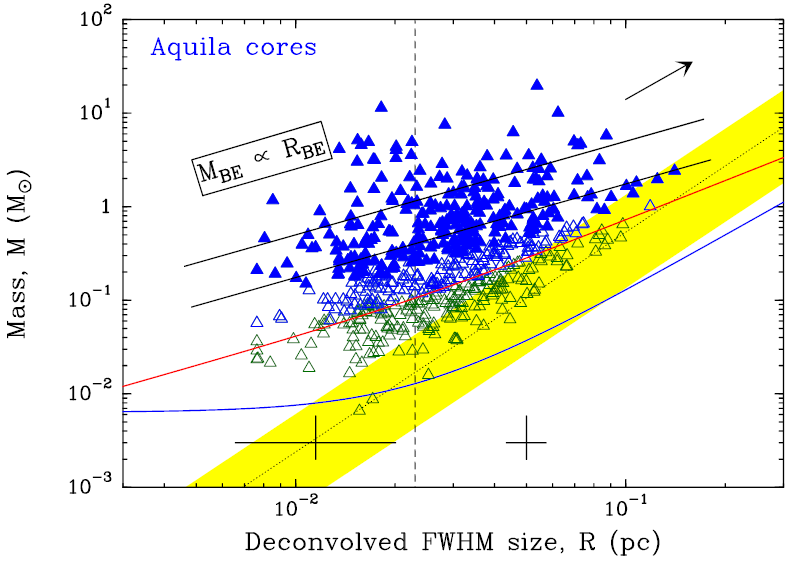}}
 \end{minipage}
   \end{center}
   \caption{Mass versus size diagram for the entire population of 651 starless cores identified with $Herschel$ in the 
    Aquila cloud. 
    The core FWHM sizes were measured with \textsl{getsources} in the high-resolution column density map (Fig.~\ref{fig_cd})
    and deconvolved from an $18.2\arcsec $ (HPBW) Gaussian beam; the vertical dashed line marks the corresponding physical $\overline{HPBW}$ 
    resolution at $d = 260$~pc. 
    The core masses were derived as explained in Sect.~\ref{sec:deriv_core_prop}. 
    Typical error bars are shown for both well-resolved and marginally-resolved cores (to the right and the left of the vertical dashed line, respectively). 
    The 292 {\it robust} prestellar cores (for which $\alpha_{\rm BE} \leq 2$ -- see text) are shown as filled blue triangles, 
    the other (candidate) prestellar cores as open blue triangles, and the rest of the starless cores as open green triangles.  
    The red curve shows the empirical lower envelope used to select the 446 {\it candidate} prestellar
    cores in the diagram 
    (i.e., $\alpha_{\rm BE} \leq 5 \times (HPBW_{\rm N_{H_2}}/FWHM_{\rm N_{H_2}})^{0.4}$ -- see text), based on the Monte-Carlo 
    simulations described in Sect.~\ref{sec:completeness}. 
    For comparison, models of critical isothermal Bonnor-Ebert spheres at $T = 7$~K and $T = 20$~K 
    are plotted as black solid lines.
    The mass--size correlation observed for diffuse CO clumps \citep[]{ElmegreenFalgarone1996} is  displayed as a shaded yellow band. 
    The blue curve marks a column density level corresponding to $5 \times N_{\rm H_2, rms} $, where $N_{\rm H_2, rms} $ is the 
    typical rms level of cirrus noise fluctuations at $A_{\rm V} \sim 7$ in the column density map (see Fig.~B.1 in Appendix~B).
    The arrow at the upper right indicates how the $Herschel$ cores (and the blue curve marking the cirrus noise level) would move in the diagram 
    using a distance of 415~pc instead of 260~pc (see Appendix~\ref{sec:appendix_distance}). 
   }
              \label{fig_massSize}%
    \end{figure}

In practice, we used the positions of the \textsl{Herschel} cores in a mass versus size diagram (Fig.~\ref{fig_massSize})
to distinguish between candidate prestellar cores and 
unbound starless cores, after deriving a reasonable lower 
envelope for self-gravitating cores in such a diagram. 
In our first-look papers \citep[][]{Konyves+2010, Andre+2010}, the criterion adopted to define this lower envelope was
simply $\alpha_{\rm BE} \leq 2$, 
by analogy with the usual criterion to select self-gravitating objects based 
on the virial mass ratio ($\alpha_{\rm vir} = M_{\rm vir} / M_{\rm obs}  \leq 2 $  -- e.g., \citealt{BertoldiMcKee1992}). 
Adopting the same criterion here led to a first sample of 292 {\it robust} prestellar cores, shown 
as filled blue triangles in Fig.~\ref{fig_massSize}. 
However, the Monte-Carlo simulations performed in Sect.~\ref{sec:completeness} below 
to assess the completeness of the survey suggest that 
this criterion may be too restrictive, in the sense that it selects only $\sim 85\% $ of the simulated 
BE cores detected by \textsl{getsources} after source classification. 
For this reason, we also derived a less restrictive lower envelope (shown as a red curve in Fig.~\ref{fig_massSize}) 
based on the results of our Monte-Carlo simulations.
This second, empirical lower envelope contains $> 95\% $ of the simulated BE cores after \textsl{getsources} extraction, 
and corresponds to the following, size-dependent limiting BE mass ratio: 
$\alpha_{\rm BE} \leq 5 \times (HPBW_{\rm N_{H_2}}/FWHM_{\rm N_{H_2}})^{0.4}$, where 
$FWHM_{\rm N_{H_2}}$ is the measured FWHM source diameter in the high-resolution column density map 
and $HPBW_{\rm N_{H_2}} = 18.2\arcsec $ is the HPBW resolution of the map. 
The limiting BE mass ratio varies from $\sim 2$ for well-resolved cores with $\overline{FWHM}_{\rm N_{H_2}} \sim 0.1$~pc
to $\sim 5$ for unresolved cores with $\overline{FWHM}_{\rm N_{H_2}} \sim \overline{HPBW}_{\rm N_{H_2}}$. 
The reason why one has to be more flexible and 
use a larger limiting BE mass ratio for unresolved or marginally resolved 
cores is that the intrinsic core radius (and therefore the intrinsic BE mass) is more uncertain for such cores.

Based on the latter criterion, 446 of the 651 starless cores in the Aquila entire field were 
classified as {\it candidate} prestellar cores. 
All of the 292 {\it robust} prestellar cores belong to the wider sample of 446 {\it candidate} prestellar cores.
These two samples of cores reflect the uncertainties in the classification of detected starless cores as gravitationally bound 
or unbound objects, which are fairly large for marginally-resolved cores. 
In the absence of higher-resolution observations, the status of the 155  {\it candidate} prestellar cores 
which do not match the first criterion ($\alpha_{\rm BE} \leq 2$) is more uncertain 
(these cores are marked as ``tentative bound'' in the last column of online Table~\ref{tab_der_cat_cores}). 
We will thus consider both samples of prestellar cores in the discussion presented in Sect.~\ref{sec:discuss} below.

The mass vs. size distribution of the entire population of selected starless 
cores (Fig.~\ref{fig_massSize}) shows a spread of deconvolved 
$\overline{FWHM}$ sizes between $\sim 0.01$~pc and $\sim 0.1$~pc 
and a range in core mass between $\sim 0.03\, M_\odot$ and  $\sim 10\, M_\odot$. 
The high fraction of self-gravitating cores ($\sim 45\%$ or $\sim 69\%$, depending on whether the {\it robust} or the {\it candidate} sample is adopted) 
is reflected in the locations of the Aquila starless cores in this mass vs. size diagram.
The selected {\it robust} prestellar cores are clustered around (or above) the mass--size relations expected for critical BE isothermal spheres with 
gas temperatures between 7~K and 20~K (parallel black solid lines in Fig.~\ref{fig_massSize}). 
Besides, they are more than an order of magnitude denser than typical CO clumps (yellow band in Fig.~\ref{fig_massSize}), 
which are mostly unbound structures \citep[e.g.,][]{ElmegreenFalgarone1996,Kramer+1998}. 

   \begin{figure}[!h]
   \begin{center}
 \begin{minipage}{1.0\linewidth}
    \resizebox{1.0\hsize}{!}{\includegraphics[angle=0]{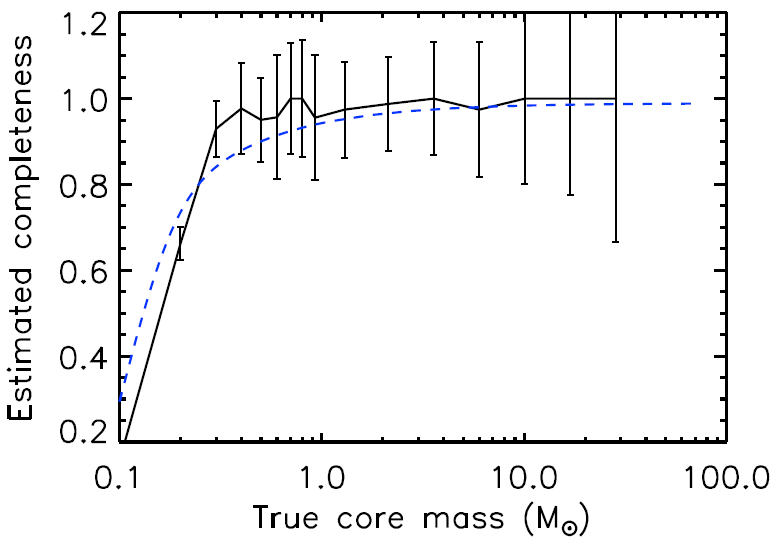}}
 \end{minipage}
   \end{center}
   \caption{
   Completeness curve of our {\it Herschel} sample of {\it candidate} prestellar cores 
   as a function of true core mass (solid line), as estimated from the Monte-Carlo simulations
   described in Sect.~\ref{sec:completeness}. For comparison, the dashed line shows the global completeness curve predicted by the model discussed 
   in Appendix~\ref{sec:appendix_model}.
   }
              \label{fig_complete}%
    \end{figure}

\subsection{Completeness of the prestellar core survey}\label{sec:completeness}

To estimate the completeness of our census of prestellar cores in Aquila, we performed Monte-Carlo simulations 
(see Appendix~\ref{sec:appendix_simulations}). 
We first constructed clean maps of the background emission at all {\it Herschel} wavelengths (including a column density plane), 
by subtracting the emission of the compact sources identified with \textsl{getsources}. 
We then inserted a population of $\sim 5600$ model Bonnor-Ebert-like\footnote{We use the term ``Bonnor-Ebert-like''  
because the model cores were given the density structure of critical isothermal Bonnor-Ebert spheres, but their dust temperature 
distributions resulted from radiative transfer calculations and were thus not strictly isothermal, in agreement 
with detailed observational studies of individual cores \citep[see, e.g.,][for the example of B68]{Roy+2014}.}
cores throughout the clean-background images to generate a full set of synthetic {\it Herschel} and column density images of the region. 
The model cores were given a flat mass distribution (d$N$/dlog$M$ $\propto$ $M^{-0.7}$) 
from $0.02\, M_\odot$ to $\sim 30\, M_\odot$ 
and were assumed to follow a $M \propto R$ mass versus size relation appropriate for isothermal spheres. 
The dust continuum emission from the synthetic Bonnor-Ebert cores in all {\it Herschel} bands 
was simulated using an extensive grid of spherical dust radiative transfer models constructed by us with the MODUST code 
\citep[e.g.,][]{Bouwman+2000, Bouwman2001PhDT}.
Compact source extraction for several sets of such synthetic skies was performed with \textsl{getsources} in the same way as for the observed images. 

Based on the results of these simulations (see  Appendix~\ref{sec:appendix_simulations}) for further details), 
we estimate that our {\it Herschel} census of {\it candidate} prestellar cores is $> 90\% $ 
complete above a true core mass of $\sim$0.3~$M_\odot$ which corresponds 
to an observed core mass of $\sim$0.2~$M_\odot$ on average, given that observed masses are typically 
underestimated by $\sim \, $20--30\%  due to the internal temperature structure of starless cores 
(see end of Sect.~\ref{sec:deriv_core_prop} and Fig.~\ref{fig_simu_mass_temp} in Appendix~\ref{sec:appendix_completeness}).  
Likewise, our sample of {\it robust} prestellar cores is estimated to be $\sim 80\% $ complete 
above a true core mass of $\sim$0.3~$M_\odot$ or an observed core mass of $\sim$0.2~$M_\odot$.
The completeness curve of the Aquila core survey as a function of true core mass is 
plotted in Fig.~\ref{fig_complete}.  
   
In reality, the completeness level of the core survey is expected to be background dependent. 
In an effort to assess the magnitude of this dependence, we constructed a simple model of the prestellar core population
and core extraction process described in Appendix~\ref{sec:appendix_model}. This model shows that the completeness of prestellar core 
extractions does decrease as background cloud column density and cirrus noise increase (see Fig.~\ref{fig_completeness}) 
but suggests that the global completeness curve of the prestellar core survey in Aquila is consistent with that inferred from our 
Monte-Carlo simulations (compare the dashed and the solid line in Fig.~\ref{fig_complete}).

Armed with a good understanding of the completeness of the core survey,  
we discuss in Sect.~\ref{sec:discuss} below the global properties of the dense core population  
and their connection with the filamentary structure of the cloud complex 
on the basis of statistically representative observational results.

\section{Discussion}\label{sec:discuss}

\subsection{Lifetimes of Herschel prestellar cores}\label{sec:lifetime}

As our {\it Herschel} survey provides an essentially complete census of prestellar cores in the Aquila cloud, 
the core statistics can be used to set constraints on the typical lifetime of prestellar cores and the timescale 
of the core formation process. 
Following a technique introduced by \citet{Beichman+1986} in the context of {\it IRAS} data, 
a rough estimate of the lifetime of prestellar cores can be obtained by comparing  
the number of starless cores found with {\it Herschel} to the number of Class~II YSOs 
detected by {\it Spitzer} in the same region. 
The underlying assumptions are 1) that all starless cores will evolve into YSOs in the future, and 
2) that  star formation proceeds at a roughly constant rate, at least when averaged over an entire cloud. 
In the $\sim 3^\circ \times 3^\circ $ field covered by {\it Herschel}, and excluding the dubious, small area 
with higher LSR velocities in the eastern corner of the column density map (see Fig.~\ref{fig_cd}  and Sect.~\ref{sec:mass_distrib}), 
our survey revealed a total of 651 starless cores, including 446 {\it candidate} and 292 {\it robust} prestellar cores,  while the combined c2d 
and Gould Belt {\it Spitzer} surveys detected 622 Class~II YSOs \citep[][Allen et al., in prep.]{Dunham+2013}.
Adopting a reference lifetime of 2~Myr for Class II YSOs \citep{Evans+2009}, these numbers lead  
to typical lifetimes of $\sim 2$~Myr, $\sim 1.4$~Myr, and $\sim 0.9$~Myr for the global populations of {\it Herschel} starless 
cores\footnote{The lifetime estimate quoted for starless cores is given under the assumption that {\it presently} unbound starless cores are still 
growing in mass and will become gravitationally bound and prestellar in the future \citep[cf.][]{Simpson+2011,Belloche+2011,KirkJ+2013}.}, 
{\it candidate} prestellar cores, and {\it robust} prestellar cores, respectively. 
Given the large sizes of the populations of starless cores and YSOs in Aquila, the main sources of error in these estimates 
come from the fact that some starless or even {\it candidate} prestellar cores may never evolve 
into YSOs, as they may be ``failed cores'' that will disperse before collapsing \citep[e.g.][]{Vazquez+2005}, 
and from the uncertainty in the number and lifetime of Class~II YSOs in Aquila. 
Combining the constraints coming from the two samples of observed prestellar cores, our best estimate of 
the global lifetime of the prestellar core phase is $t_{\rm pre} = 1.2 \pm 0.3$~Myr. 

\begin{figure}[h!]
\begin{center}
\includegraphics[scale=0.35,angle=0]{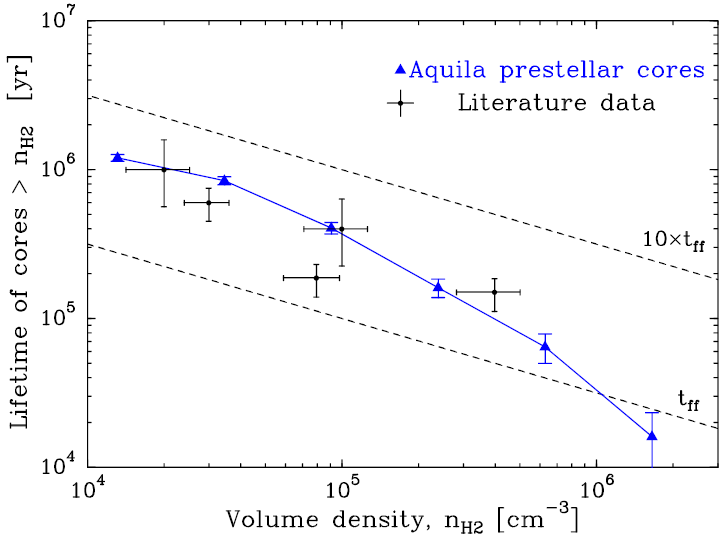}
\end{center}
\caption{Plot of estimated lifetime against minimum average volume density (blue solid line and filled triangles) 
  for the population of 446 {\it candidate} prestellar cores identified with $Herschel$ in the Aquila cloud (blue triangles), 
  similar to the ``JWT'' plot introduced by \citet{JessopWT2000}. 
  The error bars only reflect $\sqrt{N}$ counting uncertainties. 
  Literature data from 
  \citet{Ward+2007} are shown as black crosses for comparison. 
  The two parallel dashed lines correspond to the free-fall timescale ($t_{\rm ff}$) and a rough approximation 
  of the ambipolar diffusion timescale  ($10 \times t_{\rm ff}$).
  }
\label{fig_lifetime}       
\end{figure}

\begin{figure}[h!]
\begin{center}
\includegraphics[scale=0.35,angle=0]{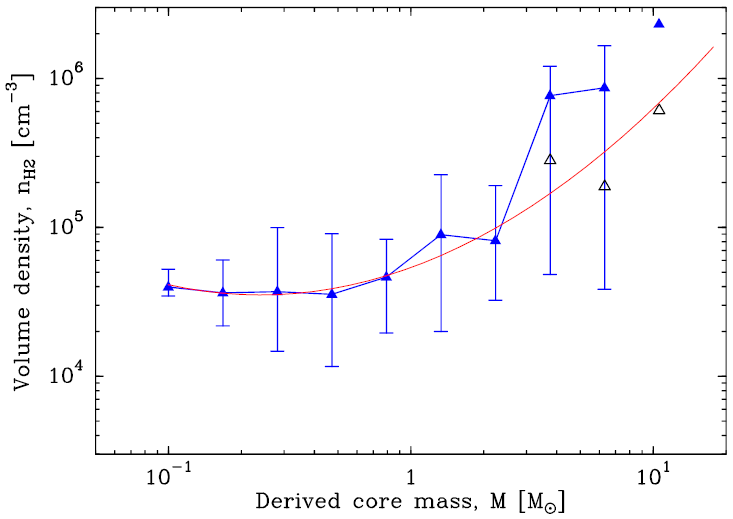}
\end{center}
\caption{Plot of average volume density versus observed core mass 
  for the sample of 446 {\it candidate} prestellar cores. The blue solid triangles mark 
  the median deconvolved volume density for each mass bin. 
  (For comparison, the black open triangles show the median 36.3\arcsec -beam-averaged densities 
  for the upper three mass bins.)
  The error bars correspond to the interquartile range of densities in each mass bin. 
  The data points become very uncertain at the high-mass end due to the small number 
  of cores in the higher mass bins (see core mass function in  Fig.~\ref{fig_CMF} below). 
  (No interquartile range can be plotted for the last bin which contains only two cores.)
  Note the weak, but significant, correlation between core density and core mass 
  above $\sim \, $2--3$\, M_\odot $. 
  The red curve represents a parabolic fit to the data points.
  }
\label{fig_dens_vs_mass}       
\end{figure}

   \begin{figure*}[!ht]
   \begin{center}
 \begin{minipage}{0.44\linewidth}
 \hspace{-0.2cm}
  \resizebox{1.05\hsize}{!}{\includegraphics[angle=0]{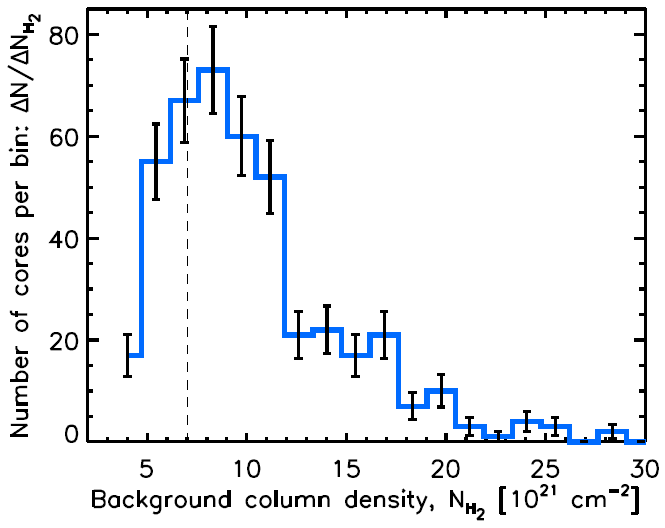}} 
 \end{minipage}
 \hspace{0.4cm}
 \begin{minipage}{0.50\linewidth}
     \resizebox{1.0\hsize}{!}{\includegraphics[angle=0]{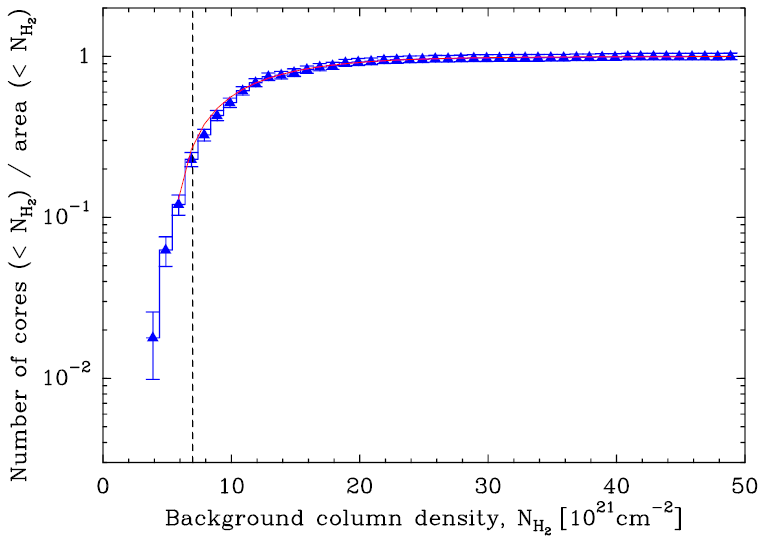}}
 \end{minipage} 
   \end{center}
   \caption{{\it Left:} Distribution of background cloud column densities for the population of 446 candidate prestellar cores identified with 
   {\it Herschel} in the whole Aquila field. The vertical dashed line marks a fiducial threshold at an equivalent visual extinction level 
   $A_{\rm V}^{\rm bg} \sim$ 7 mag \citep[cf.][]{Heiderman+2010,Lada+2010,Andre+2010,Andre+2014}. 
   {\it Right:} Normalized ``probability'' of finding a {\it Herschel} prestellar core as a function of background column density (blue histogram with error bars), 
   obtained by dividing the number of prestellar cores detected with {\it Herschel} below a given background column density level by the total 
   surface area covered by the HGBS survey below the same level. The red curve shows a simple fit of the form $P_{\rm core}(A_{\rm V})  = 1\, -\, {\rm exp}(a\times A_{\rm V} + b) $ 
   with $a= -0.17$ and $b= 0.86$. The vertical dashed line marks $A_{\rm V}^{\rm bg} \sim$ 7 mag as in the left panel.
   }
              \label{fig_bgCD}%
    \end{figure*}

We have a large enough sample of cores in Aquila to
investigate a possible trend between core lifetime and core density. 
Figure~\ref{fig_lifetime} shows a plot of estimated core lifetime versus average volume density, similar to that introduced by  
\citet{JessopWT2000}, but for the sample of $Herschel$-identified candidate prestellar cores in Aquila.  
In this plot, the Aquila data are represented by blue triangles and compared to literature data (black crosses) 
from \citet{Ward+2007}. The blue solid line and filled triangles represent the estimated trend between core lifetime 
and average core density, where the latter quantity was obtained by dividing the observed mass of each core 
by the deconvolved estimate of its volume [i.e., core density reported in col.~(16) of online Table~A.2]. 
As can be seen, the plot suggests that the typical lifetime of prestellar cores decreases 
from $\sim$1.4~Myr for cores with average volume density $\simgt 10^4\, {\rm cm}^{-3} $
to a few times $10^4$~yr for cores with average volume density $\simgt 10^5$-$10^6\, {\rm cm}^{-3} $. 
Moreover, the estimated core lifetimes lie between one free-fall time ($t_{\rm ff}$, lower dashed line in Fig.~\ref{fig_lifetime}), 
the timescale expected in free-fall collapse, and $10 \times t_{\rm ff}$ (upper dashed line in Fig.~\ref{fig_lifetime}), roughly the timescale 
expected for highly subcritical cores undergoing ambipolar diffusion \citep[e.g.,][]{Mouschovias1991}. 
At the median average volume density $\sim 4\times 10^4\, {\rm cm}^{-3} $ of the candidate prestellar cores identified with {\it Herschel},  
the estimated core lifetime is $\sim 0.75$~Myr or $\sim \, $4--5$\,  t_{\rm ff}$. 
The densest cores in our sample, which have beam-averaged volume densities $\simgt 2 \times 10^5\, {\rm cm}^{-3} $ at the 
resolution of the 500~$\mu$m data and average deconvolved volume densities $\simgt 10^6\, {\rm cm}^{-3} $, have a much  
shorter lifetime $\sim \, $0.02--0.05~Myr or $\sim t_{\rm ff}$, suggesting they may evolve essentially on a free-fall timescale. 
Indeed, the tentative presence of a power-law tail in the distribution of beam-averaged core densities above $\sim 10^5\, {\rm cm}^{-3} $ 
(see Fig.~\ref{fig_dens}b) suggests that these cores may be undergoing nearly free-fall collapse.

In this context, it is worth pointing out that density may not be the only relevant parameter and 
that core evolution may also be mass dependent as suggested by, e.g., \citet{Hatchell+2008}. 
Indeed, we observe a weak positive correlation between core density and core mass above $\sim \, $2--3$\, M_\odot $ 
(see Fig.~\ref{fig_dens_vs_mass}), indicating that the most massive prestellar cores  in our sample 
tend to be the densest objects. Assuming that the lifetime of a core is proportional to its free-fall time,  
the correlation in Fig.~\ref{fig_dens_vs_mass} suggests that prestellar cores more massive than $\sim \, $2--3$\, M_\odot $ 
may evolve on significantly shorter timescales than the majority of the cores in our sample,  
which have masses $\sim \, $0.1--2$\, M_\odot $ (see Fig.~\ref{fig_CMF} below). 
This finding would be consistent with the results of earlier searches for high-mass prestellar cores 
(i.e., precursors to stars $> 8\, M_\odot $) in massive star-forming regions which have shown that  
such cores, if they exist at all, are extremely rare with lifetimes comparable to (or shorter than) the free-fall timescale 
\citep[][]{Motte+2007}.

\subsection{Evidence of a column density threshold for prestellar core formation }\label{sec:threshold}

Figure~\ref{fig_bgCD}a shows the distribution of  background cloud column densities ($N_{\rm H_{2}}^{\rm bg}$) 
for the entire population of 446 candidate prestellar cores identified with {\it Herschel} in the Aquila cloud 
(see Sects.~\ref{sec:getsources} \& \ref{sec:bound_selection}).  
This distribution shows a steep rise above $A_{\rm V}^{\rm bg} \sim 5$ 
and is such that most ($\sim$90\%)\footnote{More precisely, $88\% $ of the {\it candidate} prestellar cores and $92\% $ of  the {\it robust} 
prestellar cores lie at $A_{\rm V}^{\rm bg} > 7$.}
prestellar cores
are found above a background column density corresponding to $A_{\rm V}^{\rm bg} \sim 7$ 
and a background gas surface density $\Sigma^{\rm bg} \sim 150~M_{\odot}$/pc$^2$. 
As already emphasized by \citet{Andre+2010,Andre+2014}, the shape of the distribution shown in Fig.~\ref{fig_bgCD}a 
strongly supports the existence of a column density threshold for the formation of prestellar cores. 
The existence of such a threshold had been suspected for a long time, based on the results of ground-based millimeter 
and submillimeter surveys for cores in, e.g., the Taurus, Ophiuchus, and Perseus clouds 
\citep[e.g.,][]{Onishi+1998,Johnstone+2004,KirkH+2006}. These early claims, however, were not completely convincing 
due to the limited column density sensitivity and spatial dynamic range of ground-based observations, 
hence their limited capability to probe prestellar cores and the parent background cloud simultaneously. 
The {\it Herschel} results presented in this paper provide a much stronger case for a (column) density threshold. 
We stress that the distribution of cloud mass as a function of column density 
(see Fig.~\ref{fig_cdPDF}b in Sect.~\ref{sec:mass_distrib})
and the background-dependent completeness level of our survey for prestellar cores 
make the threshold even more significant than Fig.~\ref{fig_bgCD}a suggests. 
Indeed, $\sim 85\% $ of the mass in the Aquila cloud is at column densities lower than $A_{\rm V} \sim 7$ 
(see Fig.~\ref{fig_cdPDF}b) and $\sim 95\% $ of the surface area covered by the {\it Herschel} survey is 
below $A_{\rm V} \sim 7$. 
Furthermore, the completeness level of our {\it Herschel} census for prestellar cores is not limited  
by sensitivity (as was typically the case for earlier ground-based surveys), but by ``cirrus confusion noise'' 
(see Appendix~B), and is better in $A_{\rm V} < 7 $ areas than in $A_{\rm V} > 7 $ areas (see Fig.~\ref{fig_completeness}). 
Therefore, if prestellar cores were distributed randomly in the cloud, we would be much more likely to detect prestellar 
cores in $A_{\rm V} < 7 $ areas than in higher column density regions. 
Figure~\ref{fig_bgCD}a already shows that this is clearly not the case. 
To further strengthen the point, we plot in Fig.~\ref{fig_bgCD}b a probability function of finding a prestellar core as a function of 
background column density, obtained by normalizing the number of prestellar cores detected below a given 
background column density by the total surface area imaged by {\it Herschel} below the same background column 
density level (for a related probability function in the case of the submm continuum cores detected by SCUBA in Perseus, see \citealp{Hatchell+2005}). 
The probability function, $P_{\rm core}^{\rm obs}(A_{\rm V}) $, shown in Fig.~\ref{fig_bgCD}b increases by more than an order of magnitude between 
$A_{\rm V} \sim 4$ and $A_{\rm V} \sim 10$, and looks like a smooth step function. 
It is very well fit by the simple exponential step function $P_{\rm core}(A_{\rm V})  = 1\, -\, {\rm exp}(-0.17\times A_{\rm V} + 0.86) $. 

\subsubsection{Comparison with models of the star formation rate}\label{sec:sfr}

\begin{figure}[]
   \begin{center}
\begin{minipage}{1.0\linewidth}
   \resizebox{0.95\hsize}{!}{\includegraphics[angle=0]{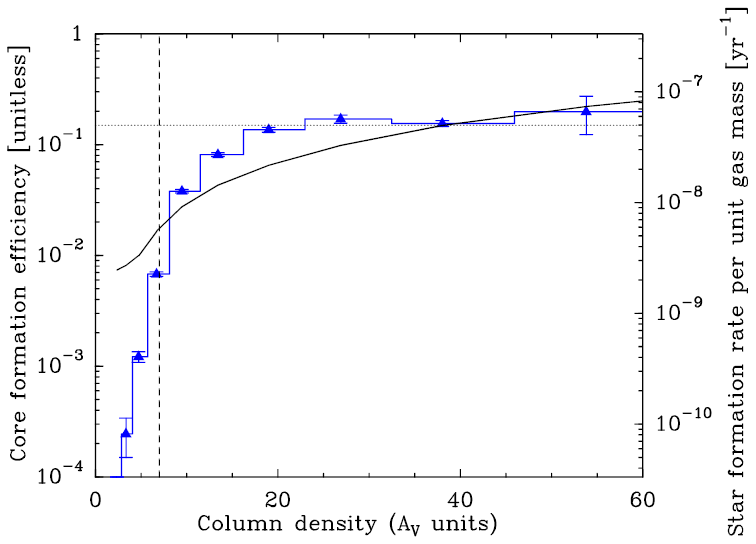}}
  \end{minipage}
   \end{center}
   \caption{Plot of the observed differential core formation efficiency (CFE) as a function of background column density expressed in $A_{\rm V}$ units 
   (blue histogram with error bars), 
   obtained by dividing the mass in the form of prestellar cores 
   in a given column density bin by the cloud mass observed in the same column density bin. 
   The right axis gives the corresponding star formation rate per unit gas mass (in units of yr$^{-1}$), estimated assuming 
   a local star formation efficiency $\epsilon_{\rm core} = 0.4 $ 
   at the core level and a prestellar core lifetime $t_{\rm pre} = 1.2$~Myr (see text).
   For comparison, the black curve shows the prediction of the multi-freefall version
   of the turbulence regulated model of the star formation rate initially proposed by \citet{KrumholzMcKee2005} \citep[see][]{HennebelleChabrier2011}. 
   The vertical dashed line marks the same fiducial threshold at     
   $A_{\rm V}^{\rm bg} \sim$ 7 as in Fig.~\ref{fig_bgCD}. 
   The horizontal dotted line marks the rough asymptotic value of the CFE  $\sim 15\% $ at $A_{\rm V} >  15$, corresponding 
   to a star formation rate per unit gas mass   $\sim 5 \times 10^{-8} \, {\rm yr}^{-1}$. 
   }
              \label{fig_cfe}%
   \end{figure}

There is some debate in the literature as to whether the kind of results shown in Fig.~\ref{fig_bgCD} reflect a true column density threshold 
for star formation or whether the efficiency of the star formation process simply increases gradually with (column) density \citep[cf.][]{Hatchell+2005}.
Starting with the work of \citet{KrumholzMcKee2005}, 
a number of theoretical models of the star formation rate (SFR) in 
molecular clouds have been proposed based on the general idea that star formation is regulated by interstellar turbulence and 
that clouds typically convert $\epsilon_{\rm ff} \sim 1\% $ of their molecular gas mass into stars per (local) free-fall time 
(e.g., \citealp{PadoanNordlund2011,HennebelleChabrier2011,Krumholz+2012} -- see also \citealp{FederrathKlessen2012} 
and \citealp{Padoan+2014} for overviews and comparisons of the models). 
In the ``multi-freefall'' versions of these theoretical models, which are most appropriate 
to fit real observations\footnote{In the initial, ``single-freefall'' model of \citet{KrumholzMcKee2005}, the relevant timescale is 
the free-fall time evaluated at the mean density of the cloud, $t_{\rm ff}(\rho_0)$, and there is no density dependence at all. 
\citet{HennebelleChabrier2011} and \citet{FederrathKlessen2012} have shown that this model generally underestimates the SFRs determined 
by \citet{Heiderman+2010} in nearby clouds.} 
\citep[cf.,][]{HennebelleChabrier2011, FederrathKlessen2012}, there is not necessarily any sharp (column) density 
threshold, but the SFR drops significantly at low densities because of a significant increase in the local free-fall time 
(see also the related discussion by \citealp{Burkert2013}).
In Fig.~\ref{fig_cfe}, we compare the observed core formation efficiency (CFE) as a function of background column density 
with the prediction of the simplified multi-freefall model of \citet{HennebelleChabrier2011}. 
Here, we define the observed core formation efficiency as $ {\rm CFE_{obs}}(A_{\rm V}) = \Delta M_{\rm cores}(A_{\rm V})/\Delta M_{\rm cloud}(A_{\rm V}) $ 
where $ \Delta M_{\rm cores}(A_{\rm V}) $ is the mass 
of the prestellar cores\footnote{Both $ \Delta M_{\rm cores}(A_{\rm V}) $ and $  \Delta M_{\rm cloud}(A_{\rm V}) $ 
represent {\it observed} masses directly estimated from the {\it Herschel} data using SED dust temperatures and the dust opacity 
of assumptions given in Sect.~\ref{sec:res_analys}.
$ \Delta M_{\rm cores}(A_{\rm V}) $ was {\it not} corrected 
for the small $\sim 25\%$ effect due to the fact that the SED mass values 
tend to slightly underestimate the intrinsic core masses  
according to our simulations (see Sect.~\ref{sec:deriv_core_prop} and Appendix~\ref{sec:appendix_simulations}).
}
identified with {\it Herschel} in a given bin of background $A_{\rm V}$ values and 
$  \Delta M_{\rm cloud}(A_{\rm V}) $ is the cloud mass estimated from the  {\it Herschel} column density map in the same $A_{\rm V}$ bin. 
In the multi-freefall model, the fraction of gas mass 
converted into core mass per 
unit time is simply $\frac{\epsilon_{\rm ff}}{\epsilon_{\rm core}} \times \frac{1} {t_{\rm ff}(\rho)} $, where 
$\epsilon_{\rm ff} \sim 1\% $ (see above), 
$\epsilon_{\rm core} \sim 40\% $ 
is the star formation efficiency at the level of an individual prestellar core (see Sect.~\ref{sec:CMF} below), 
and $t_{\rm ff}(\rho) $ is the {\it local} 
free-fall time at the local gas density $\rho $. 
Over the typical lifetime of prestellar cores $t_{\rm pre} \sim 1$~Myr (see Sect.~\ref{sec:lifetime}), 
the expected core formation efficiency is thus: 
$$ {\rm CFE_{mff}}(\rho) = \frac{\epsilon_{\rm ff}}{\epsilon_{\rm core}} \times \frac{t_{\rm pre}} {t_{\rm ff}(\rho)}. $$ 
In order to use this formula, we had to estimate the local gas density and free-fall time in the Aquila cloud.
To do so, we made use of the fact that the cloud surface area above a given column density level $S(> N_{\rm H_2}) $ scales 
as the column density PDF shown in Fig.~\ref{fig_cdPDF}a and in particular features a well-defined power-law tail 
$S(> N_{\rm H_2}) \propto N_{\rm H_2}^{-2.9}$ at high column densities ($A_{\rm V} > \, $5--7). 
In spherical geometry, this is indicative\footnote{In the case of a cloud with a spherical radial density distribution, $\rho \propto r^{-\alpha}$, it is easy
to show that both the column density PDF, $dN$/$d$log$N_{\rm H_2}$, and the surface area, $S(> N_{\rm H_2}) $, scale 
as ${N_{\rm H_2}}^m$, where $m = \frac{2}{1-\alpha}$ \citep[see, e.g.,][]{FederrathKlessen2013}.} 
of a power-law density distribution
$\rho \propto r^{-1.7}$ for the dense gas 
and is consistent with large-scale cloud contraction above $A_{\rm V} > \, $5--7. 
Under the assumption of a roughly spheroidal ambient cloud, we then derived the effective volume density, 
$n_{\rm H_2}(A_{\rm V}) $, and effective free-fall time, $t_{\rm ff}(A_{\rm V}) $, of the gas as a function of 
background cloud density expressed in $A_{\rm V}$ units. 
Applying the above multi-freefall formula, this allowed us to obtain the core formation efficiency,
$ {\rm CFE_{mff}}(A_{\rm V}) = \frac{\epsilon_{\rm ff}}{\epsilon_{\rm core}} \times \frac{t_{\rm pre}} {t_{\rm ff}(A_{\rm V})} $, 
predicted by the multi-freefall model as a function of $A_{\rm V}$, for direct comparison with $ {\rm CFE_{obs}}(A_{\rm V}) $.
As can be seen in Fig.~\ref{fig_cfe}, the {\it Herschel} observations indicate a much sharper transition than the multi-freefall model does, 
between a regime of negligible prestellar core formation efficiency at $A_{\rm V} <  5$ and a regime 
of roughly constant CFE $\sim 15\% $ at $A_{\rm V} >  15$. 
Furthermore, we stress that differential completeness between low and high column density areas 
(see Fig.~\ref{fig_completeness} in Appendix~B) implies that the real transition between the two regimes 
is in fact somewhat sharper than indicated by the blue histogram in Fig.~\ref{fig_cfe}. 
On this basis, we argue for the presence of a true {\it physical} threshold for prestellar core formation around a fiducial 
value $A_{\rm V} \sim  7$, although the observed transition is clearly not infinitely sharp like a true Heaviside step function.

%

   \begin{figure*}[!ht]
   \begin{center}
 \begin{minipage}{1.0\linewidth}
    \resizebox{1.0\hsize}{!}{\includegraphics[angle=0]{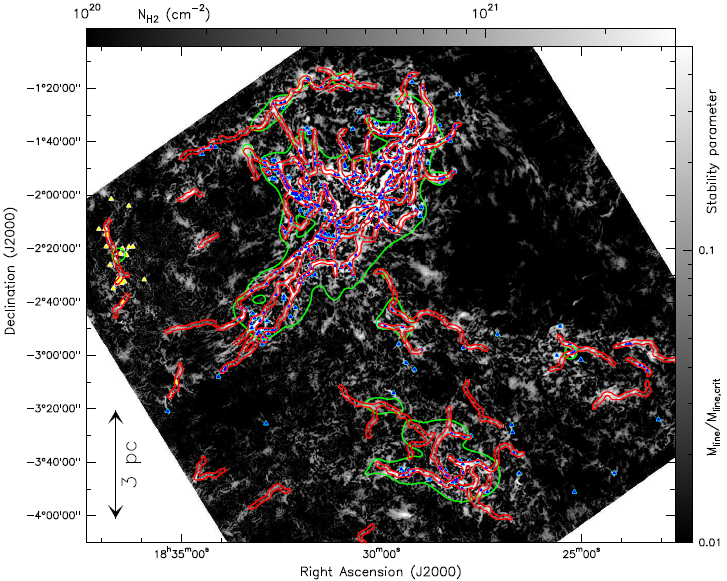}}
 \end{minipage}
   \end{center}
   \caption{ Comparison of the spatial distribution of the prestellar core population identified in Sect.~\ref{sec:bound_selection} using \textsl{getsources} 
   (blue triangles) with the footprints of all the filaments traced with DisPerSE (red contours), both overlaid on the curvelet component 
   of the high-resolution column density map (cf. Fig.~\ref{fig_fil-cur}). The gray scale corresponds to the color scale of Fig.~\ref{fig_fil-cur}. 
   The red contours outline 0.1~pc--wide footprints around the crests of filaments. The green contours correspond to $A_V = 7$ in the column density map
   smoothed to a resolution of $5\arcmin$.
   } 
              \label{fig_cores-critFilCont}%
    \end{figure*}

Interestingly, a very similar extinction threshold at $A_V^{\rm back} \sim 7$ has {\it independently} been observed with {\it Spitzer} 
in the spatial distribution of YSOs in nearby clouds  (\citealp{Heiderman+2010}; \citealp{Lada+2010}; \citealp{Evans+2014} -- see also Sect.~\ref{sec:SFR} below).
Following  \citet{Andre+2010,Andre+2014}, we interpret this star formation threshold in terms of the 
quasi-universal filamentary structure of molecular clouds in Sect.~\ref{sec:budget} below.

   \begin{figure}[!ht]
   \centering
 \begin{minipage}{1.0\linewidth}
    \resizebox{1.0\hsize}{!}{\includegraphics[angle=0]{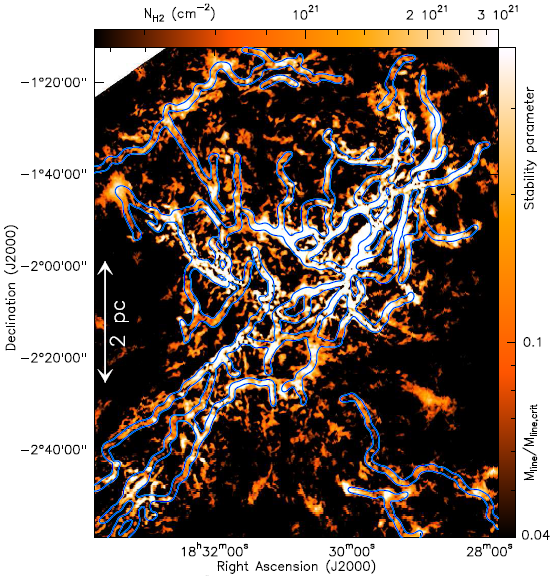}}
    \resizebox{1.0\hsize}{!}{\includegraphics[angle=0]{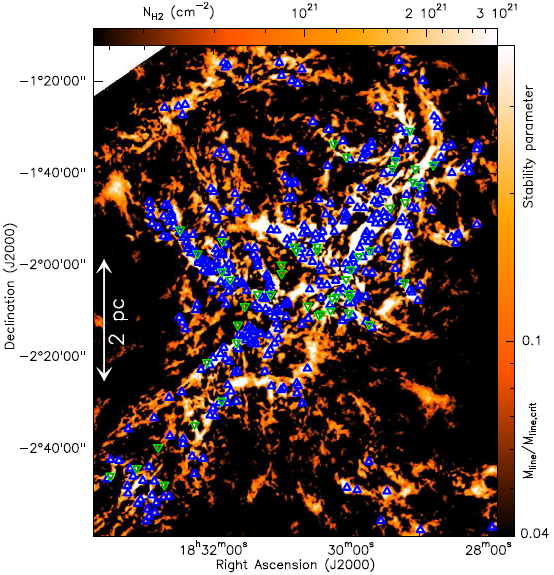}}
 \end{minipage}
   \caption{
	   {\it Upper:} Curvelet 
	   component \citep[cf.][]{Starck+2003} of a portion of the {\it Herschel} high-resolution column density map 
           shown in Fig.~\ref{fig_cd}. 
           Given the typical filament width of $\sim$0.1~pc \citep{Arzoumanian+2011}, 
           this map is equivalent to a map of the 
	   mass per unit length along the filaments \citep[cf.][]{Andre+2010}, as indicated by the color bar on the right. The white areas highlight 
	   regions of the map where the filaments have a mass per unit length larger than half the critical value $M_{\rm line, crit} = 2 c^2_{\rm s} /G$ 
	   \citep[cf.][]{InutsukaMiyama1997} and are thus likely to be gravitationally unstable (see Sect.~\ref{sec:filam} and Fig.~\ref{fig_fil-cur}). 
	   The contours overlaid in blue outline the 0.1~pc--wide footprints of the filaments traced with DisPerSE in Sect.~\ref{sec:filam}
	   (cf. red contours in Fig.~\ref{fig_cores-critFilCont}).
           {\it Lower:} Same map as in the {\it upper} panel with the locations of candidate prestellar cores and protostellar cores overlaid as 
	   blue triangles and green triangles, respectively.
           }
              \label{fig_curvelet}%
    \end{figure}

\subsection{Spatial distribution of Herschel cores and connection with filaments}\label{sec:spatial_distrib}

As already pointed out in earlier HGBS papers \citep[e.g.,][]{Andre+2010,Menshchikov+2010}, 
there is a very close correspondence between the spatial distribution of compact dense cores and 
the network of filaments identified in the {\it Herschel} column density map of the Aquila cloud. 
Furthermore, candidate prestellar cores and embedded protostars are preferentially found within the 
densest filaments with supercritical masses per unit length 
(i.e., $M_{\rm line} > M_{\rm line, crit} \equiv  2\, c_s^2/G $ -- see Sects.~\ref{sec:intro} \& ~\ref{sec:filam}) 
\citep[e.g.,][]{Andre+2010,Andre+2014}. 

The connection between cores and filaments is illustrated in Figs.~\ref{fig_cores-critFilCont} \& ~\ref{fig_curvelet} 
and can be quantified in detail based on the census of cores 
presented in Sects.~\ref{sec:getsources}, ~\ref{sec:core_selection}, \& ~\ref{sec:bound_selection} 
and the census of filaments described in Sect.~\ref{sec:filam}. 
To this end, a mask image of the filament ``footprints'' was constructed by convolving the filamentary skeleton traced with 
DisPerSE and shown in Fig.~\ref{fig_fil-cur} and Fig.~\ref{fig_fil-gf} 
with a Gaussian kernel corresponding to a typical filament inner width $\sim 0.1$~pc \citep{Arzoumanian+2011}, 
i.e., an angular width $\sim 80\arcsec $ at the distance of Aquila. 
An alternative mask image of the filaments, similar to that shown in Fig.~\ref{fig_fil-gf}, was created 
by considering all transverse angular scales up to $ 80\arcsec $ in the multi-scale decomposition performed by \textsl{getfilaments} (see Sect.~\ref{sec:filam}). 
The core positions were then compared with these two sets of 0.1-pc filament footprints to estimate the fraction of cores associated with filaments.
The results of this comparison, summarized in Table~\ref{tab_ONfil1}, 
indicate that a very high fraction ($\mathop{75\%}_{\text{-5\%}}^{\text{+15\%}}$) of prestellar cores are closely associated with filaments, i.e., 
lie within 0.1-pc filament footprints. 
This correspondence is illustrated in Fig.~\ref{fig_cores-critFilCont}, where the 0.1-pc footprints of the filaments 
traced with DisPerSE (cf. Fig.~\ref{fig_fil-cur})
are outlined by red contours and the population of 446 candidate prestellar cores identified with \textsl{getsources} in Sect.~\ref{sec:res_analys} 
are superimposed as blue triangles. 
It can be seen that most ($\sim 70\% $) of the {\it candidate} prestellar cores lie within the red filament footprints. 
Likewise, $\sim 80\% $ of the {\it robust} prestellar cores lie within the red filament footprints. 
A more detailed view of this connection in the Aquila ``main subfield'', including Serpens-South and W40, 
is provided by Fig.~\ref{fig_curvelet} which shows the locations of candidate prestellar cores 
overlaid on the curvelet component  \citep[cf.][]{Starck+2003} of the high-resolution column density map  \citep[see][for an early version of the same view]{Andre+2010}. 
It is important to stress that the connection between cores and filaments does not strongly depend 
on the precise definition adopted for a filament or on the algorithm used to trace filaments. 
In particular, as can be seen in Table~\ref{tab_ONfil1}, 
the values found for the fractions of cores associated with filaments using \textsl{getfilaments} footprints are very similar to the values 
found using DisPerSE footprints.

Table~\ref{tab_ONfil1} also reports the fractions of cores found within supercritical portions of filaments. 
For the sake of simplicity, in the present paper focusing primarily on {\it cores}, 
our classification of filament segments as either supercritical or subcritical relies on the assumption of a constant filament 
width $\sim 0.1$~pc \citep{Arzoumanian+2011} and is based on the local column density measured in the clean background 
column density image (after subtracting the contribution of cores with \textsl{getsources}). 
To take into account the fact that the transverse column density profiles of supercritical filaments feature power-law wings 
which extend beyond the 0.1~pc inner width \citep{Arzoumanian+2011,Palmeirim+2013}, 
we also considered 0.2~pc-wide filament footprints and provide corresponding core fractions in 
Table~\ref{tab_ONfil1}. 
For example, the well-studied Serpens South filament and Taurus--B211/B213 filament have 
transverse column density profiles which extend up to $\sim$0.4--0.5~pc in radius on average 
and equivalent widths of $\sim 0.2$~pc \citep{Hill+2012,Palmeirim+2013}. 
While a detailed discussion of the radial column density profiles of the present filament sample is out of the scope of this paper, 
simple comparison of the line masses obtained by integration over the filament profiles with the line masses derived from the 
central column densities of the filaments using a characteristic inner width of 0.1~pc suggests that the 
equivalent width\footnote{Here, we define the equivalent width of a filament as the effective width $W_{\rm eff}$ 
such that the line mass integrated over the filament profile is $M_{\rm line}^{\rm int} = \Sigma_0 \times W_{\rm eff}$, 
where $ \Sigma_0 $ is the central surface density of the filament.} 
of the supercritical filaments traced here with DisPerSE (see end of Sect.~\ref{sec:filam}) is also typically $\sim 0.2$~pc.

\begin{table}\small\setlength{\tabcolsep}{4.5pt}
\caption{Fractions of cores associated with 
filaments in Aquila.}
\label{tab_ONfil1}      
{\renewcommand{\arraystretch}{1.1}        
\begin{tabular}{l|c|c|c|c}     
\hline\hline       
                     &  \multicolumn{2}{c|}{DisPerSE}  &  \multicolumn{2}{c}{\textsl{getfilaments}} \\ 
\hline
                     &  0.1~pc      &   0.2~pc        &   0.1~pc            &   0.2~pc              \\  
\hline
\multicolumn{5}{l}{\bf{All filaments:}}                                                             \\                       
  prestellar ON-fil. & 71\%--78\%   &  81\%--88\%     &  83\%--87\%         &  84\%--89\%           \\  
  starless ON-fil.   & 60\%         &  72\%           &  75\%               &  77\%                 \\
\hline            
\multicolumn{5}{l}{\bf{Supercritical segments:}}                                                    \\                        
  prestellar ON-fil. & 66\%--75\%   &  76\%--84\%     &  76\%--81\%         &  77\%--83\%           \\  
  starless ON-fil.   & 55\%         &  66\%           &  67\%               &  69\%                 \\
\hline                 
\end{tabular}
}
\tablefoot{The upper part of this table gives the fractions of prestellar/starless cores 
found inside the 0.1~pc and 0.2~pc--wide filament footprints constructed with DisPerSE and \textsl{getfilaments}
over the Aquila entire field (see text). 
The lower part of the table provides similar cores fractions when only supercritical portions of the filaments are 
considered. Here, for the sake of simplicity, a portion of a filament was classified as either supercritical or subcritical 
based on whether the local column density in the clean background column density image (after subtracting the contribution 
of cores with \textsl{getsources}) was equivalent to $A_V^{\rm cl, back} > 7$ or $A_V^{\rm cl, back} < 7$, respectively, 
assuming a constant filament width $\sim 0.1$~pc (see Sect.~\ref{sec:filam} and Sect.~\ref{sec:budget}).
The lower fractions of the ranges quoted for prestellar cores correspond to {\it candidate} prestellar cores, 
the higher fractions to {\it robust} prestellar cores.} 
\end{table}

\subsection{Mass budget in the  cloud and interpretation of the star formation threshold in terms of the filamentary structure}\label{sec:budget}

\begin{figure}[!h]
   \begin{center}
 \begin{minipage}{1.0\linewidth}
    \resizebox{1.0\hsize}{!}{\includegraphics[angle=0]{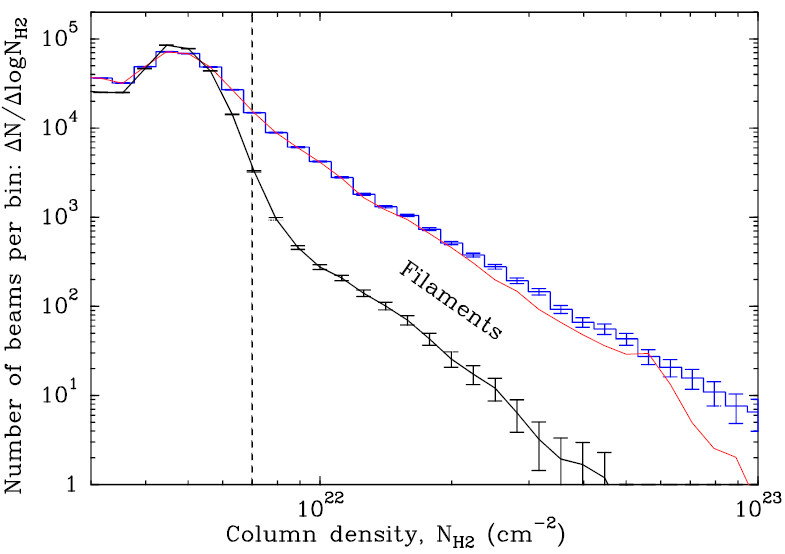}}
 \end{minipage}
   \end{center}
   \caption{Comparison of the global column density PDF in the Aquila cloud (blue histogram with statistical error bars, identical to that shown in Fig.~\ref{fig_cdPDF}) 
  to the column density PDF measured after subtraction of dense cores (red solid line) 
  and the PDF measured after subtraction of both dense cores {\it and} filaments (black solid line with statistical error bars). 
  The vertical dashed line marks the same fiducial threshold at  $A_{\rm V}^{\rm back} \sim$ 7 as in Fig.~\ref{fig_bgCD}. 
  This plot illustrates that filaments make up a dominant ($\sim 50\%$--$75\% $) fraction of the dense gas mass at $A_V > 7$--10, 
  and that dense cores contribute only a small ($\simlt 15\% $) fraction of the dense gas (except perhaps at the very highest column densities).
	   }
	      \label{fig_budget_pdf}%
    \end{figure}

Our {\it Herschel} census of prestellar cores and filaments allows us to derive a detailed mass budget in the Aquila cloud. 
Below the fiducial column density threshold at $A_V^{\rm back} \sim 7$, $\sim 10\%$--$20\% $ of the gas mass is in the form of 
(mostly subcritical) filaments and $< 1\% $ of the cloud mass is in the form of prestellar cores. 
Above $A_V^{\rm back} \sim 7$, $\sim  50\%$--$60\% $ of the cloud mass is in the form of (mostly supercritical) 
filaments and a fraction $f_{\rm pre} \sim 15\% \pm 5\% $ of the mass is in the form of prestellar cores. 
We note that $f_{\rm pre}$ roughly corresponds to the asymptotic core formation efficiency 
value reached at $A_V^{\rm back} > 15$ in Fig.~\ref{fig_cfe}. 
The fraction of cloud mass in the form of filaments reaches a very high value $\sim  75\%$ above $A_V^{\rm back} \sim 10$. 
In attempt to quantify further the relative contributions of cores and filaments to the cloud material 
as a function of column density, we compare in Fig.~\ref{fig_budget_pdf} the column density PDFs 
observed for the cloud before any component subtraction (blue histogram, identical to the PDF shown in Fig.~\ref{fig_cdPDF}),
after subtraction of dense cores (red solid line), and after subtraction of both dense cores and filaments (black solid line). 
To generate this plot, we used \textsl{getsources} to create a column density map of the cloud after subtracting the contribution 
of all compact cores, and \textsl{getfilaments} to construct another column density map after also subtracting the contribution 
of filaments. Although there are admittedly rather large uncertainties involved in this two-step subtraction process, the 
result clearly suggests that filaments dominate the mass budget of the Aquila cloud at high densities.

Since filaments appear to make up a dominant fraction of the dense gas material at $A_V^{\rm back} \ge 7$
within which the vast majority of prestellar cores are observed (see Fig.~\ref{fig_bgCD}), and since 
the spatial distribution of prestellar cores is strongly correlated with filaments (see Sect.~\ref{sec:spatial_distrib}), 
it is tempting to interpret the 
star formation threshold discussed in Sect.~\ref{sec:threshold} in terms of the quasi-universal 
filamentary structure of molecular clouds \citep[cf.][]{Andre+2014}.
Given the typical width $W_{\rm fil} \sim 0.1$~pc measured for 
filaments (Arzoumanian et al. 2011) 
and the relation $M_{\rm line} \approx \Sigma_0 \times W_{\rm fil}$ between the central 
gas surface density $\Sigma_0$ 
and the mass per unit length $M_{\rm line}$ of a filament \citep[cf. Appendix~A of][]{Andre+2010}, 
the threshold at $A_V^{\rm back} \sim 7$ or $\Sigma_{\rm gas}^{\rm back} \sim $~150~$M_\odot \, {\rm pc}^{-2} $ 
corresponds to within 
a factor\footnote{Strictly speaking, the formal agreement between $\Sigma_{\rm gas}^{\rm back}  \times W_{\rm fil}$ and $M_{\rm line, crit}$ 
is even better than 10\%. For several reasons (e.g., factor of $\sim 2$ spread in filament width and distribution of filament inclination angles), 
however, the column density threshold is not a sharp boundary but a smooth transition \citep[see discussion in Sect.~6.2 of ][]{Andre+2014}, 
as also observed in Fig.~\ref{fig_bgCD}b.} 
of $< 2$ to the critical mass per unit length $M_{\rm line, crit} = 2\, c_s^2/G \sim 16\, M_\odot \, {\rm pc}^{-1} $  
of nearly isothermal, long cylinders (see Inutsuka \& Miyama 1997)  
for a typical gas temperature $T \sim 10$~K.
Thus, the prestellar core formation threshold approximately corresponds to the {\it threshold above which 
interstellar filaments become gravitationally unstable} \citep{Andre+2010}.

\subsection{Prestellar CMF and link with the IMF}\label{sec:CMF}

\begin{figure}[h!]
   \begin{center}
 \begin{minipage}{1.0\linewidth}
    \resizebox{1\hsize}{!}{\includegraphics[angle=0]{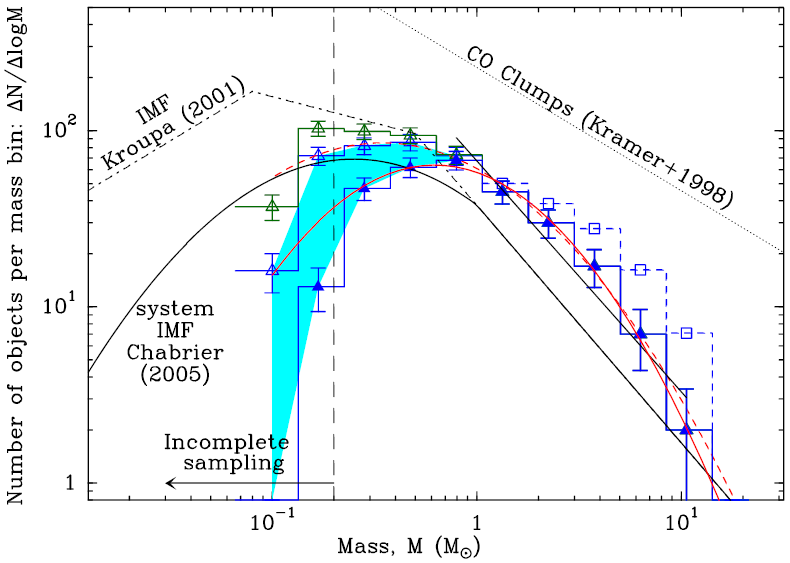}} 
 \end{minipage}
   \end{center}
   \caption{
  Differential core mass function (d$N$/dlog$M$) of the 651 starless cores  (dark green histogram), 
  446 {\it candidate} prestellar cores (upper blue histogram and open triangles), and 292 {\it robust} prestellar cores (lower blue histogram and filled triangles) 
  identified with {\it Herschel} in the whole Aquila field. 
  The error bars correspond to $\sqrt{N}$ statistical uncertainties. 
  The shaded area in light blue reflects the uncertainties in the prestellar CMF arising from the uncertain classification 
  of observed starless cores as gravitationally bound or unbound objects (see Sect.~\ref{sec:bound_selection}). 
  The dashed blue histogram and open squares show how the prestellar CMF would change at the high-mass end after 
  correction for a possible differential timescale bias (see text). 
  The 90\% completeness level of the prestellar core sample 
  is indicated by the vertical dashed line 
  (see Sect.~\ref{sec:completeness} and Appendix~B).
  Lognormal fits to the CMF of  {\it robust}  and {\it candidate} prestellar cores (solid and dashed red curves, respectively), 
  as well as  a power-law fit to the high-mass end of the CMF (black solid line) are superimposed. 
  The two lognormal fits peak at $0.62\, M_\odot $ and $0.34\, M_\odot $, and have standard deviations 
  of $\sim 0.47 $ and $\sim 0.57 $ in log$_{10}M$, respectively. 
  The power-law fit has a slope of $-1.33 \pm 0.06$ (compared to a Salpeter slope of $-1.35$ in this format).
  The IMF of single stars \citep[corrected for binaries -- e.g.,][]{Kroupa2001},  the IMF of multiple systems \citep[e.g.,][]{Chabrier2005}, 
  and the typical mass distribution of CO clumps \citep[e.g.,][]{Kramer+1998} are also shown for comparison.
  }
 \label{fig_CMF}%
 \end{figure}

The prestellar core mass function (CMF) derived from the samples of 446 {\it candidate} and 292 {\it robust} 
prestellar cores 
identified in the whole Aquila cloud (see Sect.~\ref{sec:bound_selection}), excluding the CO high-$V_{\rm LSR}$ 
area in the eastern corner of the field (see Sects.~\ref{sec:mass_distrib} \& \ref{sec:core_selection}), is shown in the form of a 
differential mass distribution in Fig.~\ref{fig_CMF} (see dark blue histograms and light blue shade). 
The mass distribution of the wider sample of 651 starless cores selected in Sect.~\ref{sec:core_selection} 
is plotted as a green histogram for comparison. 
The 90\% completeness level of our {\it Herschel} census of prestellar cores, as estimated both 
from Monte-Carlo simulations (Sect.~\ref{sec:completeness}) and the simple model described in Appendix~\ref{sec:appendix_model}, 
is marked by the vertical dashed line. 
We stress that the differential CMF presented here \citep[see][for preliminary versions of this CMF]{Konyves+2010, Andre+2010} 
is based on a core sample $\sim \, $2--9 times larger than the CMFs derived from earlier 
ground-based studies \citep[e.g.,][]{Motte+1998,Johnstone+2000,Stanke+2006,Alves+2007,Enoch+2008}
and that its shape is therefore much more robustly defined. In particular, it suffers very little from the arbitrary choice of mass bins, a well-known 
disadvantage of differential mass functions \citep[e.g.,][]{ReidWilson2006b}, except perhaps at the very high mass end (e.g. at $M \simgt 5\, M_\odot $, 
where the number of cores per mass bin drops to less than 10 in Fig.~\ref{fig_CMF}). 
Note also that, while we preferred to display the differential form of the CMF in Fig.~\ref{fig_CMF} because it is more intuitive and easier to 
compare with the IMF, we used the cumulative form -- which is independent of binning and thus amenable to cleaner statistical tests -- 
to quantify the resemblance of the observed CMF to several well-known functional forms. 

As can be seen in Fig.~\ref{fig_CMF}, the Aquila prestellar CMF is well fit by a lognormal distribution (solid and dashed red curves 
for the samples of {\it robust}  and {\it candidate} prestellar cores, respectively) 
and very similar in shape to the system IMF advocated by \citet{Chabrier2005}.  
Performing a non-parametric Kolmogorov-Smirnov (K-S) test 
\citep[see, e.g.][]{Press+1992}
on the corresponding cumulative mass distributions 
N($>$$M$) 
indicates that the observed prestellar 
CMF is statistically indistinguishable at the 97\% confidence level 
from a lognormal mass function with central mass $0.45 \pm 0.2 M_\odot $ and 
standard deviation $0.52\pm 0.05$ above the completeness mass limit $\sim 0.2\, M_\odot $. 
For comparison, the lognormal part of the \citet{Chabrier2005} system IMF has a central mass of $0.25\, M_\odot $ 
and a standard deviation of 0.55 in log$_{10}M$.
The error on the two parameters of the lognormal fit to the prestellar CMF (i.e., central mass and standard deviation) 
are mainly driven by the uncertain classification of observed starless cores as gravitationally bound 
or unbound objects, which leads to two slightly different CMF shapes 
for the samples of {\it robust}  and {\it candidate} prestellar cores (blue shaded area in Fig.~\ref{fig_CMF}).
The high-mass end of the Aquila CMF above $1\, M_\odot $ is also consistent with a power-law 
mass function, d$N$/dlog$M$ $\propto$ $M^{-1.33 \pm 0.06}$, at a K-S significance level of 87\%. 
Here, the error bar on the power-law exponent was derived from the range of values for which the K-S significance level is larger than 68\% 
(corresponding to $1\, \sigma $ in Gaussian statistics). 
This function is very similar to the Salpeter power-law IMF which is d$N$/dlog$M$ $\propto$ $M^{-1.35}$ in this format. 
(We note, however, that given the limited range of core masses probed by our data 
a power law does not provide a significantly better fit to the high-mass end of the CMF than a pure lognormal fit.)
In contrast, the CMF observed above $1\, M_\odot $ differs from the shallower power-law mass distribution 
of CO clumps and clouds \citep[d$N$/dlog$M$ $\propto$ $M^{-0.7}$ -- e.g.][]{Blitz1993,Kramer+1998} 
at a very high confidence level. The probability that the CMF can be consistent with 
d$N$/dlog$M$ $\propto$ $M^{-0.7}$ is only P$_{\rm K-S} \sim 7.7\times 10^{-7} $.

A possible caveat to the similarity between the Salpeter IMF and the prestellar CMF at the high-mass end 
should be mentioned, however.
As pointed out by \citet{Clark+2007}, if cores of different mass evolve on different timescales then 
the observed CMF may not be representative of the intrinsic prestellar core mass function.
This is because an observer is more likely to detect long-lived cores than short-lived cores. 
Therefore, if there is a correlation between core lifetime and core mass, then 
the observed CMF can be significantly distorted compared to the ``initial'' prestellar core mass distribution. 
In the present sample of prestellar cores, there is essentially no correlation between core density 
and core mass below $\sim 2\, M_\odot $ but a weak positive correlation above $\sim \, $2--3$\, M_\odot $ (Fig.~\ref{fig_dens_vs_mass}), 
suggesting that prestellar cores more massive than $\sim \, $2--3$\, M_\odot $ may evolve to protostars 
somewhat faster than lower mass cores do (see end of Sect.~\ref{sec:lifetime}).   
To quantify the importance of this potential differential timescale bias on the CMF, we have 
overplotted in Fig.~\ref{fig_CMF} a weighted version of the observed CMF (blue open 
squares and dashed histogram), 
obtained by weighting the number of prestellar cores observed in each mass bin by a factor inversely 
proportional to a mass-dependent free-fall time. The latter was estimated for each mass bin by using 
the parabolic fit to the observed correlation between core density and core mass 
shown by the red curve in Fig.~\ref{fig_dens_vs_mass}. 
As can be seen in Fig.~\ref{fig_CMF}, the weighted CMF is indistinguishable from the unweighted CMF 
for $M < 2\, M_\odot $, but somewhat shallower above $\sim 2\, M_\odot $. 
(A K-S analysis indicates that the high-mass end of the weighted CMF above $2\, M_\odot $ is consistent with a power-law 
mass function, d$N$/dlog$M$ $\propto$ $M^{-1.0 \pm 0.2}$, at a K-S significance level of 90\%.)  
The main effect of the differential timescale correction is to broaden the prestellar CMF, 
leaving the peak mass at $\sim 0.5\, M_\odot $ essentially unchanged.

As already discussed by \citet{Andre+2010} and \citet{Konyves+2010} \citep[see also][]{Alves+2007}, the observed CMF is 
consistent with an essentially one-to-one mapping between prestellar core mass and 
stellar system mass\footnote{As pointed out by a number of authors \citep[e.g.][]{Delgado+2003,Goodwin+2008,Hatchell+2008}, sub-fragmentation of prestellar cores 
into binary or multiple systems complicates the direct mapping of the prestellar CMF onto the IMF of {\it individual} stars. 
Lacking sufficient spatial resolution to probe core multiplicity with the present {\it Herschel} observations, we do not enter 
this debate here and concentrate on the relationship between the prestellar CMF and the {\it system} IMF.}, 
i.e., $M_{\star \rm sys} = \epsilon_{\rm core} \times M_{\rm core} $), where  $ \epsilon_{\rm core}$ 
represents the efficiency of the conversion process from core mass to stellar system mass, i.e., the star formation 
efficiency within an individual prestellar core.  
The peak of the prestellar CMF is at $0.45 \pm 0.2\, M_\odot $ in observed core mass, suggesting 
a real peak at $0.6 \pm 0.2\, M_\odot $ in terms of intrinsic prestellar core mass, after correcting the 
observed masses upward by $\sim 25\%$ due to the fact that the SED mass values 
tend to slightly underestimate the intrinsic core masses  
according to our simulations (see Sect.~\ref{sec:deriv_core_prop} and Appendix~\ref{sec:appendix_simulations}). 
Our data therefore suggest that $ \epsilon_{\rm core} \sim  0.4^{+0.2}_{-0.1}$.

   \begin{figure}[h!]
   \begin{center}
 \begin{minipage}{1.0\linewidth}
    \resizebox{1\hsize}{!}{\includegraphics[angle=0]{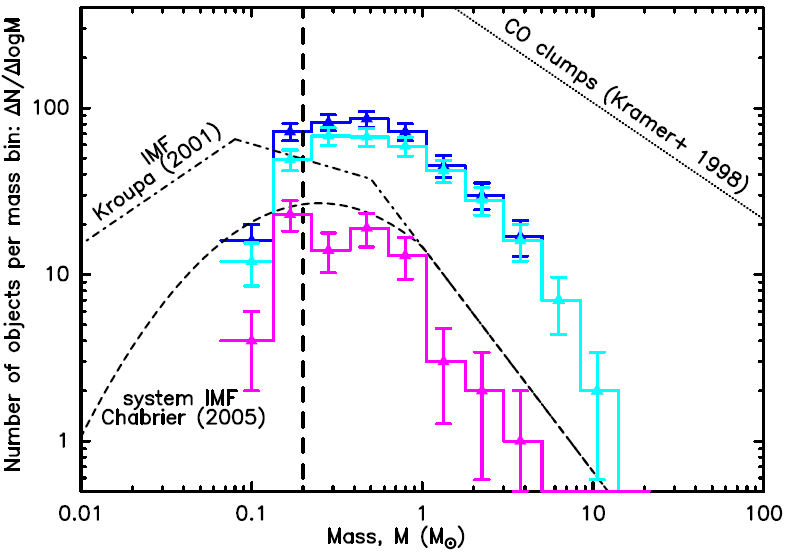}}
 \end{minipage}
   \end{center}
   \caption{Comparison of the CMF observed for the majority ($\sim$81\%) of candidate prestellar cores lying within the 0.2~pc-wide footprints of the DisPerSE-detected
   filaments (light blue histogram) to that observed for the minority $\sim$19\%) of prestellar cores lying outside these filaments (magenta histogram). 
   The upper dark blue histogram and the other lines are the same as in Fig.~\ref{fig_CMF}.
  }
    \label{fig_CMF_onoff}%
    \end{figure}

It is also interesting to investigate possible variations in the CMF as a function of local cloud environment, 
in particular depending on whether the cores lie within or outside dense filaments. 
In L1641 (Orion~A), for instance, \citet{Polychroni+2013} reported that the cores lying on filaments were generally more 
massive than those lying off filaments.
Figure~\ref{fig_CMF_onoff} compares the CMF derived for the candidate prestellar cores lying on filaments (light blue 
histogram) to the CMF of the candidate prestellar cores lying off filaments (magenta histogram)  and 
to the global prestellar CMF in Aquila (upper dark blue histogram). 
It can be seen that the prestellar CMF observed on filaments is very similar to the global prestellar CMF. 
A two-sample K-S test confirms that these two CMFs are indistinguishable at a $> 95\% $ confidence level.
On the other hand, there is a marginal indication that the prestellar CMF observed off filaments may peak at a somewhat lower mass. 
A two-sample K-S test indicates that the probability that the CMFs observed on and off filaments above $0.2\, M_\odot $ 
are drawn from the same intrinsic distribution 
is only $\sim 2\% $ (equivalent to a $\sim 2.3\, \sigma $ result 
in Gaussian statistics). 
This is not a very strong conclusion, however. 
First, there are only 54 candidate prestellar cores with masses $> 0.2\, M_\odot $ lying outside the 0.2~pc-wide filament footprints, 
implying that our estimate of the prestellar CMF off filaments suffers from small-number statistics. 
In fact, we cannot even exclude the possibility that some of the prestellar cores presently classified as lying off filaments  
may be associated with 
faint filaments not identified with DisPerSE in Sect.~\ref{sec:filam}. 
Second, the median background cloud column density observed off filaments 
is lower ($A_V^{\rm back} \sim 4$) than the median background cloud column density observed on filaments
($A_V^{\rm back} \sim 7.5$).  Accordingly, 
the completeness level of our {\it Herschel} survey for prestellar cores is expected to be somewhat better off filaments than on filaments 
(see Fig.~\ref{fig_completeness} in Appendix~B), which may slightly bias the direct comparison of the two CMFs.

\subsection{A quasi-universal efficiency of the star formation process in dense gas?}\label{sec:SFR}

Our {\it Herschel} results on the prestellar core formation efficiency (CFE) as a function of column density in the Aquila cloud 
(see Sect.~\ref{sec:threshold} and Fig.~\ref{fig_cfe}) connect very well with recent near-/mid-infrared studies of the star formation rate (SFR) 
as a function of gas surface density in nearby molecular clouds 
\citep[e.g.,][]{Heiderman+2010,Lada+2010,Lada+2012,Evans+2014}. 
These infrared studies show that the global SFR derived from direct YSO counting (as opposed to the prestellar core counting used in the present study)
tends to be linearly proportional to the mass of dense gas above a surface density threshold corresponding to $A_V^{\rm back} \sim \, $7--8, 
and drops to much lower values below the threshold. 
This column density threshold is essentially the {\it same} as that found with $Herschel$ for the formation of prestellar cores in the Aquila cloud 
(cf. Figs.~\ref{fig_bgCD} \& \ref{fig_cfe}). 
Moreover, the star formation rate per unit mass of dense gas above the threshold found by infrared studies of nearby clouds, namely 
${\rm SFR}/M_{\rm dense} \sim 4.6 \times 10^{-8}\,  {\rm yr}^{-1} $ \citep{Lada+2010,Lada+2012} or 
${\rm SFR}/M_{\rm dense} \sim 2.5^{+1.7}_{-1} \times 10^{-8}\,  {\rm yr}^{-1} $ \citep{Evans+2014}, is entirely 
consistent with the roughly constant prestellar CFE derived for $A_{\rm V} >  7$ in Aquila, which corresponds to  
${\rm SFR}/M_{\rm dense} \sim 5^{+2}_{-2} \times 10^{-8}\,   {\rm yr}^{-1} $ 
(see horizontal dotted line in Fig.~\ref{fig_cfe}) 
adopting a typical prestellar core lifetime $t_{\rm pre} = 1.2$~Myr (see Sect.~\ref{sec:lifetime})
and a local star formation efficiency $\epsilon_{\rm core} = 0.4 $ at the core level (see Sect.~\ref{sec:CMF}). 

As pointed out by \citet{Lada+2010,Lada+2012}, the nearby cloud value of the ``efficiency'' of the star formation process in dense gas 
is also very similar to the efficiency value ${\rm SFR}/M_{\rm dense} \sim 2 \times 10^{-8}\, {\rm yr}^{-1} $ 
found by \citet{GaoSolomon2004} for external galaxies, using HCN observations of dense gas and far-infrared ({\it IRAS}) estimates 
of the SFR in galaxies. While direct comparison between the Galactic and extragalactic values is affected by large uncertainties 
because different tracers of dense gas and star formation were used by \citet{Lada+2010} on the one hand and \citet{GaoSolomon2004} on 
the other, these results suggest that there may be a quasi-universal ``star formation law'' within dense gas above the (column) density threshold. 
Equivalently, in terms of a concept often used in the extragalactic community, this means that there may be a quasi-universal depletion time, 
$t_{\rm dep} \equiv M_{\rm dense}/{\rm SFR}  \sim \,$20-50~Myr, for the dense gas above the threshold. 
This ``star formation law'' is not strictly universal since its does not seem to apply to the extreme environmental conditions of the central 
molecular zone near the Galactic center, for instance, where star formation is observed to be more inefficient above the density threshold, 
by more than an order of magnitude \citep{Longmore+2013}.

Our $Herschel$ findings in the Aquila cloud allow us to go one step further and link this quasi-universal efficiency of the 
star formation process in dense gas to three parameters characterizing 
the physics of prestellar cores, i.e., the core formation efficiency in supercritical filaments,  
$ f_{\rm pre} $, the lifetime of prestellar cores, $t_{\rm pre}$, and the efficiency of the conversion from prestellar core mass 
to stellar system mass, $\epsilon_{\rm core}$, i.e.,
$${\rm SFR}/M_{\rm dense} = f_{\rm pre} \times \epsilon_{\rm core}/t_{\rm pre} = \frac{0.15^{+0.05}_{-0.05}  \times 0.4^{+0.2}_{-0.1} }{1.2^{+0.3}_{-0.3}  \times 10^6} = 5^{+2}_{-2} \times 10^{-8}\,  {\rm yr}^{-1}$$
\citep[see][]{Andre+2014}.

\section{Summary and conclusions}\label{sec:conclusions}

We used the SPIRE and PACS parallel-mode maps taken as part of the {\it Herschel} Gould Belt survey to obtain an extensive 
census of dense cores and their connection with molecular cloud structure in the Aquila star-forming region. 
Our main results and conclusions may be summarized as follows:
         
\begin{enumerate}
  \item The high-resolution ($\sim 18\arcsec $ or $\sim 0.02$~pc) column density map that we derived from the {\it Herschel} photometric data shows that the 
  Aquila cloud is highly filamentary and features a column density probability density function (PDF) with a prominent power-law tail above $A_V \sim 5$--7. 
  About $10\%$--$20\% $  of the gas mass is in the form of filaments below $A_V \sim 7$, while as much as $\sim  50\%$--$75\% $
  of the gas mass  is in the form of filamentary structures above $A_V \sim 7$--10. 
  \item In the $\sim$11~deg$^2$ field imaged with both SPIRE and PACS at five wavelengths from 70~$\mu$m to 500~$\mu$m, we identified 651 starless cores, 
  446 {\it candidate} and 292 {\it robust} prestellar cores, 
  and 58 protostellar cores (such as Class~0 objects), based on multi-scale, multi-wavelength core extraction with the \textsl{getsources} algorithm.
  The samples of {\it candidate} and {\it robust} prestellar cores were estimated to be $\sim 90\% $ and $\sim 80\% $ complete, respectively, down to an observed core mass $\sim 0.2\, M_\odot $. 
  The {\it candidate} prestellar cores have estimated median mass $\sim 0.45\, M_\odot $, median deconvolved FWHM diameter $\sim 0.03$~pc,  median average column density $\sim 7\times 10^{21}\, {\rm cm}^{-3}$
  and median average volume density $\sim 4\times 10^4\, {\rm cm}^{-3}$.
  \item  The typical lifetime of the {\it Herschel} prestellar cores was estimated to be $t_{\rm pre} = 1.2 \pm 0.3$~Myr 
  or $\sim \, $4 free-fall times ($ t_{\rm ff}$) 
  and to decrease from $t_{\rm pre} \sim$1.4~Myr for cores with average volume density $\simgt 10^4\, {\rm cm}^{-3} $
  to a few times $10^4$~yr for cores with average volume density $\simgt 10^6\, {\rm cm}^{-3} $.  The densest prestellar cores in the sample appear to 
  have a lifetime comparable to their free-fall timescale and may be collapsing.
  \item There is strong evidence of a column density threshold  for the formation of prestellar cores, at an equivalent visual extinction level $A_{\rm V}^{\rm bg} \sim 7$, 
  in the sense that the probability function of finding a prestellar core increases by more than an order of magnitude from $A_{\rm V}^{\rm bg} \sim 4$ to
  $A_{\rm V}^{\rm bg} \sim 10$ and is well fit by a smooth exponential step function. Likewise, the prestellar core formation efficiency (CFE) or fraction of cloud mass in the form of prestellar cores  
  was found to increase by about two orders of magnitude between $A_{\rm V}^{\rm bg} \sim 5$ and $A_{\rm V}^{\rm bg} \sim 15$ and to reach a roughly constant value 
  ${\rm CFE_{max}}  \equiv  f_{\rm pre} \sim 15\% $ at higher column densities. 
  This reflects a significantly sharper transition than predicted by ``multi-freefall'' models of the star formation rate in molecular clouds, and argues for the presence of a true physical (column) 
  density threshold for prestellar core formation.
  \item The compact dense cores are closely associated with the filamentary structure, and preferentially the densest filaments.
  In particular, a very high fraction ($\mathop{75\%}_{\text{-5\%}}^{\text{+15\%}}$) of prestellar cores were found to lie within supercritical filaments with masses per unit length 
  $M_{\rm line} > M_{\rm line, crit} $, where $M_{\rm line, crit} \equiv  2\, c_s^2/G \sim 16\, M_\odot$/pc is the critical mass per unit length of nearly isothermal, long cylinders  
  at $T \sim 10$~K \citep[see][]{InutsukaMiyama1997}.
  \item The prestellar core mass function (CMF) derived using the samples of 446 {\it candidate} and 292 {\it robust} prestellar cores is well fit by a lognormal distribution, peaks at $\sim \,$0.4--0.6$\, M_\odot $, and is very similar 
  in shape to the system IMF. This CMF is consistent with an essentially one-to-one mapping between prestellar core mass and stellar system mass  
  with a local star formation efficiency $ \epsilon_{\rm core} \sim  0.4^{+0.2}_{-0.1}$ within an individual prestellar core.
  \item Our $Herschel$ findings in the Aquila cloud connect very well with recent {\it Spitzer} studies of the star formation rate in nearby molecular clouds. 
  They support the view that there may be a quasi-universal ``efficiency'' of the star formation process in dense gas, ${\rm SFR}/M_{\rm dense} \sim 5^{+2}_{-2} \times 10^{-8}\,  {\rm yr}^{-1}$, 
  and that this quasi-universal ``efficiency'' may be closely linked to the physics of prestellar core formation within filaments: ${\rm SFR}/M_{\rm dense} = f_{\rm pre} \times \epsilon_{\rm core}/t_{\rm pre}$. 
\end{enumerate}

\begin{acknowledgements}
SPIRE has been developed by a consortium of institutes led by Cardiff Univ. (UK) 
and including: Univ. Lethbridge (Canada); NAOC (China); CEA, LAM (France); 
IFSI, Univ. Padua (Italy); IAC (Spain); Stockholm Observatory (Sweden); 
Imperial College London, RAL, UCL-MSSL, UKATC, Univ. Sussex (UK); 
and Caltech, JPL, NHSC, Univ. Colorado (USA). This development has been 
supported by national funding agencies: CSA (Canada); NAOC (China); 
CEA, CNES, CNRS (France); ASI (Italy); MCINN (Spain); SNSB (Sweden); 
STFC, UKSA (UK); and NASA (USA). 
PACS has been developed by a consortium of institutes led by MPE
(Germany) and including UVIE (Austria); KUL, CSL, IMEC (Belgium); CEA,
OAMP (France); MPIA (Germany); IFSI, OAP/AOT, OAA/CAISMI, LENS, SISSA
(Italy); IAC (Spain). This development has been supported by the funding
agencies BMVIT (Austria), ESA-PRODEX (Belgium), CEA/CNES (France),
DLR (Germany), ASI (Italy), and CICT/MCT (Spain). 
This work has received support from the European Research Council 
under the European Union's Seventh Framework Programme 
(ERC Advanced Grant Agreements no. 291294 --  `ORISTARS'  -- and no. 267934 -- `MISTIC') 
and from the French National Research Agency (Grant no. ANR--11--BS56--0010 -- `STARFICH').
\end{acknowledgements}

\bibliographystyle{aa}
\bibliography{aql_1stgen_paper_accepted_konyves_etal_2015}

\newpage

\begin{appendix} 

\section{A catalog of dense cores identified with {\it Herschel} in the Aquila cloud complex}\label{sec:appendix_catalog}

Based on our {\it Herschel} SPIRE/PACS parallel-mode imaging survey of the Aquila cloud complex, 
we identified a total of 749 dense cores, including 685 starless cores and 64 protostellar cores. 
(Among these, 34 starless cores shown as yellow triangles in Fig.~\ref{fig_cd}, as well as 6 protostellar cores, were excluded from the scientific discussion of Sect.~\ref{sec:discuss} 
due to likely contamination by more distant, background objects -- see Sect.~\ref{sec:mass_distrib}.)
The master catalog listing the observed properties of all of these {\it Herschel} cores is available 
in online Table~\ref{tab_obs_cat}. A template of this online catalog is provided below to illustrate its form and content.

The derived properties 
(physical radius, mass, SED dust temperature, peak column density at the resolution of the 500~$\mu$m data, average column density, 
peak volume density, and average density) are given in online Table \ref{tab_der_cat_cores} for each core. 
A portion of this online table is also provided below.
The derived properties of the {\it Herschel}-detected protostars and YSOs will be published 
in a forthcoming paper.

\clearpage

\begin{sidewaystable*}[htbf]\tiny\setlength{\tabcolsep}{2.5pt}
\caption{Catalog of  dense cores identified in the HGBS maps of the Aquila complex (template, full catalog only provided online).} 
\label{tab_obs_cat}
{\renewcommand{\arraystretch}{0.5}
\begin{tabular}{|r|c|c c c c c c c c c c} 
\toprule[1.0pt]\toprule[0.5pt]  
 rNO      & Core name         &  RA$_{\rm 2000}$ &  Dec$_{\rm 2000}$        & Sig$_{\rm 070}$ &  $S^{\rm peak}_{\rm 070}$ &  $S^{\rm peak}_{\rm 070}$/$S_{\rm bg}$ &  $S^{\rm conv,500}_{\rm 070}$ &  $S^{\rm tot}_{\rm 070}$ &  FWHM$^{\rm a}_{\rm 070}$ &  FWHM$^{\rm b}_{\rm 070}$ &  PA$_{\rm 070}$  \\ 
          & HGBS\_J*          &  (h m s)         &  (\degr~\arcmin~\arcsec) &                 & (Jy/beam)                 &                                        & (Jy/beam$_{\rm 500}$)         &  (Jy)                    &  (\arcsec)                &  (\arcsec)                &  (\degr)         \\         
 (1)      & (2)               &  (3)             &  (4)                     &  (5)            &      (6) ~ $\pm$ ~ (7)    &  (8)                                   & (9)                           &    (10) ~ $\pm$ ~ (11)   &  (12)                     &  (13)                     &  (14)             \\         
\toprule[0.8pt] 
$\cdots$  &                   &                  &                          &                 &                           &                                        &                               &                          &                           &                           &                   \\
  79      & 182754.3-034237   & 18:27:54.36      &  --03:42:37.2            &   136.9         &    1.01e+00 ~ 6.5e-02     &   1.87                                 &   4.82e+00                    &   6.38e+00  ~ 1.8e-01    &  21                       &  18                       &  127             \\
$\cdots$  &                   &                  &                          &                 &                           &                                        &                               &                          &                           &                           &                   \\
  88      & 182805.4-034021   & 18:28:05.48      &  --03:40:21.4            &     0.3         &  --2.09e-03 ~ 1.5e-02     &  --0.03                                &  --8.54e-03                   &   2.19e-02  ~ 1.8e-02    &  62                      &  8                     &	 59             \\
$\cdots$  &                   &                  &                          &                 &                           &                                        &                               &                          &                           &                           &                   \\
 303      & 183005.4-014833   & 18:30:05.46      &  --01:48:33.6            &     5.9         &  --3.90e-02 ~ 1.5e-02     &  --0.58                                &  --3.26e-01                   &  --2.05e-01 ~ 2.0e-02    &  56                      &  20                      &	 34              \\
$\cdots$  &                   &                  &                          &                 &                           &                                        &                               &                          &                           &                           &                   \\
\midrule[0.5pt]
\end{tabular}
}
\scalebox{1.2}{$\sim$}
\vspace{0.2cm}

\scalebox{1.2}{$\sim$}
{\renewcommand{\arraystretch}{0.5}
\begin{tabular}{c c c c c c c c c  c c c c c c c} 
\toprule[1.0pt]\toprule[0.5pt]  
 Sig$_{\rm 160}$ &  $S^{\rm peak}_{\rm 160}$    &  $S^{\rm peak}_{\rm 160}$/$S_{\rm bg}$ &  $S^{\rm conv,500}_{\rm 160}$ &  $S^{\rm tot}_{\rm 160}$ &  FWHM$^{\rm a}_{\rm 160}$ &  FWHM$^{\rm b}_{\rm 160}$ &  PA$_{\rm 160}$  &  Sig$_{\rm 250}$ &  $S^{\rm peak}_{\rm 250}$ &  $S^{\rm peak}_{\rm 250}$/$S_{\rm bg}$ &  $S^{\rm conv,500}_{\rm 250}$ &  $S^{\rm tot}_{\rm 250}$ &  FWHM$^{\rm a}_{\rm 250}$ &  FWHM$^{\rm b}_{\rm 250}$ &  PA$_{\rm 250}$  \\ 
                 &  (Jy/beam)                   &                                        & (Jy/beam$_{\rm 500}$)         &  (Jy)		    &  (\arcsec)		&  (\arcsec)		    &  (\degr)         &		  & (Jy/beam)		      & 				       & (Jy/beam$_{\rm 500}$)         &  (Jy)  		  &  (\arcsec)  	      &  (\arcsec)		  &  (\degr)	     \\       
  (15)           &      (16) ~ $\pm$ ~ (17)     &   (18)                                 &  (19)   			 &    (20) ~ $\pm$ ~ (21)   &  (22)                     & (23)                      & (24)             & (25)             &    (26) ~ $\pm$ ~ (27)    &  (28)                                  &  (29)                         &   (30) ~ $\pm$ ~ (31)    &  (32)                     &  (33)                     &  (34)            \\        
\toprule[0.8pt]
$\cdots$         &                              &                                        &                               &			    &				&			    &		       &		  &			      & 				       &                               &			  &			      & 			  &		     \\
  258.7          &  8.22e+00 ~ 3.5e-01          &  1.73				         &  1.31e+01			 &  1.97e+01 ~ 5.3e-01      &  21			&  14		            &  30	       &   231.0	  &  1.37e+01 ~ 6.1e-01       &    1.73				       &  1.58e+01		       &  2.09e+01 ~ 6.5e-01      &  24			      &  18		          &  193             \\
$\cdots$         &                              &                                        &                               &			    &				&			    &		       &		  &			      & 				       &                               &			  &			      & 			  &		     \\
  0.0            &  1.23e-01 ~ 3.4e-02          &  0.07				         &  4.74e-01			 &  7.84e-01 ~ 5.6e-02      &  43			&  26		            &  100	       &   6.2	          &  3.54e-01 ~ 1.0e-01       &    0.11				       &  5.15e-01		       &  6.59e-01 ~ 1.1e-01      &  37			      &  18		          &   87             \\
$\cdots$         &                              &                                        &                               &			    &				&			    &		       &		  &			      & 				       &                               &			  &			      & 			  &		     \\
  0.0            &  7.44e-02 ~ 7.0e-02          &  0.05				         &  4.10e-01			 &  1.01e+00 ~ 1.8e-01      &  62		        &  45		            &  202	       &   6.2	          &  6.07e-01 ~ 1.1e-01       &    0.20				       &  1.20e+00		       &  2.13e+00 ~ 2.0e-01      &  35			      &  31		          &   83             \\
$\cdots$         &                              &                                        &                               &			    &				&			    &		       &		  &			      & 				       &                               &			  &			      & 			  &		     \\
\midrule[0.5pt]
\end{tabular}
}
\scalebox{1.2}{$\sim$}
\vspace{0.2cm}

\scalebox{1.2}{$\sim$}
{\renewcommand{\arraystretch}{0.5}
\begin{tabular}{c c c c c c c c c c c c c c c c}  
\toprule[1.0pt]\toprule[0.5pt] 
Sig$_{\rm 350}$  &  $S^{\rm peak}_{\rm 350}$ &  $S^{\rm peak}_{\rm 350}$/$S_{\rm bg}$ &  $S^{\rm conv,500}_{\rm 350}$ &  $S^{\rm tot}_{\rm 350}$  &  FWHM$^{\rm a}_{\rm 350}$ &  FWHM$^{\rm b}_{\rm 350}$ &  PA$_{\rm 350}$   &  Sig$_{\rm 500}$ &  $S^{\rm peak}_{\rm 500}$ &  $S^{\rm peak}_{\rm 500}$/$S_{\rm bg}$ &  $S^{\rm tot}_{\rm 500}$ &  FWHM$^{\rm a}_{\rm 500}$ &  FWHM$^{\rm b}_{\rm 500}$  &  PA$_{\rm 500}$   \\
                 &  (Jy/beam)                &                                        & (Jy/beam$_{\rm 500}$)         &  (Jy)			  &  (\arcsec)  	      &  (\arcsec)		  &  (\degr)	      & 		 & (Jy/beam)		     &  				      &  (Jy)                    &  (\arcsec)                &  (\arcsec)                 &  (\degr)          \\	       
  (35)           &   (36) ~ $\pm$ ~ (37)     &  (38)                                  &  (39)                         &    (40) ~ $\pm$ ~ (41)    &  (42)                     &  (43)                     &  (44)             &  (45)            &    (46) ~ $\pm$ ~ (47)    &  (48)                                  &    (49) ~ $\pm$ ~ (50)   &  (51)                     &  (52)                      &  (53)             \\	
\toprule[0.8pt] 
$\cdots$         &                           &                                        &                               & 			  &			      & 			  &		      & 		 &			     &  				      &                          &                           &                            &                   \\
187.2		 & 1.16e+01 ~ 6.3e-01        &   1.54  			              &  1.27e+01		      &  1.42e+01 ~ 6.3e-01       &  30			      &  25			  &  184	      &     110.5	 &  7.85e+00 ~ 4.5e-01       &  1.33                                  &  9.38e+00 ~ 4.5e-01      &  43			     &  36			  &  174  	      \\
$\cdots$         &                           &                                        &                               & 			  &			      & 			  &		      & 		 &			     &  				      &                          &                           &                            &                   \\
  6.2		 & 3.91e-01 ~ 1.6e-01        &   0.11  			              &  4.91e-01		      &  5.30e-01 ~ 1.6e-01       &  46			      &  25			  &  105	      &      0.0         &  4.14e-01 ~ 1.6e-01       &  0.11                                  &  3.24e-01 ~ 1.6e-01      &  36			     &  36			  &  118  	      \\
$\cdots$         &                           &                                        &                               & 			  &			      & 			  &		      & 		 &			     &  				      &                          &                           &                            &                   \\
 14.8		 & 1.20e+00 ~ 1.5e-01        &   0.34  			              &  1.56e+00		      &  2.53e+00 ~ 2.1e-01       &  35			      &  32			  &   91	      &      16.9	 &  1.39e+00 ~ 1.6e-01       &  0.38                                  &  1.62e+00 ~ 1.6e-01      &  39			     &  36			  &   58  	      \\
$\cdots$         &                           &                                        &                               & 			  &			      & 			  &		      & 		 &			     &  				      &                          &                           &                            &                   \\
\midrule[0.5pt]
\end{tabular}
}
\scalebox{1.2}{$\sim$}
\vspace{0.2cm}

\scalebox{1.2}{$\sim$}
{\renewcommand{\arraystretch}{0.5}
\begin{tabular}{c c c c c c c c c c c c c c|}  
\toprule[1.0pt]\toprule[0.5pt] 
Sig$_{\rm N_{H_2}}$& $N^{\rm peak}_{\rm H_2}$ &  $N^{\rm peak}_{\rm H_2}$/$N_{\rm bg}$ &  $N^{\rm conv,500}_{\rm H_2}$ &  $N^{\rm bg}_{\rm H_2}$  &  FWHM$^{\rm a}_{\rm N_{H_2}}$ &  FWHM$^{\rm b}_{\rm N_{H_2}}$ & PA$_{\rm N_{H_2}}$  &  N$_{\rm SED}$  &  CSAR  &  Core type     & SIMBAD                  & {\it Spitzer}          & Comments    \\
                   & (10$^{21}$ cm$^{-2}$)    &                                        & (10$^{21}$ cm$^{-2}$)         &  (10$^{21}$ cm$^{-2}$)   &  (\arcsec)  	          &  (\arcsec)                    &  (\degr)            &                 &        &                &                         &                        &             \\ 
 (54)              & (55)                     &  (56)                                  & (57)                          &  (58)                    &  (59)                         &  (60)                         &  (61)               &    (62)         &  (63)  & (64)           & (65)                    & (66)                   & (67)        \\
\toprule[0.8pt] 																									            
$\cdots$           &                          &                                        &                               &                          &                               &                               &                     &                 &        &                &                         &                        &             \\
206.8              & 36.0		      &   1.95  			       &   13.5		               &  18.5	                  &  32			          &  19			          &  172	        &		5 &    1   & protostellar   & IRAS 18252-0344         & SSTc2d J1827547-034239 &             \\
$\cdots$           &                          &                                        &                               &                          &                               &                               &                     &                 &        &                &                         &                        &             \\
 10.4              &  1.8		      &   0.18  			       &    0.7		               &  10.1	                  &  33			          &  18			          &  109		&		2 &    0   & starless       &                         &                        &             \\
$\cdots$           &                          &                                        &                               &                          &                               &                               &                     &                 &        &                &                         &                        &             \\
 48.0              &  9.3		      &   0.73  			       &    4.7		               &  12.8	                  &  35			          &  29			          &   96		&		3 &    2   & prestellar     &                         &                        &             \\
$\cdots$           &                          &                                        &                               &                          &                               &                               &                     &                 &        &                &                         &                        &             \\
\midrule[0.5pt]
\end{tabular}
}
\tablefoot{Catalog entries are as follows: 
{\bf(1)} Core running number;
{\bf(2)} Core name $=$ HGBS\_J prefix directly followed by a tag created from the J2000 sexagesimal coordinates; 
{\bf(3)} and {\bf(4)}: Right ascension and declination of core center; 
{\bf(5)}, {\bf(15)}, {\bf(25)}, {\bf(35)}, and {\bf(45)}: Detection significance from monochromatic single scales, in the 70, 160, 250, 350, and 500~$\mu$m maps, respectively. 
(NB: the detection significance has the special value of $0.0$ when the core is not visible in clean single scales); 
{\bf(6)}$\pm${\bf(7)}, {\bf(16)}$\pm${\bf(17)} {\bf(26)}$\pm${\bf(27)} {\bf(36)}$\pm${\bf(37)} {\bf(46)}$\pm${\bf(47)}: Peak flux density and its error in Jy/beam as estimated by \textsl{getsources};
{\bf(8)}, {\bf(18)}, {\bf(28)}, {\bf(38)}, {\bf(48)}: Contrast over the local background, defined as the ratio of the background-subtracted peak intensity to the local background intensity ($S^{\rm peak}_{\rm \lambda}$/$S_{\rm bg}$); 
{\bf(9)}, {\bf(19)}, {\bf(29)}, {\bf(39)}: Peak flux density measured after smoothing to a 36.3$\arcsec$ beam; 
{\bf(10)}$\pm${\bf(11)}, {\bf(20)}$\pm${\bf(21)}, {\bf(30)}$\pm${\bf(31)}, {\bf(40)}$\pm${\bf(41)}, {\bf(49)}$\pm${\bf(50)}: Integrated flux density and its error in Jy as estimated by \textsl{getsources}; 
{\bf(12)}--{\bf(13)}, {\bf(22)}--{\bf(23)}, {\bf(32)}--{\bf(33)}, {\bf(42)}--{\bf(43)}, {\bf(51)}--{\bf(52)}: Major \& minor FWHM diameters of the core (in arcsec), respectively, 
as estimated by \textsl{getsources}. (NB: the special value of $-1$ means that no size measurement was possible); 
{\bf(14)}, {\bf(24)}, {\bf(34)}, {\bf(44)}, {\bf(53)}: Position angle of the core major axis, measured east of north, in degrees; 
{\bf(54)} Detection significance in the high-resolution column density image;  
{\bf(55)} Peak H$_{2}$ column density in units of $10^{21}$ cm$^{-2}$ as estimated by \textsl{getsources} in the high-resolution column density image; 
{\bf(56)} Column density contrast over the local background, as estimated by \textsl{getsources} in the high-resolution column density image;
{\bf(57)} Peak column density measured in a 36.3$\arcsec$ beam; 
{\bf(58)} Local background H$_{2}$ column density as estimated by \textsl{getsources} in the high-resolution column density image; 
{\bf(59)}--{\bf(60)}--{\bf(61)}: Major \& minor FWHM diameters of the core, and position angle of the major axis, respectively, as measured in the high-resolution column density image; 
{\bf(62)} Number of {\it Herschel} bands in which the core is significant (Sig$_{\rm \lambda} >$ 5) and has a positive flux density, excluding the column density plane; 
{\bf(63)} 'CSAR' flag: 2 if the \textsl{getsources} core has a counterpart detected by the CSAR source-finding algorithm \citep{KirkJ+2013} within 6$\arcsec$ of its peak position,
1 if no close CSAR counterpart exists but the peak position of a CSAR source lies within the FWHM contour of the \textsl{getsources} core in the high-resolution column density map, 
0 otherwise;
{\bf(64)} Core type: starless, prestellar, or protostellar; 
{\bf(65)} Closest counterpart found in SIMBAD, if any, up to 6$\arcsec$ from the {\it Herschel} peak position;
{\bf(66)} Closest {\it Spitzer}-identified YSO from the c2d survey \citep[][Allen et al., in prep.]{Dunham+2013} within 6$\arcsec$ of the {\it Herschel} peak position, if any. When present, the {\it Spitzer} source name 
has the form of (SSTc2d $+$) JHHMMSSs$\pm$DDMMSS \citep{Dunham+2013};
{\bf(67)} Comments. }
\end{sidewaystable*}

\clearpage

\begin{sidewaystable*}[htbf]\tiny\setlength{\tabcolsep}{6.8pt}
\caption{Derived properties of the dense cores identified in the HGBS maps of the Aquila region (template, full table only provided online).}
\label{tab_der_cat_cores}
{\renewcommand{\arraystretch}{0.9}
\begin{tabular}{|r|c|c c c c c c c c c c c c|} 
\toprule[1.0pt]\toprule[0.5pt]  
 rNO     & Core name         &  RA$_{\rm 2000}$ &  Dec$_{\rm 2000}$        & $R_{\rm core}$        &  $M_{\rm core}$   &  $T_{\rm dust}$   &  $N^{\rm peak}_{\rm H_2}$ &  $N^{\rm ave}_{\rm H_2}$   &   $n^{\rm peak}_{\rm H_2}$  &  $n^{\rm ave}_{\rm H_2}$   &  $\alpha_{\rm BE}$  & Core type     &  Comments \\
         & HGBS\_J*          &  (h m s)         &  (\degr~\arcmin~\arcsec) & (pc)                  &  ($M_\odot$)      &  (K)              &  (10$^{21}$ cm$^{-2}$)    &  (10$^{21}$ cm$^{-2}$)     &   (10$^{4}$ cm$^{-3}$)      &  (10$^{4}$ cm$^{-3}$)      &                     &		     &		 \\   
 (1)     & (2)               &  (3)             &   (4)                    & (5) ~~~~~~ (6)        &  (7) $\pm$ (8)    &  (9) $\pm$ (10)   &  (11)                     &  (12) ~~~~ (13)            &   (14)                      &  (15) ~~~ (16)             &         (17)        &    (18)       &  (19)     \\   
\toprule[0.8pt] 																				 
$\cdots$ &                   &                  &                          &                       &                   &                   &                           &                            &	                          &                            &                     &               &		 \\ 
  79     & 182754.3-034237   & 18:27:54.36      &  --03:42:37.2            &  2.0e-02 ~~~ 3.1e-02  &  2.06 ~~~ 0.26    &  12.3 ~~~ 0.5     &  42.4                     &  24.8 ~~~ 64.2             &    25.5                     &  17.3 ~~~ 72.1             &         0.2         &  protostellar &	         \\
$\cdots$ &                   &                  &                          &                       &                   &                   &                           &                            &	                          &                            &                     &               &		 \\ 
  88     & 182805.4-034021   & 18:28:05.48      &  --03:40:21.4            &  2.0e-02 ~~~ 3.1e-02  &  0.09 ~~~ 0.04    &  11.5 ~~~ 4.5     &   1.2                     &   1.1 ~~~ 2.6              &     0.7                     &   0.8 ~~~  2.9             &         4.5         &  starless     & no SED fit  \\
$\cdots$ &                   &                  &                          &                       &                   &                   &                           &                            &	                          &                            &                     &               &		 \\ 
 303     & 183005.4-014833   & 18:30:05.46      &  --01:48:33.6            &  3.3e-02 ~~~ 4.0e-02  &  1.01 ~~~ 0.28    &   9.2 ~~~~ 0.6    &  24.6                     &   7.4 ~~~ 11.0             &    13.2                     &   4.0 ~~~  7.3             &         0.6         &  prestellar   &	         \\
$\cdots$ &                   &                  &                          &                       &                   &                   &                           &                            &	                          &                            &                     &               &		 \\ 
\midrule[0.5pt]
\end{tabular}
}
\tablefoot{Table entries are as follows: {\bf(1)} Core running number; {\bf(2)} Core name $=$ HGBS\_J prefix directly followed by a tag created from the J2000 sexagesimal coordinates; 
{\bf(3)} and {\bf(4)}: Right ascension and declination of core center; 
{\bf(5)} and {\bf(6)}: Geometrical average between the major and minor FWHM sizes of the core (in pc), as measured in the high-resolution column density map 
after deconvolution from the 18.2$\arcsec$ HPBW resolution of the map and before deconvolution, respectively.
(NB: Both values provide estimates of the object's outer {\it radius} when the core can be approximately described by a Gaussian distribution, as is the case 
for a critical Bonnor-Ebert spheroid); 
{\bf(7)} Estimated core mass ($M_\odot$) assuming the dust opacity law advocated by \citet{Roy+2014}; 
{\bf(9)} SED dust temperature (K); {\bf(8)} \& {\bf(10)} Statistical errors on the mass and temperature, respectively, including calibration uncertainties, but excluding dust opacity uncertainties; 
{\bf(11)} Peak H$_2$ column density, at the resolution of the 500$~\mu$m data, derived from a graybody SED fit to the core peak flux densities measured in a common 36.3$\arcsec$ beam at all wavelengths; 
{\bf(12)} Average column density, calculated as $N^{\rm ave}_{\rm H_2} = \frac{M_{\rm core}}{\pi R_{\rm core}^2} \frac{1}{\mu m_{\rm H}}$, 
          where $M_{\rm core}$ is the estimated core mass (col. {\bf 7}), $R_{\rm core}$ the estimated core radius prior to deconvolution (col. {\bf 6}), and $\mu = 2.8$;
{\bf(13)} Average column density calculated in the same way as for col. {\bf 12} but using the deconvolved core radius (col. {\bf 5}) instead of the core radius measured prior to deconvolution;  
{\bf(14)} Beam-averaged peak volume density at the resolution of the 500~$\mu$m data, derived from the peak column density (col. {\bf 11}) assuming a Gaussian spherical distribution: 
          $n^{\rm peak}_{\rm H_2} = \sqrt{\frac{4 \ln2}{\pi}} \frac{N^{\rm peak}_{\rm H_2}}{\overline{FWHM}_{\rm 500}}$; 
{\bf(15)} Average volume density, calculated as
          $n^{\rm ave}_{\rm H_2} = \frac{M_{\rm core}}{4/3 \pi R_{\rm core}^3} \frac{1}{\mu m_{\rm H}}$, using the estimated core radius prior to deconvolution; 
{\bf(16)} Average volume density, calculated in the same way as for col. {\bf 15} but using the deconvolved core radius (col. {\bf 5}) instead of the core radius measured prior to deconvolution; 
{\bf(17)} Bonnor-Ebert mass ratio: $\alpha_{\rm BE} = M_{\rm BE,crit} / M_{\rm obs} $ (see text for details); 
{\bf(18)} Core type: starless, prestellar, or protostellar; 
{\bf(19)} Comments may be \textit{no SED fit}, \textit{tentative bound}, or \textit{CO high-V\_LSR} (see text for details). 
}

\end{sidewaystable*}

\clearpage
\newpage
   \begin{figure*}[!!!!!!!bbbb]
   \begin{center}
 \begin{minipage}{0.45\linewidth}
    \resizebox{1\hsize}{!}{\includegraphics[angle=0]{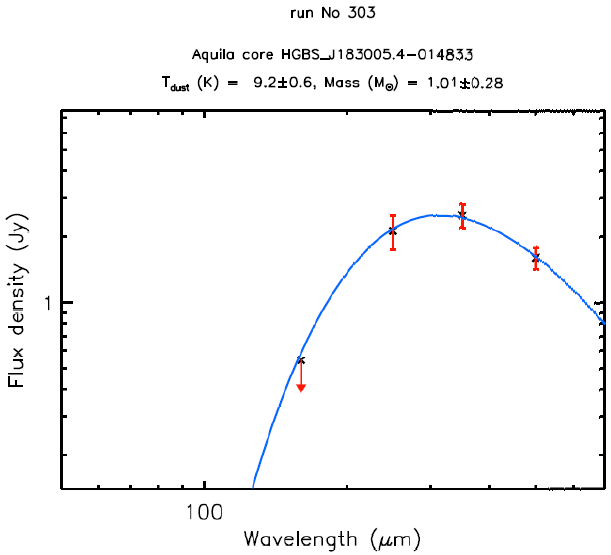}}
 \end{minipage}
 \hspace{2mm}
 \begin{minipage}{0.47\linewidth}
    \resizebox{1\hsize}{!}{\includegraphics[angle=0]{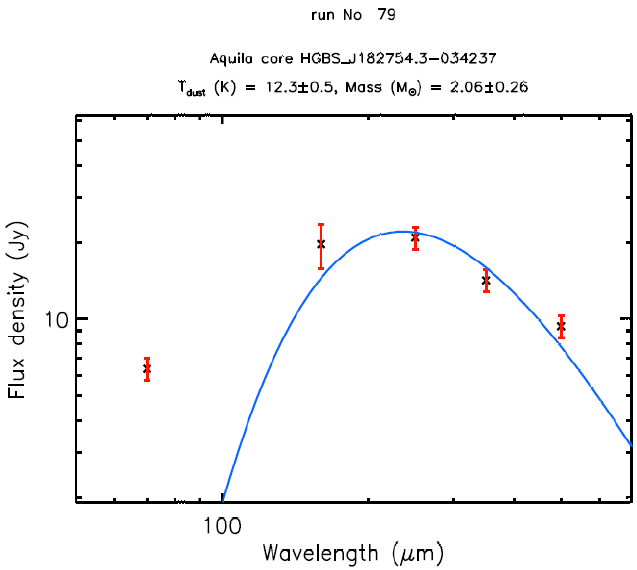}}
 \end{minipage} 
   \end{center}
   \caption{
   Examples of {\it Herschel} spectral energy distributions (SEDs) for a prestellar core ({\it left}, see Fig. \ref{fig_zooms1} for 
   the corresponding image cutouts) and a protostellar core ({\it right}, see Fig. \ref{fig_zooms2} for the corresponding image cutouts).
   These SEDs are constructed  from the background-subtracted integrated flux densities (cross symbols) measured by \textsl{getsources}. 
   A graybody fit to the SED observed longward of 160~$\mu$m is superimposed as a blue curve in both panels.
   Only upper limits are available at 70~$\mu$m and 160~$\mu$m for the prestellar core shown in the left panel. 
   Similar SED plots are provided online for all selected cores. 
   A single-temperature graybody rarely provides a good fit to the overall SED of a protostellar core
   but can nevertheless describe the SED longward of 160~$\mu$m reasonably well (cf. {\it right} panel).
   }
    \label{fig_SED}%
    \end{figure*}
%
   \begin{figure*}
   \begin{center}
 \begin{minipage}{1.0\linewidth}
    \resizebox{0.5\hsize}{!}{\includegraphics[angle=0]{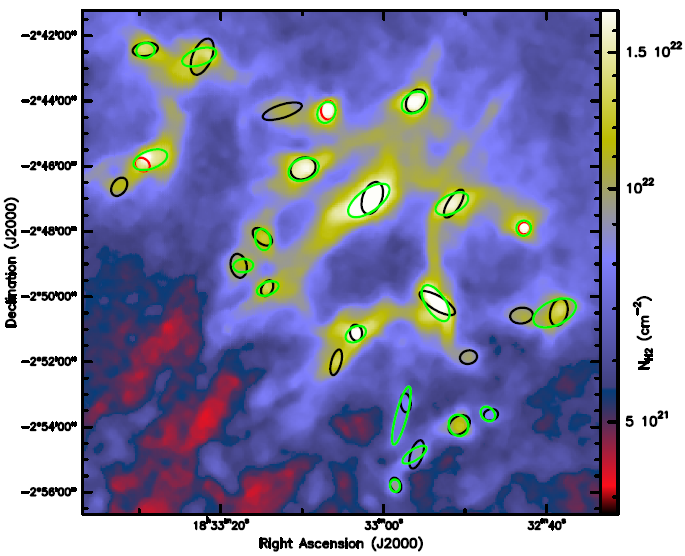}} 
    \resizebox{0.5\hsize}{!}{\includegraphics[angle=0]{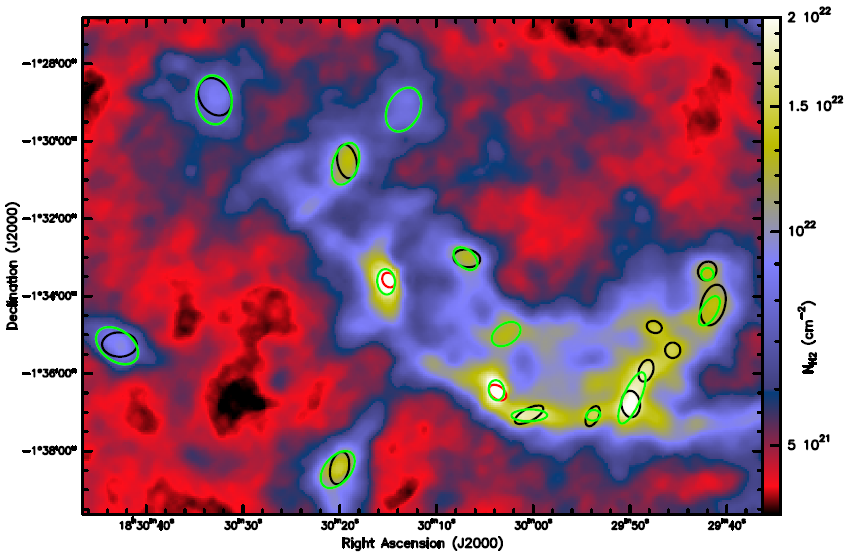}}
 \end{minipage}
   \end{center}
    \caption{Blow-up column density images of two Aquila subfields at $18.2\arcsec $ resolution. 
   Black and red ellipses mark the FWHM sizes of the starless cores and protostellar cores, respectively, selected from 
   \textsl{getsources} extractions in these two subfields. Green ellipses show the FWHM sizes of the sources independently 
   detected with CSAR \citep{KirkJ+2013} in the high-resolution column density image. 
    }	
              \label{fig_zoom_ccl}%
    \end{figure*}

\clearpage
\begin{figure*}
  \begin{center}
  \begin{minipage}{0.9\linewidth}
    \resizebox{1\hsize}{!}{\includegraphics[angle=0]{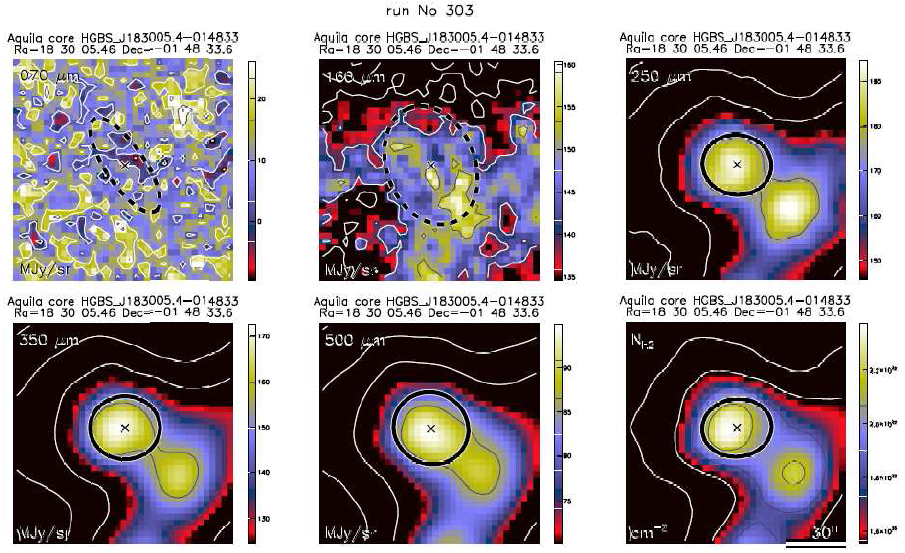}} 
  \end{minipage}
  \end{center}
   \caption{Example blow-up {\it Herschel} images at 70/160/250/350/500~$\mu$m and high-resolution column density map for a (bound) prestellar core. 
           Ellipses represent the estimated major and minor FWHM sizes of the core at each wavelength; they are shown as solid or dashed curves
           depending on whether the core is significantly detected or not, respectively, at a particular wavelength.
	   See Table \ref{tab_der_cat_cores} for the physical radius of the core and other derived properties. 
	   Similar image cutouts are provided online for all selected starless cores. 
           }
   \label{fig_zooms1}%
\end{figure*}
\begin{figure*}
  \begin{center}
  \begin{minipage}{0.9\linewidth}
    \resizebox{1\hsize}{!}{\includegraphics[angle=0]{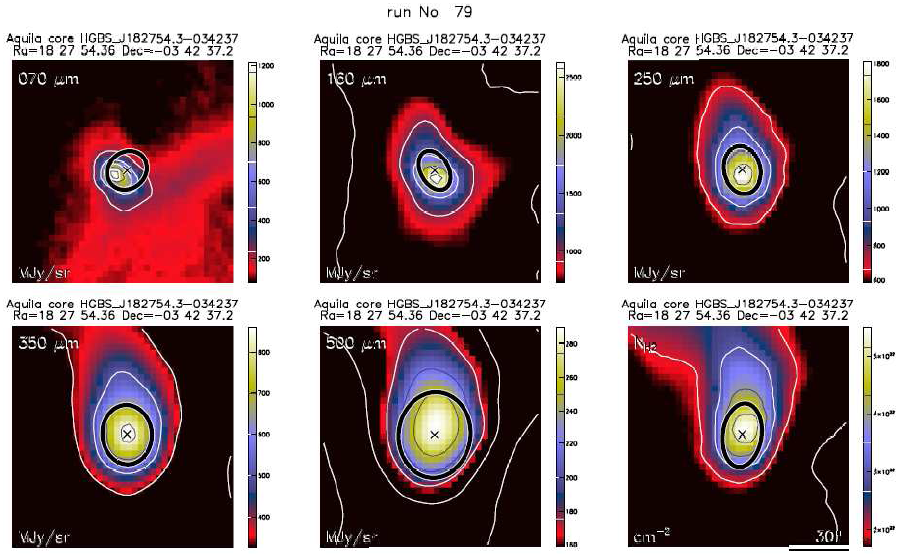}}
  \end{minipage}
  \end{center}
   \caption{Same as Fig.~\ref{fig_zooms1} for a protostellar core. 
   Similar image cutouts are provided online for all selected protostellar cores.}
   \label{fig_zooms2}%
\end{figure*}

\clearpage

\section{Completeness of HGBS prestellar core extractions in Aquila}\label{sec:appendix_completeness}

To estimate the completeness of our census of prestellar cores in Aquila, 
we used several sets of simulated data on the one hand (Sect.~\ref{sec:appendix_simulations}), 
and a simple model of the core extraction process and completeness problem  
on the other (Sect.~\ref{sec:appendix_model}).

   \begin{figure}[!h]
   \begin{center}
 \begin{minipage}{1.0\linewidth}
    \resizebox{0.9\hsize}{!}{\includegraphics[angle=0]{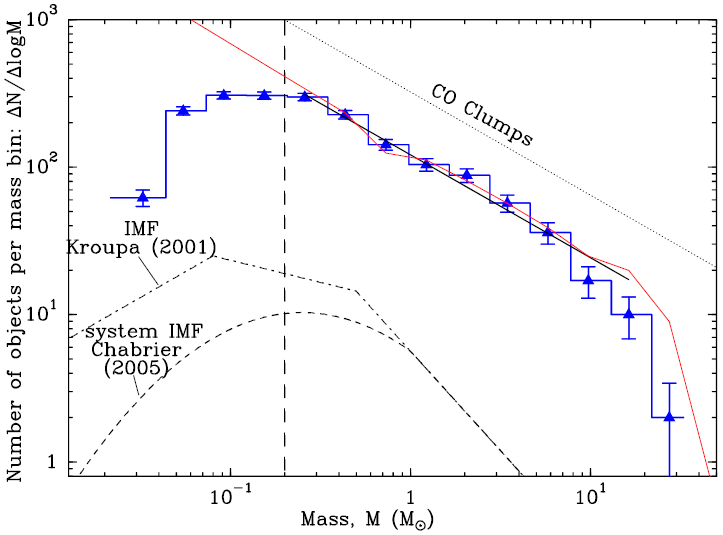}}
 \end{minipage}
   \end{center}
   \caption{Synthetic core mass function (CMF) derived from simulated source extractions (blue histogram) 
  compared to the input mass function (d$N$/dlog$M \propto M^{-0.7}$) of a population of 5622 model cores (red curve) constructed as described in the text. 
  The estimated  90\% completeness level (in observed core mass) is indicated by the vertical dashed line at $0.2\, M_\odot $. 
  The black solid line shows a power-law fit to the derived CMF above the $0.2\, M_\odot $ completeness level; it is in 
  excellent agreement with the input core mass function. 
  The drop of the synthetic CMF below the input CMF at the high-mass end is due to the fact that 
  the derived core masses tend to underestimate the true core masses by $\sim 20\% $ on average (see Fig.~\ref{fig_simu_mass_temp}a).
   }
              \label{fig_simu_CMF}%
    \end{figure}

\subsection{Monte-Carlo simulations}\label{sec:appendix_simulations} 

\begin{figure*}[!htbp]
   \begin{center}
    \hspace{-0.2cm}
 \begin{minipage}{0.46\linewidth}
     \resizebox{1.0\hsize}{!}{\includegraphics[angle=0]{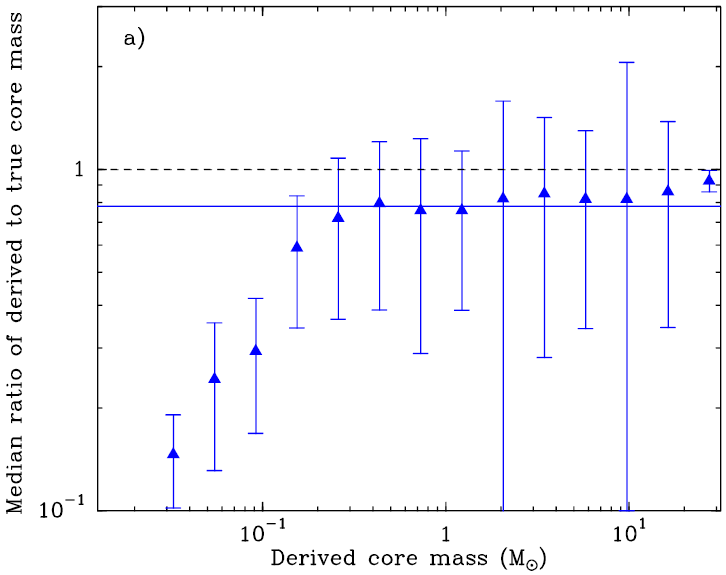}}
 \end{minipage}
 \hspace{0.5cm}
 \begin{minipage}{0.46\linewidth}
    \resizebox{1.0\hsize}{!}{\includegraphics[angle=0]{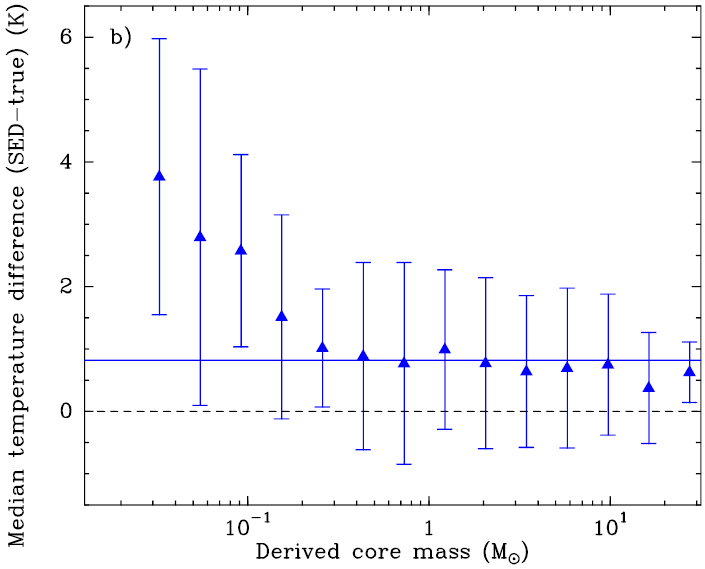}}
 \end{minipage} 
   \end{center}
   \caption{{\it Left:} Ratio of derived to intrinsic  (or `true') core mass as a function of derived core mass for the same set of simulated core extractions as used in Sect.~\ref{sec:completeness} and Fig.~\ref{fig_simu_CMF}. 
   The error bars are $\pm 1\sigma $ where $\sigma $ is the dispersion of the mass ratio in each mass bin. 
   The median mass ratio is $\sim 0.8$ above $0.4\, M_\odot $ (as indicated by the horizontal blue line) and $\sim 0.7$ close to the 90\% completeness limit of $0.2\, M_\odot $ in observed core mass.
   The horizontal dashed line marks the mass ratio of 1 expected in the case of perfect core extractions and mass estimates.
   {\it Right:} Difference between derived SED temperature and intrinsic mass-averaged dust temperature as a function of derived core mass 
   for the same set of simulated core extractions. 
   The error bars are $\pm 1\sigma $ where $\sigma $ is the dispersion of the temperature difference in each mass bin. 
   The median temperature difference is about $+0.8$~K above $0.4\, M_\odot $ (as indicated by the horizontal blue line) and $\sim 1$~K close to a derived core mass of $0.2\, M_\odot $ (completeness limit). 
   The horizontal dashed line marks the zero difference expected in the case of perfect core extractions and temperature estimates.
   }
              \label{fig_simu_mass_temp}%
    \end{figure*}

To simulate real core extractions, we first constructed 
clean maps of the background emission at all {\it Herschel} wavelengths (including a column density plane), 
by subtracting the emission of the compact cores identified with \textsl{getsources} in the observed data (cf. Sects.~\ref{sec:getsources} \& \ref{sec:core_selection}).
We then inserted several sets of model Bonnor-Ebert-like cores throughout
the clean-background images in order to generate a full set of synthetic {\it Herschel} and column density images of the region. 
In the example illustrated in Figs.~\ref{fig_simu_CMF} \& \ref{fig_simu_mass_temp}, for instance, we used a population of 5622 model 
starless cores with a flat input mass distribution (d$N$/dlog$M$ $\propto$ $M^{-0.7}$, 
similar to the mass distribution of CO clumps) 
from $0.02\, M_\odot$ to $\sim 30\, M_\odot$. 
This example is particularly useful as it allowed us to test the robustness of the conclusion that 
the observed prestellar CMF is significantly steeper than the mass distribution of CO clumps. 
The model cores had
positions in a mass versus size diagram consistent with critical Bonnor-Ebert isothermal spheres 
at effective gas temperatures $\sim \, $7--20~K.
The dust continuum emission from the synthetic Bonnor-Ebert cores in all {\it Herschel} bands 
was simulated using an extensive grid of spherical dust radiative transfer models constructed by us with the MODUST code 
\citep[e.g.,][]{Bouwman+2000, Bouwman2001PhDT}.
In particular, each of the synthetic prestellar cores was given a realistic dust temperature profile with a significant drop in dust temperature toward core center, 
as observed in the case of spatially-resolved starless cores \citep[cf.][]{Roy+2014}. 
The synthetic cores were spatially distributed randomly over the regions of the column density map where $N_{\rm H_2}^{\rm bg} \ge 5 \times 10^{21}$ cm$^{-2}$ 
(containing most, if not all, of the observed prestellar cores in the real data -- see Sect.~\ref{sec:threshold}), with no particular mass segregation.
Once satisfactory synthetic skies resembling the observed images had been generated, compact source extraction and core selection/classification 
were performed with \textsl{getsources} in the same way as for the real data (see Sects.~\ref{sec:getsources} \& \ref{sec:core_selection}). 

As mentioned in Sect.~\ref{sec:completeness} and shown in Fig.~\ref{fig_complete}, 
the results of these Monte-Carlo simulations 
suggest that our {\it Herschel} census of prestellar cores in the Aquila cloud complex is $\sim 90\% $ complete down to 
$\sim 0.3\, M_\odot $ in {\it true} core mass. 
Figure~\ref{fig_simu_CMF} further illustrates that the core mass function can be reliably determined down to the completeness mass limit.  
In this example, a Kolmogorov-Smirnov (K-S) test shows that the derived CMF is statistically indistinguishable 
(at the $\sim 90\% $ confidence level) 
from the input mass function 
above the completeness limit. In particular, the best-fit power-law function to the derived CMF (black solid line in Fig.~\ref{fig_simu_CMF}) 
is identical to the input d$N$/dlog$M$ $\propto$ $M^{-0.7}$ power law. 
This test therefore confirms that the best-fit power law to 
the observed CMF (d$N$/dlog$M$ $\propto$ $M^{-1.33 \pm 0.06}$ -- see Sect.~\ref{sec:CMF}) is significantly steeper than the typical 
mass distribution of CO clumps/clouds (d$N$/dlog$M$ $\propto$ $M^{-0.7}$ -- e.g. Blitz 1993; Kramer et al. 1998) 
and cannot be an artifact of the core extraction process.

\begin{figure}[!h]
   \begin{center}
 \begin{minipage}{1.0\linewidth}
    \resizebox{0.95\hsize}{!}{\includegraphics[angle=0]{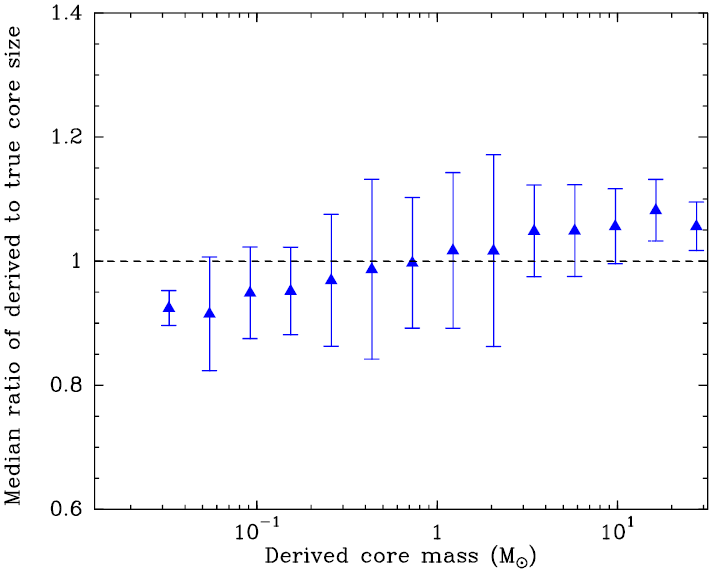}}
 \end{minipage}
   \end{center}
   \caption{Ratio of derived to true core size as a function of derived core mass for the same set of simulated core extractions as in Fig.~\ref{fig_simu_CMF} 
   and Fig.~\ref{fig_simu_mass_temp}.
   The error bars are $\pm 1\sigma $ where $\sigma $ is the dispersion of the size ratio in each mass bin. 
   The horizontal dashed line marks the size ratio of 1 expected in the case of perfect core extractions and size estimates.
   Note how the median core size measured in each mass bin remains within 5\% of the true core size above the $\sim 90\% $ completeness 
   of $\sim 0.2\ M_\odot $ in derived core mass.
   }
              \label{fig_simu_sizes}%
    \end{figure}

The same Monte-Carlo simulations were also used to assess the accuracy of the main derived parameters 
(e.g. core mass, radius, and dust temperature) by comparing the estimated values after core extraction 
to the intrinsic input values of the model cores. 
Figure~\ref{fig_simu_mass_temp}a shows that the derived core masses tend to underestimate the true core 
masses by $\sim \, $20--30\%  on average, and Fig.~\ref{fig_simu_mass_temp}b shows that the derived SED temperatures 
tend to overestimate the intrinsic mass-averaged dust temperatures of the cores by typically $\sim 1$~K. 
A similar plot for the core sizes (Fig.~\ref{fig_simu_sizes}) suggests that the derived core sizes (prior to deconvolution) are quite reliable 
and remain within $\sim 5\% $ of the true core sizes on average.  
We interpret the mass effect (Fig.~\ref{fig_simu_mass_temp}a) as a direct consequence of the temperature effect (Fig.~\ref{fig_simu_mass_temp}b) 
since overestimating the dust temperatures leads to underestimating the core masses. 
The temperature effect arises from the fact that the dust temperature derived from a global fit to the SED of a starless core
overestimates the mass-averaged dust temperature owing to a distribution of dust temperatures along the line of sight \citep[see][and Sect.~\ref{sec:deriv_core_prop}]{Roy+2014}. 

Taking the $\sim \, $20--30\% mass effect into account, we conclude that the $\sim 90\% $ completeness limit at $\sim 0.3\, M_\odot $ in {\it true} core mass 
corresponds to $\sim 0.2\, M_\odot $ in {\it observed} core mass.

\subsection{Model of the completeness problem}\label{sec:appendix_model}

The Monte-Carlo simulations described above provide an estimate of the global completeness limit of the core survey.  
The completeness level of the core extractions is, however, expected to be background dependent. 
To assess the importance of this dependence, we constructed a simplified model of the core extraction process. 

\begin{figure}[!h]
   \begin{center}
 \begin{minipage}{1.0\linewidth}
    \resizebox{1.0\hsize}{!}{\includegraphics[angle=0]{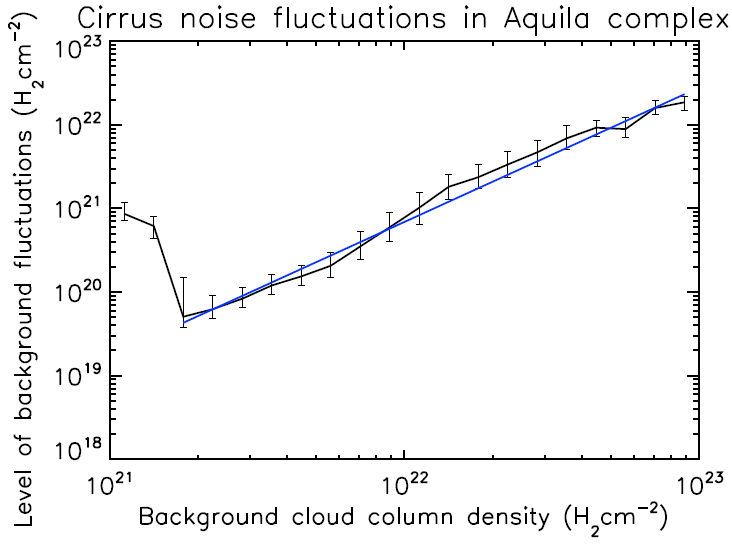}}
 \end{minipage}
   \end{center}
   \caption{Median root mean square (rms) level of background column density fluctuations as a function of 
   background cloud column density as measured in a $1\arcmin \times 1\arcmin$ sliding box 
   over the clean background image of the Aquila complex produced by \textsl{getsources} from the high-resolution column density map. 
   The error bars correspond to the interquartile range of background fluctuations about the median level in each column density bin.
   The straight line 
   represents a weighted power-law fit to the data points above $N_{\rm H_2} \sim 2 \times 10^{21}\, {\rm cm}^{-2} $.
   }
              \label{fig_cirrus}%
    \end{figure}

Owing to the high sensitivity and quality of the {\it Herschel} images, the HGBS survey is not limited by instrumental 
noise but by confusion arising from small-scale cloud structure, an effect commonly referred to as ``cirrus confusion noise'' 
in the literature \citep[e.g., see][]{Gautier+1992,Kiss+2001,Roy+2010}.
To estimate the level of such cirrus confusion noise from the {\it Herschel} data, we measured the rms level of background 
fluctuations in a sliding box $1\arcmin \times 1\arcmin$ in size\footnote{The size of the sliding box corresponds to 
$\sim 0.075\, $pc$\, \times\,  0.075$~pc at $d \sim 260$~pc, which is similar to the size scale of prestellar cores.}
over the entire column density map of the Aquila complex after subtracting the sources identified by \textsl{getsources}.
Correlating the resulting map of rms fluctuations with the input background column density map 
led to Fig.~\ref{fig_cirrus}, which clearly shows that the level of column density fluctuations increases with 
background column density approximately as a power law:  
\begin{equation}
N_{\rm H_2, rms} \sim 3.9 \times 10^{20}\, {\rm cm}^{-2} \times  \left(\frac{N_{\rm H_2, back}}{7 \times 10^{21}\, {\rm cm}^{-2} }\right)^{1.6}. 
\end{equation} 
The power-law index of 1.6 
derived here from {\it Herschel} data is very similar to that reported in earlier papers discussing 
cirrus noise (e.g. Gautier et al. 1992, Kiss et al. 2001, Roy et al. 2010). 
Since the level of background fluctuations increases with column density, one expects core extraction to be increasingly more difficult 
and thus survey completeness to decrease significantly in higher column density areas within the field.

The model we used to estimate the magnitude of this effect and get around the problem of a background-dependent completeness level
was based on the following assumptions:

$\bullet$ A dense core is defined as the immediate vicinity of a column density peak departing significantly, i.e., by more 
than $5 \times N_{\rm H_2, rms} $ from the field of background cloud fluctuations (see core selection criteria in Sect~\ref{sec:core_selection}).

$\bullet$ A prestellar core, i.e., a self-gravitating starless core, can be approximately modeled 
as a critical Bonnor-Ebert spheroid of mass $M_{\rm BE}$ and outer radius $R_{\rm BE}$, bounded by the gravitational 
pressure of the background cloud $P_{{\rm back}} \approx 0.88~G~\Sigma_{\rm back}^2$ \citep{McKeeTan2003}, where 
$\Sigma_{\rm back} = \mu m_{\rm H}\times N_{\rm H_2, back}$. The mean intrinsic column density contrast of such a model 
prestellar core is $\Sigma_{\rm BE}/\Sigma_{\rm back} \sim 1.5$, where $\Sigma_{\rm BE} \equiv M_{\rm BE}/(\pi\, R_{\rm BE}^2) $.

\begin{figure}[!h]
   \begin{center}
 \begin{minipage}{1.0\linewidth}
    \resizebox{1.0\hsize}{!}{\includegraphics[angle=0]{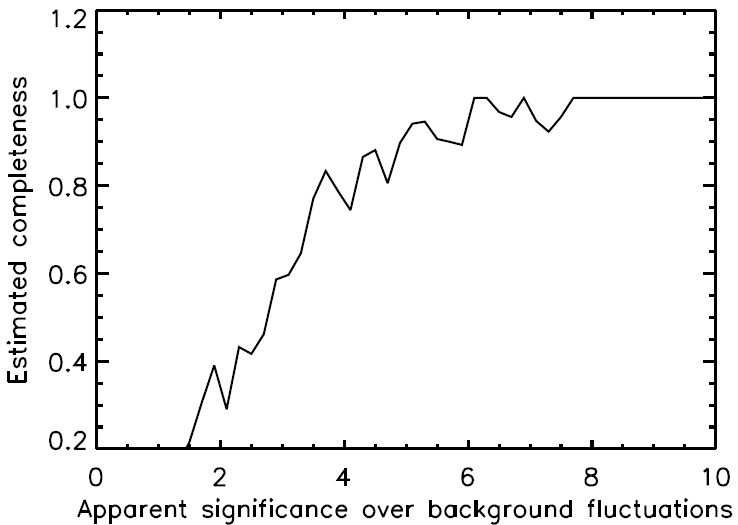}} 
 \end{minipage}
   \end{center}
   \caption{Completeness curve as a function of apparent column density significance over local background cloud 
   fluctuations derived from the Monte-Carlo simulations described in Sect.~\ref{sec:completeness}.
   }
              \label{fig_sig_completeness}%
    \end{figure}

$\bullet$ The ability to detect a core in the {\it Herschel} data depends primarily on the apparent column density significance 
of the core defined as $\Sigma_{\rm core, obs}/\Sigma_{\rm rms}$, where $\Sigma_{\rm core, obs}$ is the apparent (observed) 
column density of the core after convolution with the observing beam, i.e.,  $\Sigma_{\rm core, obs} \equiv M_{\rm core}/(\pi\, R_{\rm core, conv}^2) $, 
and $\Sigma_{\rm rms} = \mu m_{\rm H}\times N_{\rm H_2, rms}$. The Monte-Carlo simulations of 
Appendix~\ref{sec:appendix_simulations} 
are consistent with this assumption and suggest that the completeness level is $\simgt 90\% $ for cores with an apparent column density 
significance larger than 5 (see Fig.~\ref{fig_sig_completeness}). 

In outline, our simplified model of the completeness problem may be described as follows:

$\bullet$ Two effects, beam dilution and temperature dilution, can make the apparent column density contrast  
$\Sigma_{\rm BE, obs}/\Sigma_{\rm back}$ of a model core smaller than its intrinsic column density contrast of 1.5:
$$\Sigma_{\rm BE, obs}/\Sigma_{\rm back} \sim 1.5 \times (R_{\rm BE}/R_{\rm BE, conv})^2 \times [B_{\nu_{\rm eff}}(T_{\rm core})/B_{\nu_{\rm eff}}(T_{\rm back}), $$ 
where $\nu_{\rm eff}$ is a fiducial {\it Herschel} observing frequency which we take to correspond to $\lambda \sim 350\, \mu$m.
Taking advantage of the fact that the column density distribution of a Bonnor-Ebert core with outer radius $R_{\rm BE}$ is well approximated by 
a Gaussian distribution of FWHM $\sim R_{\rm BE}$, the observed radius of the core is approximately $R_{\rm BE, conv} = (R_{\rm BE}^2 + \overline{HPBW}^2)^{1/2}$
(where $\overline{HPBW}$ corresponds to the half-power beam width resolution of the column density map projected at the distance of the Aquila cloud), 
and the beam dilution factor can thus be expressed as $ (R_{\rm BE}/R_{\rm BE, conv})^2 = 1/[1+(\overline{HPBW}/R_{\rm BE})^2] $.

$\bullet$ The apparent column density significance can be written as the product of the apparent column density contrast and a cirrus noise
factor, $\Sigma_{\rm BE, obs}/\Sigma_{\rm rms} = (\Sigma_{\rm BE, obs}/\Sigma_{\rm back}) \times (\Sigma_{\rm back}/\Sigma_{\rm rms})$, 
where the cirrus noise factor is:
$$ \Sigma_{\rm back}/\Sigma_{\rm rms} = N_{\rm H_2, back}/N_{\rm H_2, rms} \sim 18 \times  \left(\frac{N_{\rm H_2, back}}{7 \times 10^{21}\, {\rm cm}^{-2} }\right)^{-0.6}, $$
according to Eq.~(B.1).

\begin{figure}[!h]
   \begin{center}
 \begin{minipage}{1.0\linewidth}
    \resizebox{1.0\hsize}{!}{\includegraphics[angle=0]{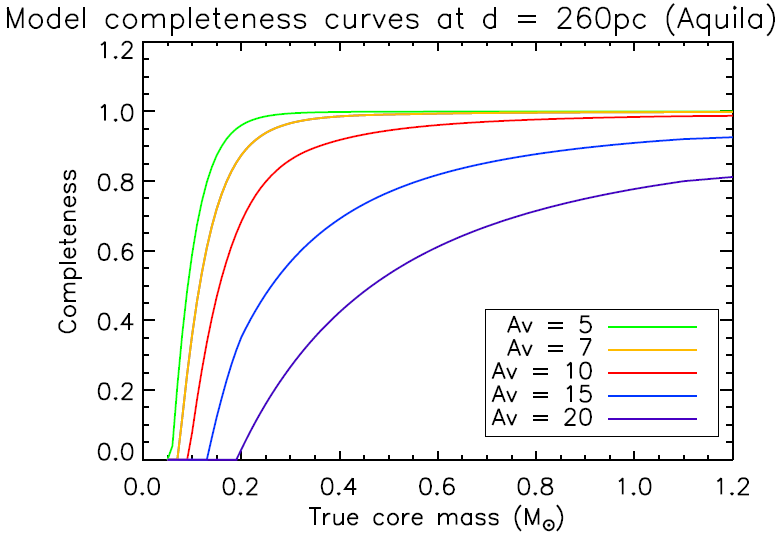}} 
 \end{minipage}
   \end{center}
   \caption{Model completeness curves of {\it Herschel} prestellar core extractions in Aquila for five values of the 
   background cloud column density expressed in units of visual extinction from $A_{\rm V, back} = 5$ to $A_{\rm V, back} = 20$.
   }
              \label{fig_completeness}%
    \end{figure}

$\bullet$ Assuming that the fundamental completeness curve  is the completeness function $\mathcal{F}(\tilde{S})$ of apparent column density significance $\tilde{S}$ 
shown in Fig.~\ref{fig_sig_completeness}, completeness can be estimated as a function of core mass and background column density as 
$ \mathcal{C}(M_{\rm BE},\Sigma_{\rm back}) = \mathcal{F}[\tilde{S}(M_{\rm BE},\Sigma_{\rm back})]$.
The corresponding function of $M_{\rm BE}$ is shown for five values of the background column density $N_{\rm H_2, back}$ in Fig.~\ref{fig_completeness}. 
Figure~\ref{fig_completeness} shows how the completeness of prestellar core extractions is expected to decrease as  background cloud column density
and cirrus noise increase.

$\bullet$ To estimate a global completeness curve for our census of prestellar cores in the Aquila complex, 
we used the observed distribution of mass in the cloud as a function of background column density (cf. Figs.~\ref{fig_cdPDF}a/b) 
and took advantage of the existence of a column density ``threshold'' at $A_{\rm V, back} \sim \,$5--7, above which the bulk of core and star formation 
is believed to occur (cf. Sect.~\ref{sec:threshold} and Fig.~\ref{fig_bgCD}) and the column density PDF is well fitted by a power-law distribution. We also assumed that the number of prestellar  
cores in the cloud scales linearly with cloud mass above the threshold. This assumption is consistent with recent infrared studies which find that the 
global star formation rate tends to be linearly proportional to the mass of dense gas above the threshold 
\citep[e.g.,][]{Heiderman+2010,Lada+2010,GaoSolomon2004}. 
It is also consistent with the roughly constant prestellar core formation efficiency 
found here above the threshold (see Fig.~\ref{fig_cfe}).
The global completeness curve was thus computed as a weighted average of the individual completeness curves at fixed
background column densities:
$$ \mathcal{GC}(M_{\rm BE}) =  \frac{1}{M_{\rm dense}}  \int_{A_{\rm V} = 5}^{+\infty} \mathcal{C}(M_{\rm BE},\Sigma_{\rm back}) \frac{dM_{\rm dense}}{d\Sigma }(A_{\rm V, back}) dA_{\rm V, back}. $$

\begin{figure}[!h]
   \begin{center}
 \begin{minipage}{0.8\linewidth}
    \resizebox{1.0\hsize}{!}{\includegraphics[angle=0]{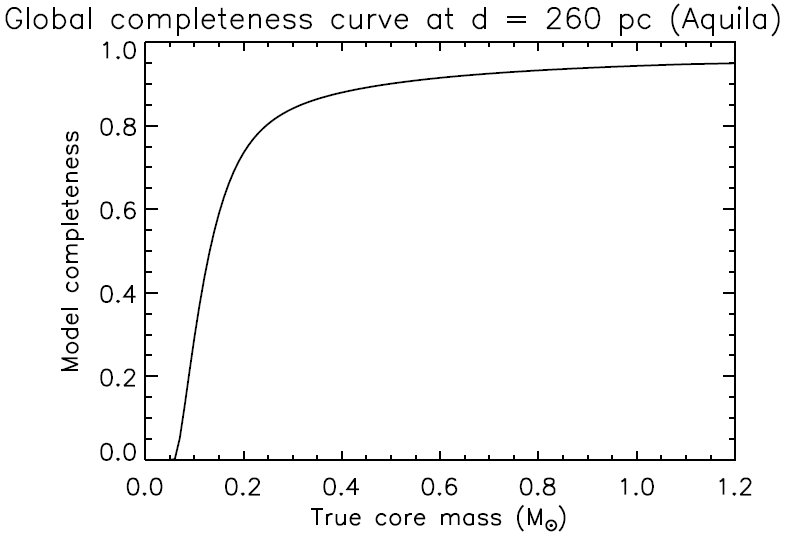}} 
 \end{minipage}
   \end{center}
   \caption{Model completeness curve of {\it Herschel} prestellar core extractions in the Aquila cloud complex 
   ($d = 260$~pc) as a function of intrinsic model core mass. 
   }
    \label{fig_global_completeness}%
    \end{figure}

The resulting global completeness curve, which represents the best estimate of the completeness of our {\it Herschel} survey for prestellar cores in 
Aquila according to our model, is shown in Fig.~\ref{fig_global_completeness}.  It can be seen that this global completeness curve is very similar 
to the individual completeness curves for background column densities close to the threshold (see $A_{\rm V, back} =\, $5--10 curves in Fig.~\ref{fig_completeness}).
It is also very similar to the empirical completeness curve derived from Monte-Carlo simulations (see Sect.~\ref{sec:completeness}). 
The model completeness curve is almost flat above a true core mass level of $0.3\, M_\odot $. 
Using this model curve to correct the observed CMFs of {\it candidate} and {\it robust} prestellar cores for incompleteness  
would only have a minimal effect in Fig.~\ref{fig_CMF} above an observed core mass level of $\sim 0.2\, M_\odot $. 
(The corrected CMFs differ from the uncorrected CMFs only below $\sim 1\, M_\odot $ and by much less than 
the uncertainty area displayed in light blue in Fig.~\ref{fig_CMF}.)

\section{Effect of distance uncertainty}\label{sec:appendix_distance}

As mentioned in Sect.~\ref{sec:AqlRift}, there is some ambiguity concerning the distance to 
the Aquila molecular cloud complex. 
A number of arguments, presented by \citet{Bontemps+2010} and summarized 
in Sect.~\ref{sec:AqlRift}, suggest that the bulk of the region studied here and shown in Fig.~\ref{fig_cd} 
corresponds to a coherent cloud complex at $d_{-} = 260$~pc 
\citep[see also][]{Gutermuth+2008}, which is the default distance adopted in the present paper. 
Other studies in the literature (see references in Sect.~\ref{sec:AqlRift}), 
however, place the complex at the larger distance, $d_{+} = 415$~pc, of the Serpens Main cloud \citep[][]{Dzib+2010}.
It is thus worth discussing how our results would be affected 
if we had adopted the larger distance estimate,  $d_{+} $, instead of $d_{-}$. 
The core mass estimates, which scale as $S_{\nu}\, d^2/[B_{\nu}(T_{\rm d})\, \kappa_{\nu}] $ where 
$S_{\nu}$ is integrated flux density and $B_{\nu}(T_{\rm d})$ is the Planck function,  
would systematically increase by a factor of 2.5. 
This would shift the CMFs shown in Fig.~\ref{fig_CMF} and Fig.~\ref{fig_CMF_onoff} to the right 
and thus lower the efficiency $ \epsilon_{\rm core}$ from $0.4^{+0.2}_{-0.1}$ 
to $0.2 \pm 0.1$. 
In comparison, the core size estimates, which scale linearly with distance $d$, would increase by only 60\%. 
The BE mass ratio $\alpha_{\rm BE} = M_{\rm BE,crit} / M_{\rm obs} $, listed in Col.~17 of online Table~A.2, 
scales as $d^{-1}$ and would {\it decrease} by 60\% for all cores. 
Accordingly, all cores would move upward as indicated by an arrow in the mass versus size diagram of Fig.~\ref{fig_massSize}, 
which would {\it increase} the fraction of prestellar cores among starless cores 
from $60\% \pm 10\%$ to $70\% \pm 10\%$. 
More precisely, the number of {\it candidate} prestellar cores would increase from 446 to 565 
and the number of {\it robust} prestellar cores would increase from 292 to 391, 
while the total number of starless cores (651) would remain the same. 
Accordingly, the estimated lifetime of {\it candidate} prestellar cores 
would also slightly increase from $\sim 1.4$~Myr to $\sim 1.8$~Myr, 
and that of {\it robust} prestellar cores from $\sim 0.9$~Myr to $\sim 1.3$~Myr (see Sect.~\ref{sec:lifetime}), 
leading to $t_{\rm pre} = 1.5 \pm 0.3$~Myr. 
The prestellar core formation efficiency (CFE) as a function of background column density (cf. Fig.~\ref{fig_cfe}), 
and in particular the roughly constant value ${\rm CFE_{max}}  \equiv  f_{\rm pre} \sim 15\% $ at high column densities, 
would not change. 
Our corresponding estimate of the ``efficiency'' of the star formation process in dense gas (cf. Sect.~\ref{sec:SFR}),  
${\rm SFR}/M_{\rm dense} = f_{\rm pre} \times \epsilon_{\rm core}/t_{\rm pre}$, would however 
decrease from $5 \times 10^{-8}\,  {\rm yr}^{-1}$ to $2 \times 10^{-8}\,  {\rm yr}^{-1}$, 
becoming closer to the efficiency value 
reported by \citet{Evans+2014} and \citet{GaoSolomon2004} 
than to the value found by \citet{Lada+2010}. 
Finally, the column density maps shown in Fig.~\ref{fig_cd},  Fig.~\ref{fig_fil-cur}, Fig.~\ref{fig_fil-gf}, Fig.~\ref{fig_cores-critFilCont}, 
and Fig.~\ref{fig_curvelet}, 
as well as the spatial correspondence between cores and filaments, 
would remain unchanged. 
The scaling of our column density maps in terms of mass per unit length along the filaments would however 
change by $\sim 60\% $ upward, since the characteristic {\it physical} width of the filaments would increase by $\sim 60\% $. 
As a consequence, the white areas which highlight supercritical filaments in Figs.~\ref{fig_fil-cur},~\ref{fig_fil-gf},~\ref{fig_cores-critFilCont}, 
and \ref{fig_curvelet} would slightly {\it expand}, improving the correspondence between 
the spatial distribution of prestellar cores/protostars and that of supercritical filaments.
To summarize, our main conclusions do not depend strongly on the adopted distance. 

\end{appendix}

\newpage
\clearpage

\Online

\onlfig{}{
\setcounter{figure}{17}

   \begin{figure}[!h]
   \begin{center}
 \begin{minipage}{1.0\linewidth}
    \resizebox{1.0\hsize}{!}{\includegraphics[angle=0]{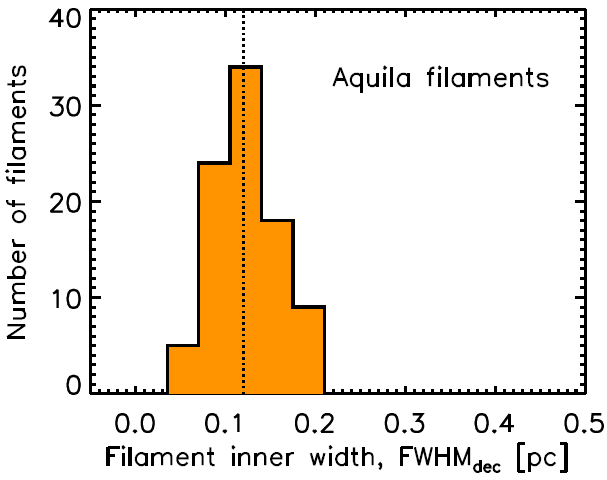}}
 \end{minipage}
   \end{center}
   \caption{Distribution of mean FWHM inner widths for the 90 filaments traced with DisPerSE 
  in the Aquila entire field (see blue skeleton in Figs.~\ref{fig_fil-cur} \&  \ref{fig_fil-gf} and Sect.~\ref{sec:filam}). 
  These widths results from a filament profile analysis similar to that described in \citet{Arzoumanian+2011}  
  and were deconvolved from the $18.2\arcsec $ HPBW resolution of the high-resolution column density map 
  used to construct the radial profiles of the filaments. 
  The median filament width is 0.12~pc, as marked by the vertical dotted line, 
  and the standard deviation of the distribution is 0.04~pc.
   }
              \label{fig_fil_width}%
    \end{figure}
%

   \begin{figure*}[!hb]
   \begin{center}
 \begin{minipage}{0.49\linewidth}
     \resizebox{1.0\hsize}{!}{\includegraphics[angle=0]{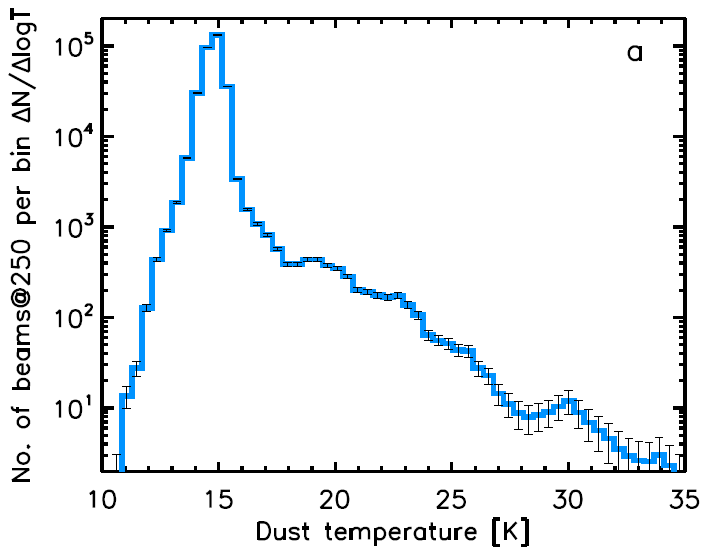}}
 \end{minipage}
 \begin{minipage}{0.49\linewidth}
    \resizebox{1.0\hsize}{!}{\includegraphics[angle=0]{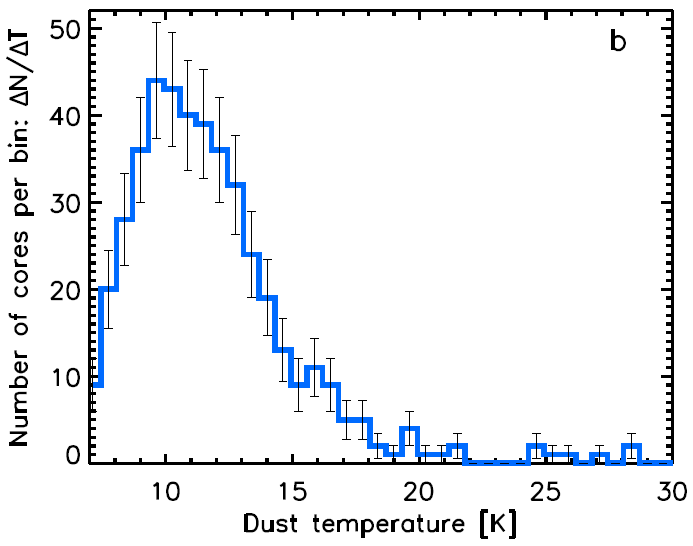}}
 \end{minipage} 
   \end{center}
   \caption{{\bf a)} Distribution of dust temperature values in the Aquila temperature map shown in Fig.~\ref{fig_temp}.
           {\bf b)} Distribution of SED dust temperatures for all selected starless cores with reliable SED fits
           (see Sect.~\ref{sec:deriv_core_prop}). Note how the distribution of core temperatures peaks at significantly 
	   lower values than the distribution of background cloud temperatures.
            }
              \label{fig_map-core_temp}%
    \end{figure*}
}

\end{document}